\documentclass[twocolumn]{aastex62}
\usepackage{color,epsfig,graphics,graphicx}
\usepackage{natbib}
\usepackage{graphicx}
\usepackage{xcolor}
\usepackage{tablefootnote}
\usepackage{natbib}
\usepackage{booktabs}
\usepackage{bbding}
\usepackage{textgreek}
\usepackage[percent]{overpic} 
\usepackage{hyperref}

\usepackage[symbol]{footmisc}

\definecolor{brown}{rgb}{0.6,0.4,0.2} 
\definecolor{purple}{rgb}{0.5,0,0.5} 

\newcommand{\oif}{[O\,I]}

\newcommand{\neiiif}{\ion{[Ne}{3}]}  
  
\newcommand{\nevf}{\ion{[Ne}{5}]}

\newcommand{\kms}{km~s$^{-1}$}

\usepackage{amsmath}

\shorttitle{$ISO$ Supernova Remnants}
\shortauthors{Millard et al.}

\newcommand{\mic}{$\mu$m} 

\newcommand{\ergsb}{erg\,s$^{-1}$\,cm$^{-2}$\,sr$^{-1}$}

\begin{document} 

\renewcommand{\thefootnote}{\fnsymbol{footnote}}

\title{A Spectroscopic Study of Supernova Remnants with the Infrared Space Observatory\footnote{Based on observations with $ISO$, an ESA project with instruments funded by ESA Member States (especially the PI countries: France, Germany, the Netherlands and the United Kingdom) and with the participation of ISAS and NASA.}}

\renewcommand*{\thefootnote}{\arabic{footnote}}

\author{Matthew J. Millard}
\affil{Box 19059, Department of Physics, University of Texas at Arlington, Arlington, TX 76019}
\correspondingauthor{Matthew J. Millard}
\email{matthew.millard@mavs.uta.edu}

\author{Aravind P. Ravi}
\affiliation{Box 19059, Department of Physics, University of Texas at Arlington, Arlington, TX 76019}

\author{Jeonghee Rho}
\affil{SETI Institute, 189 N. Bernardo Ave, Suite 200, Mountain View, CA 94043, USA}

\author{Sangwook Park}
\affiliation{Box 19059, Department of Physics, University of Texas at Arlington, Arlington, TX 76019}

\begin{abstract} 

We present far-infrared (FIR) spectroscopy of supernova remnants (SNRs) based on the archival data of the Infrared Space Observatory ($ISO$) taken with the Long Wavelength Spectrometer (LWS).  Our sample includes previously unpublished profiles of line and continuum spectra for 20 SNRs in the Galaxy and Magellanic Clouds. In several SNRs including G21.5--0.9, G29.7--0.3, the Crab Nebula, and G320.4--1.2, we find evidence for broad [O I], [O III], [N II], and [C II] lines with velocity dispersions up to a few 10\textsuperscript{3} km s\textsuperscript{--1}, indicating that they are associated with high--velocity SN ejecta. Our detection of Doppler-broadened atomic emission lines and a bright FIR continuum hints at the presence of newly formed dust in SN ejecta.
For G320.4--1.2, we present the first estimate of 
an ejecta-dust mass of  0.1 -- 0.2 M\textsubscript{\(\odot\)}, which spatially coincides with the broad line emission,  
by applying a blackbody model fit with components of the SNR and background emission. 
Our sample includes raster maps of 63, 145 \textmugreek{}m [O I] and 158 \textmugreek{}m [C II] lines toward SNRs Kes 79, CTB 109, and IC 443. Based on these line intensities, we suggest interacting shock types in these SNRs. Finally, we compare our LWS spectra of our sample SNRs with the spectra of several HII regions, and discuss their FIR line intensity ratios and continuum properties.  Follow--up observations with modern instruments (e.g. $JWST$ and $SOFIA$) with higher spatial and spectral resolution are encouraged for an extensive study of the SN ejecta and the SN dust.

\end{abstract} 


\section{Introduction}

Supernovae (SNe) enrich the interstellar medium (ISM) by dispersing metal-rich ejecta created in the SN explosion into their surrounding environment.  In the years following a CCSN (since Type Ia
SNe have not been observed to form dust), the ejecta cools quickly as it expands, allowing chemical reactions to take place, which leads to the creation of molecules and dust grains in the ejecta \citep{todini01, nozawa03, sarangi15, sluder18}. The elements comprising these grains, including Si and O, will likely remain locked in dust until they encounter a strong reverse shock created by the interaction of the blast wave and the ISM, or the shock wave from the pulsar wind nebula (PWN). The ejecta are compressed and heated by the shock, causing ionization and  destruction of the molecules and dust within. Once the ejecta cools to $\sim$ 10$^2$ -- 10$^3$ K, line emission from atomic fine-structure transitions is radiated in the far-infrared (FIR) band. These emission lines originate from a rapidly expanding shell of ejecta and therefore are broadened due to the Doppler effect. The radiation from the surviving dust grains shows a blackbody-like spectrum (the so-called ``grey-body''), where the overall flux level depends on the total mass of the dust. 

It has yet to be determined if SNe are major sources of dust in the Universe. The large quantity of dust observed in high-redshift galaxies \citep{bertoldi03, laporte17, isaak02, hirashita17} raises the fundamental astrophysical question on the origin of dust in the Universe, because the timescales of dust formation in low-mass stellar winds are longer than the age of the Universe in that era. On the other hand, CC-SNe of massive stars may create dust on short timescales of $\lesssim$ several 10\textsuperscript{6} yr \citep{morgan03,dwek08}, and thus are compelling candidate sources for significant dust production in the early Universe.

The Long Wavelength Spectrometer \citep[LWS;][]{clegg96} on board the {\it Infrared Space Observatory} \citep[{\it ISO};][]{kessler96}  operated in the 43 -- 196.8 \textmugreek{}m band, making it an ideal instrument for measuring FIR atomic fine-structure lines and continuum emission from cool dust in SN ejecta.  The {\it ISO}  mission lasted from November 1995 to May 1998, utilizing a 60--cm diameter primary mirror cooled to 1.7 K with superfluid helium, and performed over 26,000 scientific observations during its run.  $ISO$ was an important tool for the diagnosis of the physics and chemistry of a wide range of astrophysical objects.  For a comprehensive review of the achievements of $ISO$, see \citet{cesarsky06}. Although it had a made significant contributions in the field of infrared astronomy, some archival $ISO$ data sets have yet to be analyzed, including a substantial fraction of the supernova remnant (SNR) observations.

To make use of the unpublished archival {\it ISO} spectroscopic data, we probed the {\it ISO} data archive for LWS spectra of SNRs.  In these spectra, we search for Doppler-broadened atomic fine-structure emission lines due to high-speed ejecta that are also coincident with bright continuum emission from dust.  Based on these emission lines and underlying FIR continuum, we study the characteristics of the shocked gas and dust. We present the results of our FIR spectroscopic study of 18 CC-SNRs. We also present our FIR emission line profile measurements for two Type Ia SNRs (Kepler's and Tycho's SNRs) in this work. Although Type Ia SNe would unlikely contribute to the dust budget of the early Universe, our measurements of FIR fine-structure emission line profiles may be helpful to search for unshocked ejecta in these historical Type Ia SNRs (which is beyond the scope of this work). We investigate our sample by comparing [O I] and [C II] line strengths in several SNRs interacting with molecular clouds with various shock models. Finally, we present previously unpublished LWS data of HII regions to compare their emission line profiles and ratios with those for SNRs. We show that several FIR line flux ratios may be useful to differentiate SNRs from HII regions in future SNR searches.

%

\section{Observations and Data}\label{sec:obs}

\subsection{Sample Selection}
We searched the {\it ISO} data archive for LWS observations of SNRs.  We identified 31 instances of observations where the data are unpublished, including 20 individual SNRs. Table \ref{tab:obsList} lists these observations. $ISO$ LWS observations of SNRs that are not included in our sample are shown in Appendix A (Table \hyperlink{tabA1}{A.1}). For each observation in our sample, we selected the Highly Processed Data Products (HPDPs)\footnote{https:$//$www.cosmos.esa.int$/$web$/$iso$/$highly-processed-data-products} from the archive \citep{lloyd03}.  We use the average values of the forward and reverse scans in our analysis.  We also compare the unpublished LWS data from the HII regions W51, G159.6--18.5, and NGC 6334, and the planetary nebula (PN) NGC 6720, with the SNR spectra.

\subsection{The LWS Spectra}
\begin{figure*}
\plotone{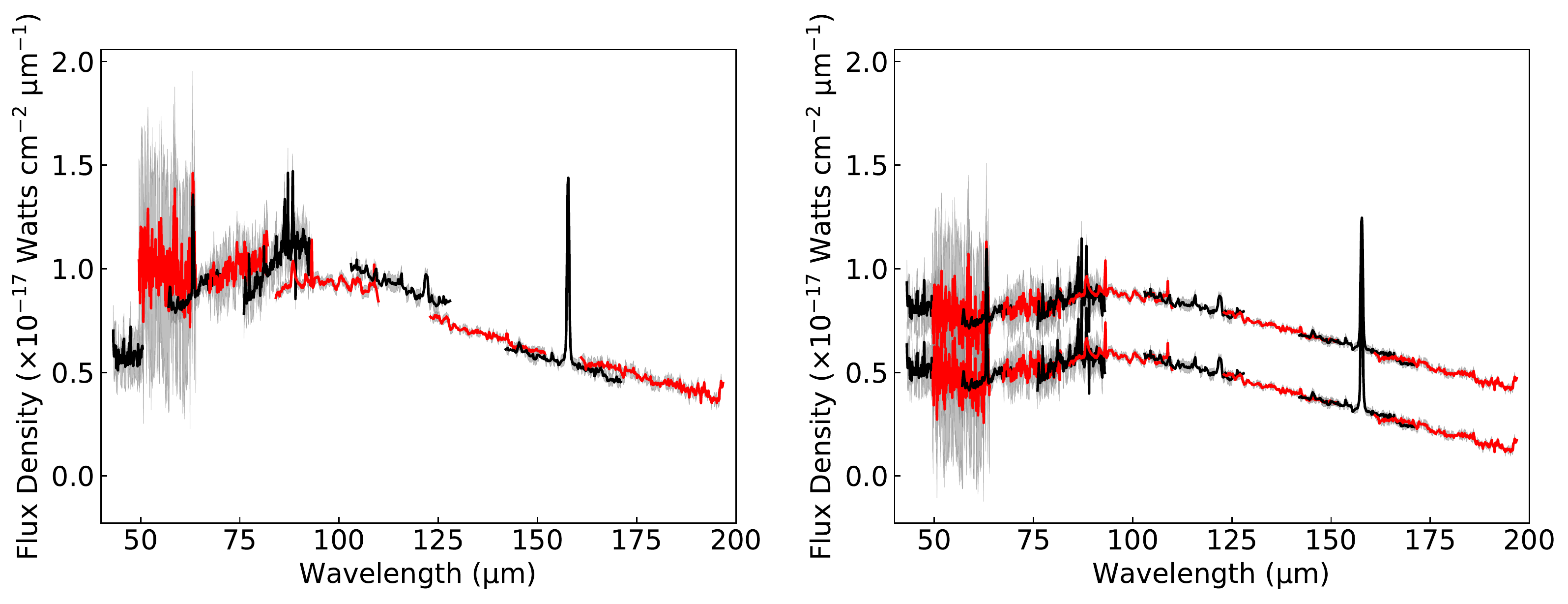}
\caption{Left Panel: The $ISO$ LWS spectrum of RCW 103 taken directly from the $ISO$ Data Archive. The alternating red and black colors represent different subdetector spectra. Right Panel: The same spectrum as the left panel with the extended source correction and subdetector matching applied. Applying relative normalizations among the individual subdetector spectra results in a range of flux levels of the overall spectrum. The upper and lower bound flux levels are shown.  The extended source corrections reduce the overall flux level by $\sim$ 30$\%$ on average.\label{fig:rcw103_before-after} }
\end{figure*}

\begin{figure*}
\centering
\begin{overpic}[width=8 cm,tics=10]{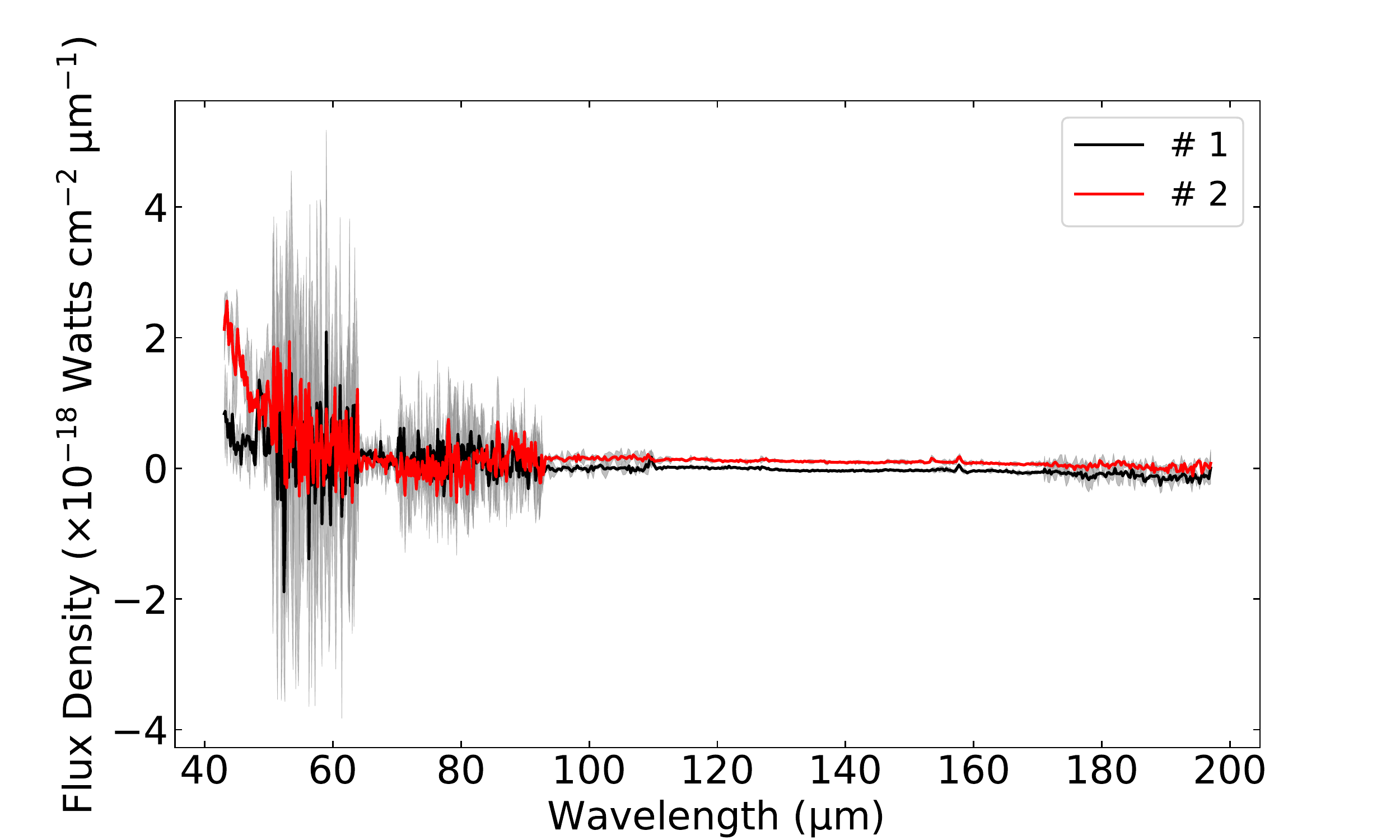}
 \put (12,55) {\Large \textbf{Kepler's SNR}}
\end{overpic}
\includegraphics[width=5.2cm]{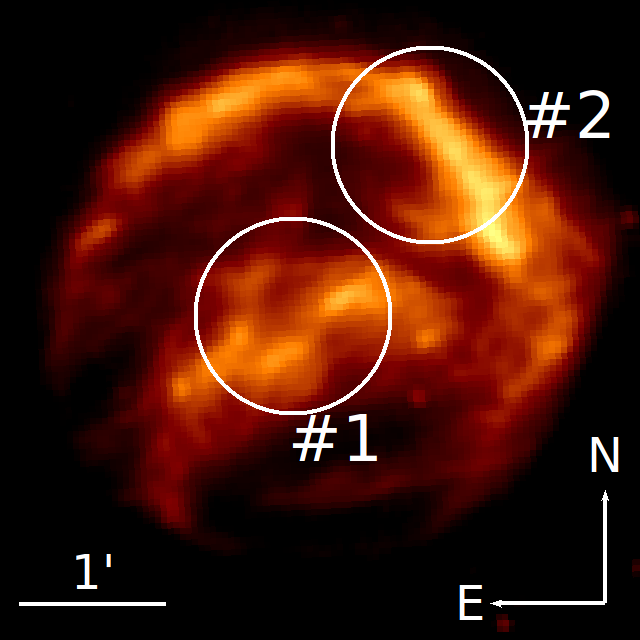}
\caption{Left: The $ISO$ LWS spectra of \hypertarget{w28-exact}{Kepler's SNR}. Right: The $Spitzer$ MIPS 24 \textmugreek{}m image of Kepler's SNR. Two $ISO$ LWS pointings are shown with white circles.}
\label{fig:kep}
%
\end{figure*}

\begin{figure*}[!b]
\centering
\begin{overpic}[width=8 cm,tics=10]{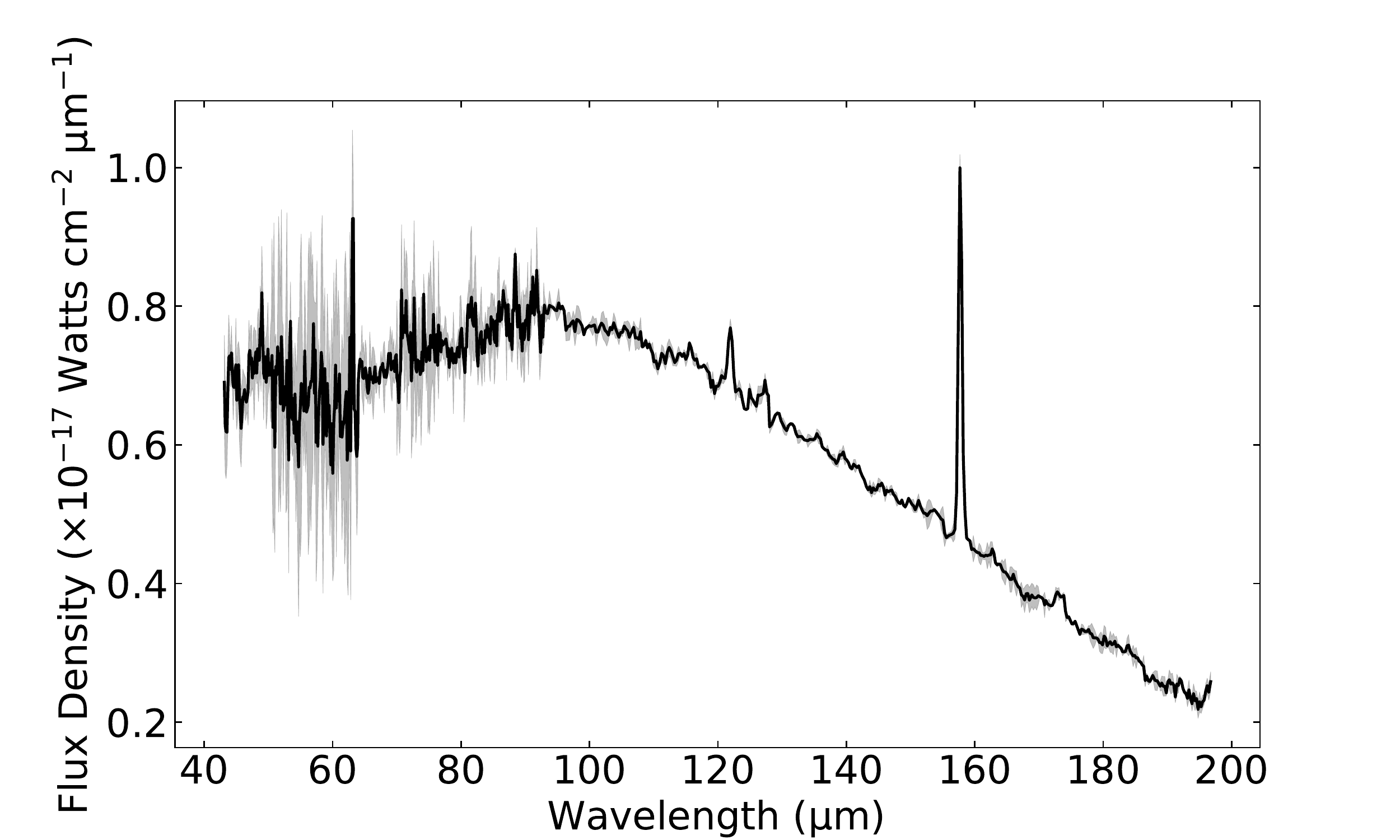}
 \put (12,55) {\Large \textbf{W 28}}
\end{overpic}
\includegraphics[width=5.2cm]{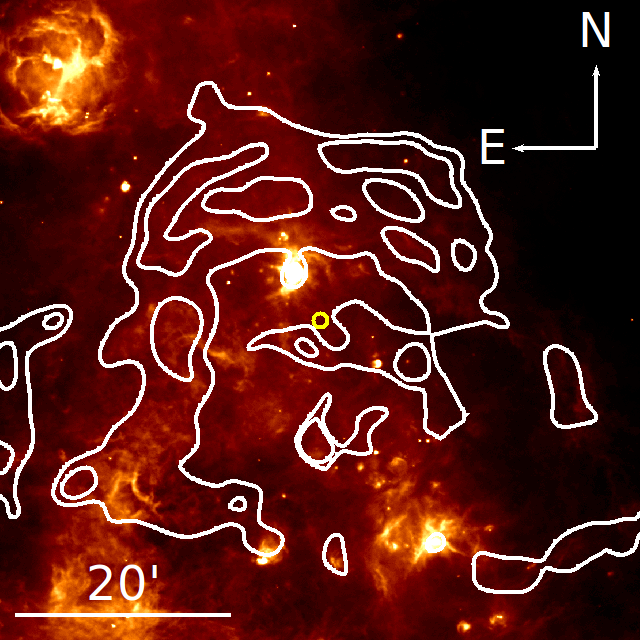}
\caption{Left: The $ISO$ LWS spectrum of \hypertarget{g21p5-exact}{W 28}. Right: The {\it Herschel} PACS 70 \textmugreek{}m image of W 28. The $ISO$ LWS pointing is shown with a yellow circle. The white contours are from a VLA 325 MHz image.  }
\label{fig:w28}
\end{figure*}

\begin{figure*}
\centering
\begin{overpic}[width=8 cm,tics=10]{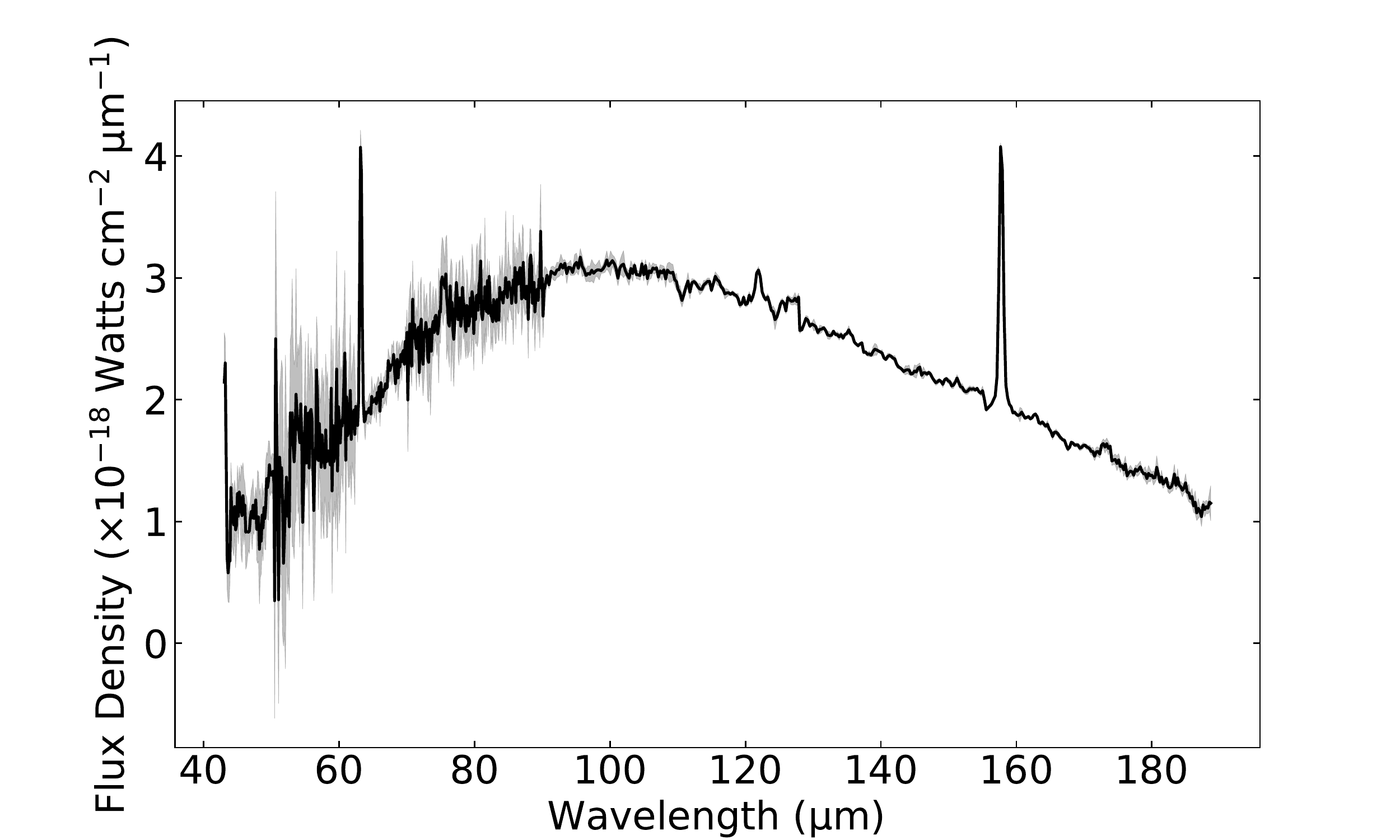}
 \put (12,55) {\Large \textbf{G21.5--0.9}}
\end{overpic}
\includegraphics[width=5.2cm]{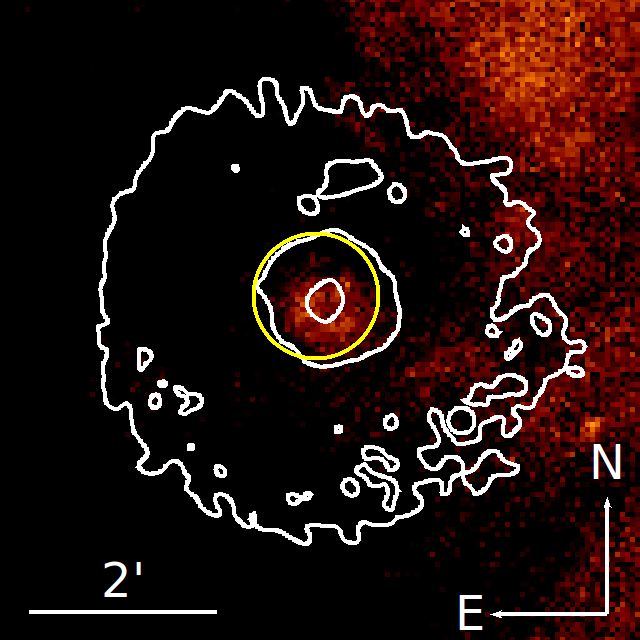}
\caption{Left: The $ISO$ LWS spectrum of \hypertarget{g29p7-exact}{G21.5--0.9}. Right: The {\it Herschel} PACS 70 \textmugreek{}m image of G21.5--0.9. The $ISO$ LWS pointing is shown with a yellow circle. The white contours are from a {\it Chandra} (0.5 -- 7.0 keV) image.} 
\label{fig:g21p5}
\end{figure*}

\begin{figure*}
\centering
\begin{tabular}{lc}
\begin{overpic}[width=8 cm,tics=10]{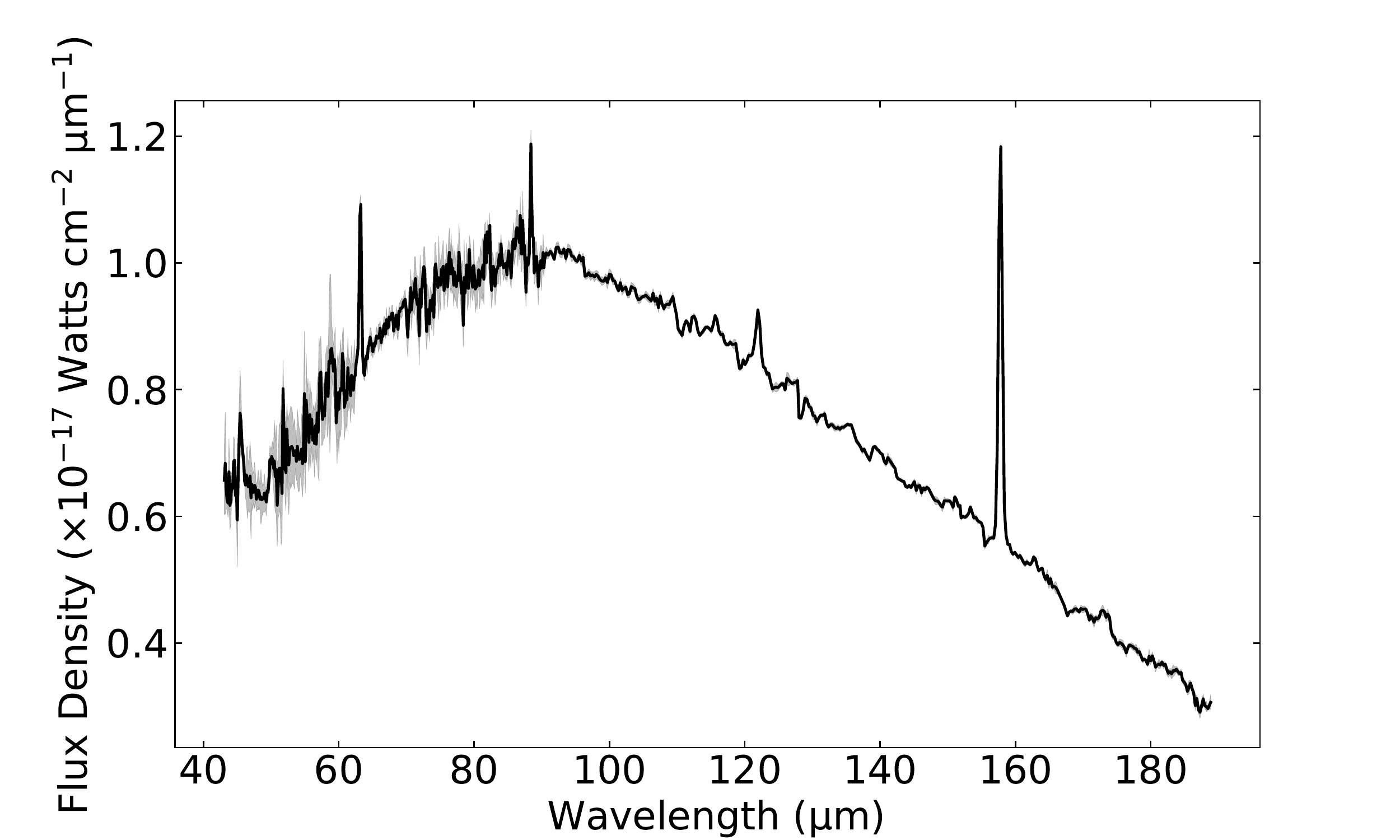}
 \put (12,55) {\Large \textbf{G29.7--0.3}}
\end{overpic}&
\includegraphics[width=5.2cm]{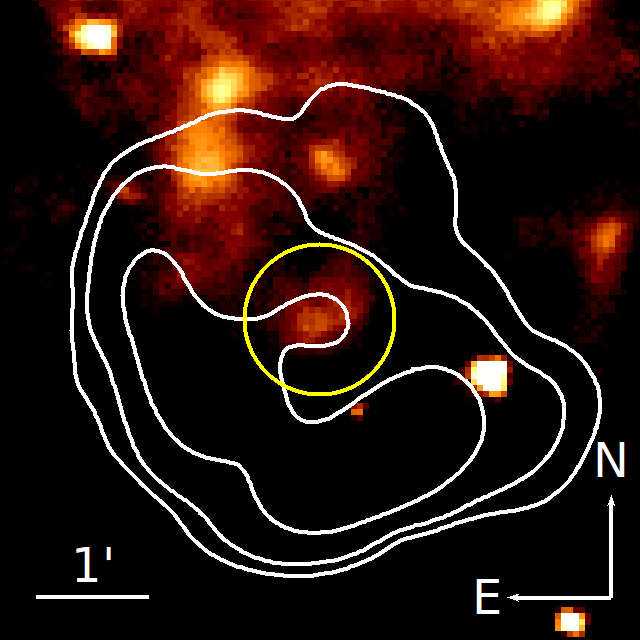}
\end{tabular}
\caption{
Left: The $ISO$ LWS spectrum of \hypertarget{}{G29.7--0.3}. Right: The {\it Herschel} PACS 70 \textmugreek{}m image of G29.7--0.3. The $ISO$ LWS pointing is shown with a yellow circle.
The white contours are from a VLA (1.4 GHz) image.}  
\label{fig:g29p7}
\end{figure*}

\begin{figure*}
\centering
\begin{tabular}{lc}
\begin{overpic}[width=8 cm,tics=10]{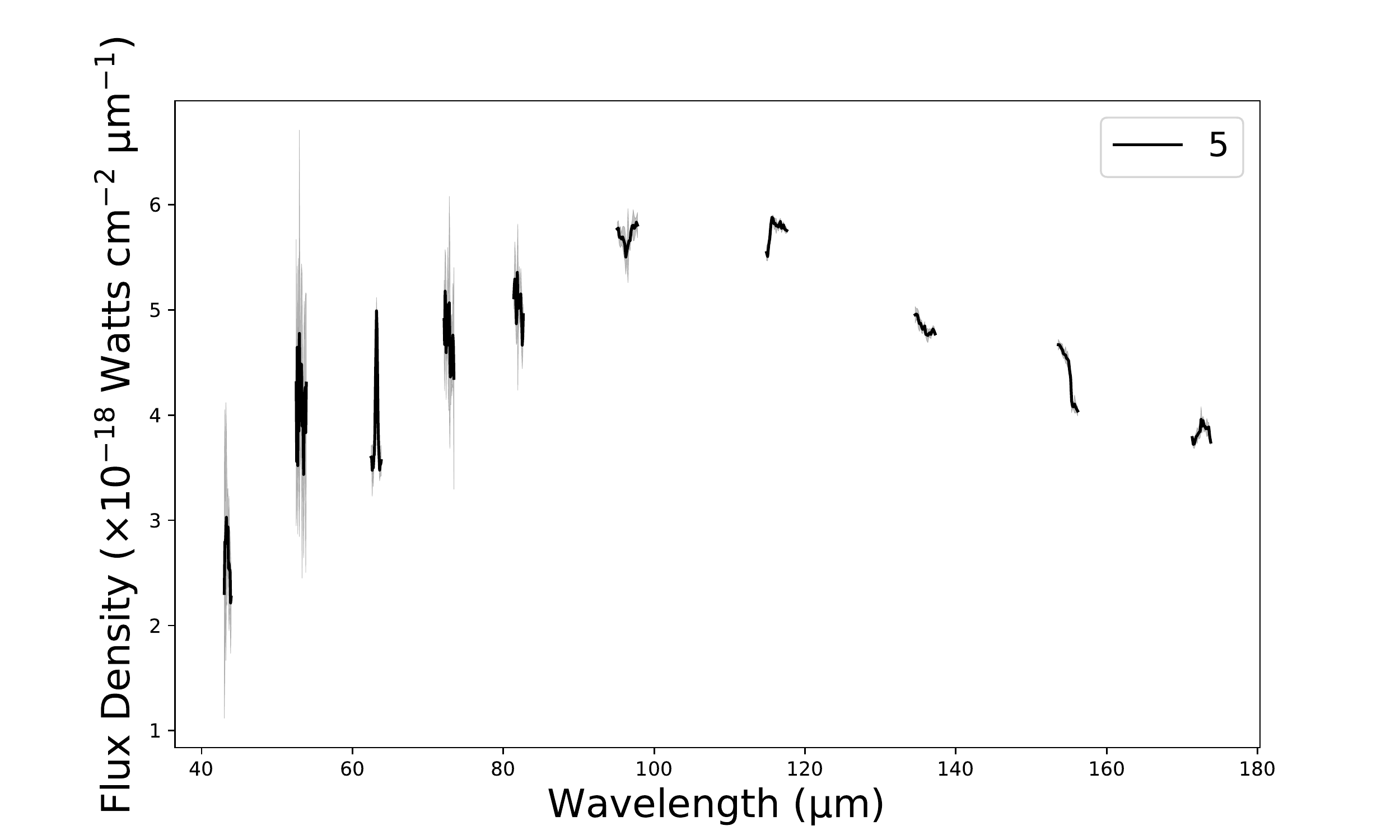}
 \put (12,55) {\Large \textbf{Kes 79}}
\end{overpic}&
\includegraphics[width=5.2cm]{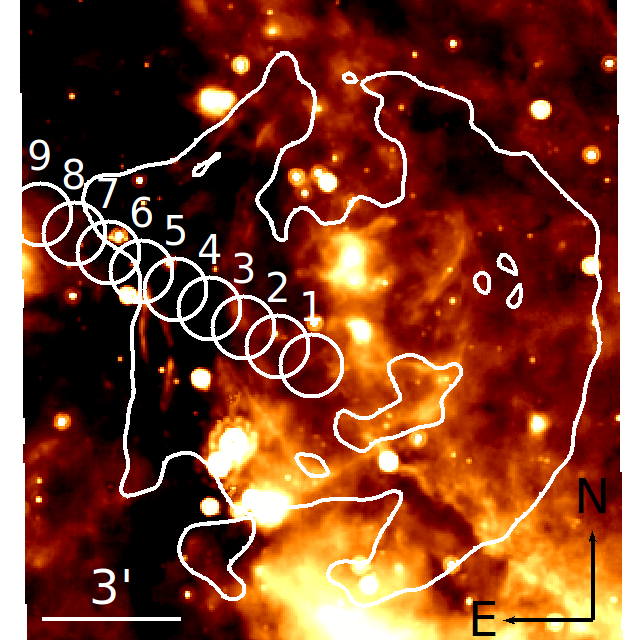}
\end{tabular}
\caption{
Left: The $ISO$ LWS spectrum of \hypertarget{g54p1-exact}{Kes 79}. Right: The $Spitzer$ MIPS 24 \textmugreek{}m image of Kes 79. The $ISO$ LWS pointings are shown with white circles. The white contours are from a VLA (1.4 GHz) image. The raster spectra are shown in Figure \ref{fig:lineRasters}.} 
\label{fig:kes79}
\end{figure*}

\begin{figure*}
\centering
\begin{tabular}{lc}
\begin{overpic}[width=8 cm,tics=10]{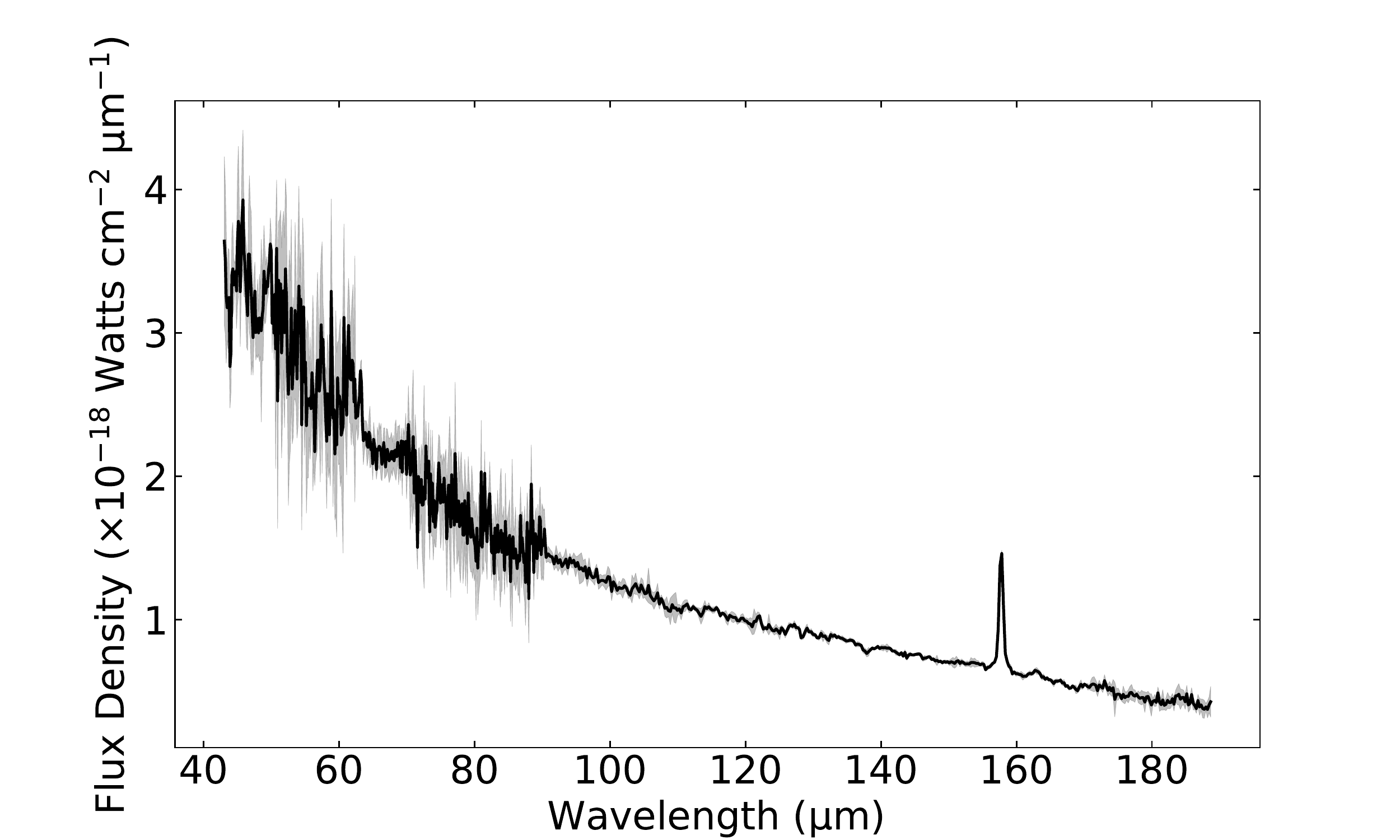}
 \put (12,55) {\Large \textbf{G54.1+0.3}}
\end{overpic}&
\includegraphics[width=5.2cm]{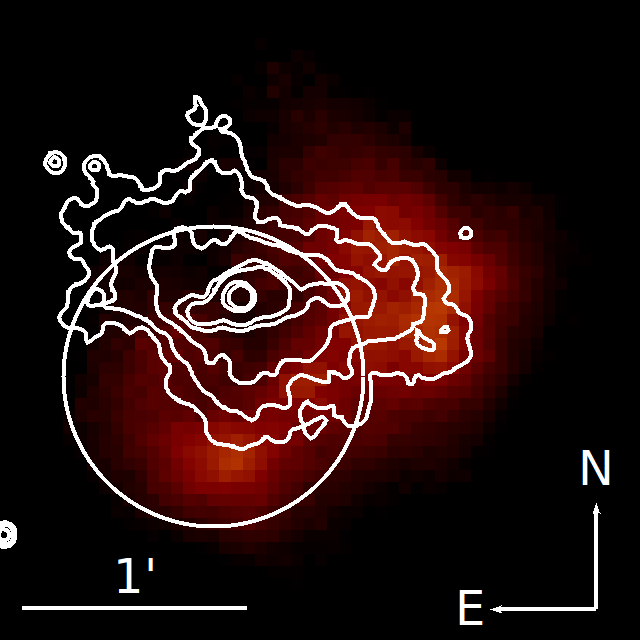} 
\end{tabular}
\caption{Left: The $ISO$ LWS spectrum of \hypertarget{ctb109-exact}{G54.1+0.3}. Right: The {\it Herschel} PACS 70 \textmugreek{}m image of G54.1+0.3. The $ISO$ LWS pointing is shown with a white circle.
The white contours are from a {\it Chandra} (0.5 -- 7.0 keV) image.}  
\label{fig:g54p1}
\end{figure*}

\begin{figure*}
\centering
\begin{tabular}{lc}
\begin{overpic}[width=8 cm,tics=10]{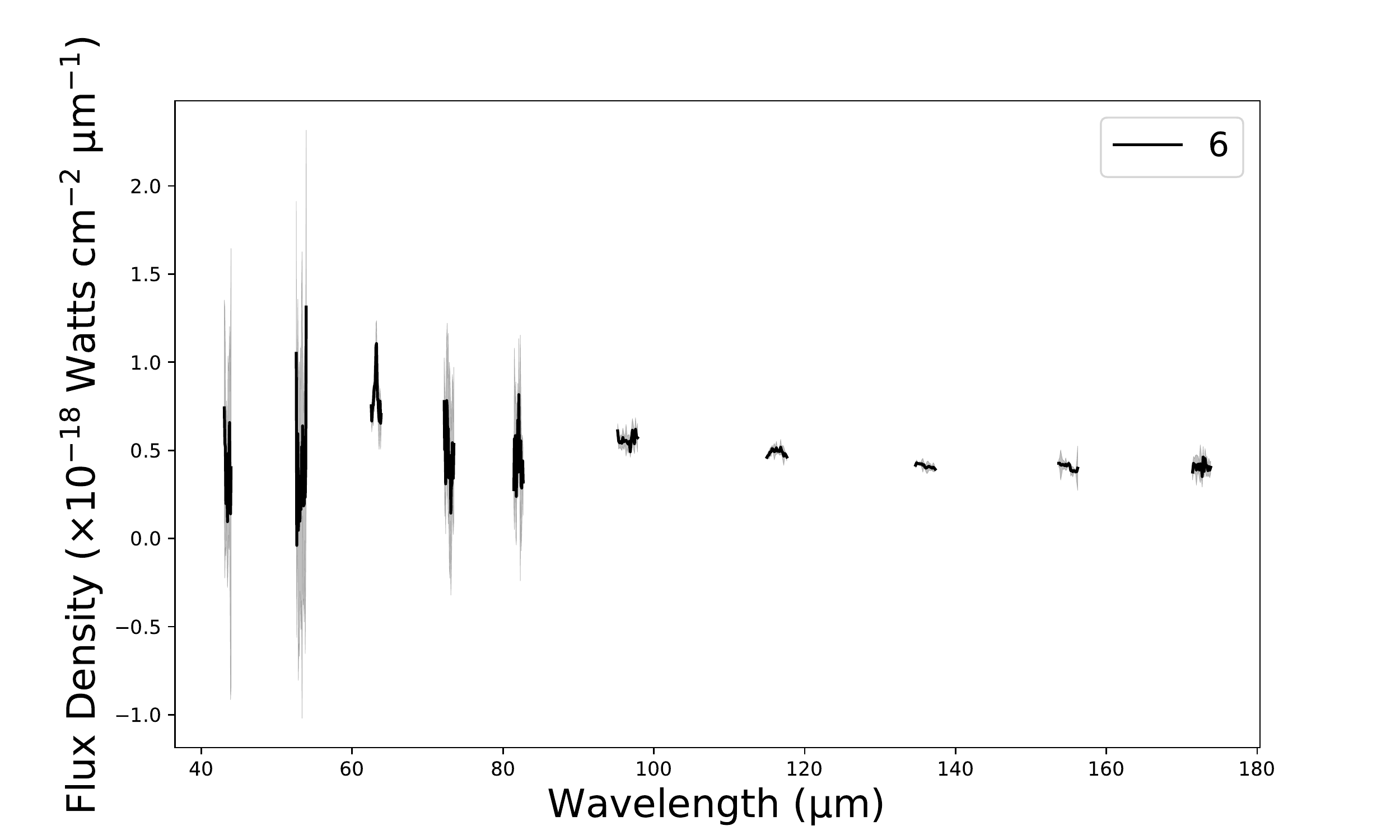}
 \put (12,55) {\Large \textbf{CTB 109}}
\end{overpic}&
\includegraphics[width=5.2cm]{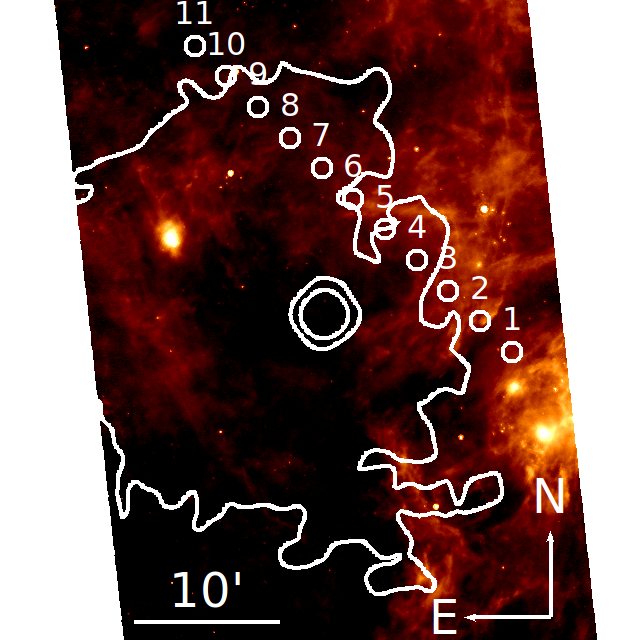}
\end{tabular}
\caption{
Left: The $ISO$ LWS spectrum of \hypertarget{tych-exact}{CTB 109}. Right: The {\it Spitzer} MIPS 24 \textmugreek{}m image of CTB 109. The $ISO$ LWS pointings are shown with white circles.
The white contours are from an {\it XMM-Newton} (2.0 -- 7.2 keV) image. The raster spectra are shown in Figure \ref{fig:lineRasters}.} 
\label{fig:ctb109}
\end{figure*}

\begin{figure*}
\centering
\begin{tabular}{lc}
\begin{overpic}[width=8 cm,tics=10]{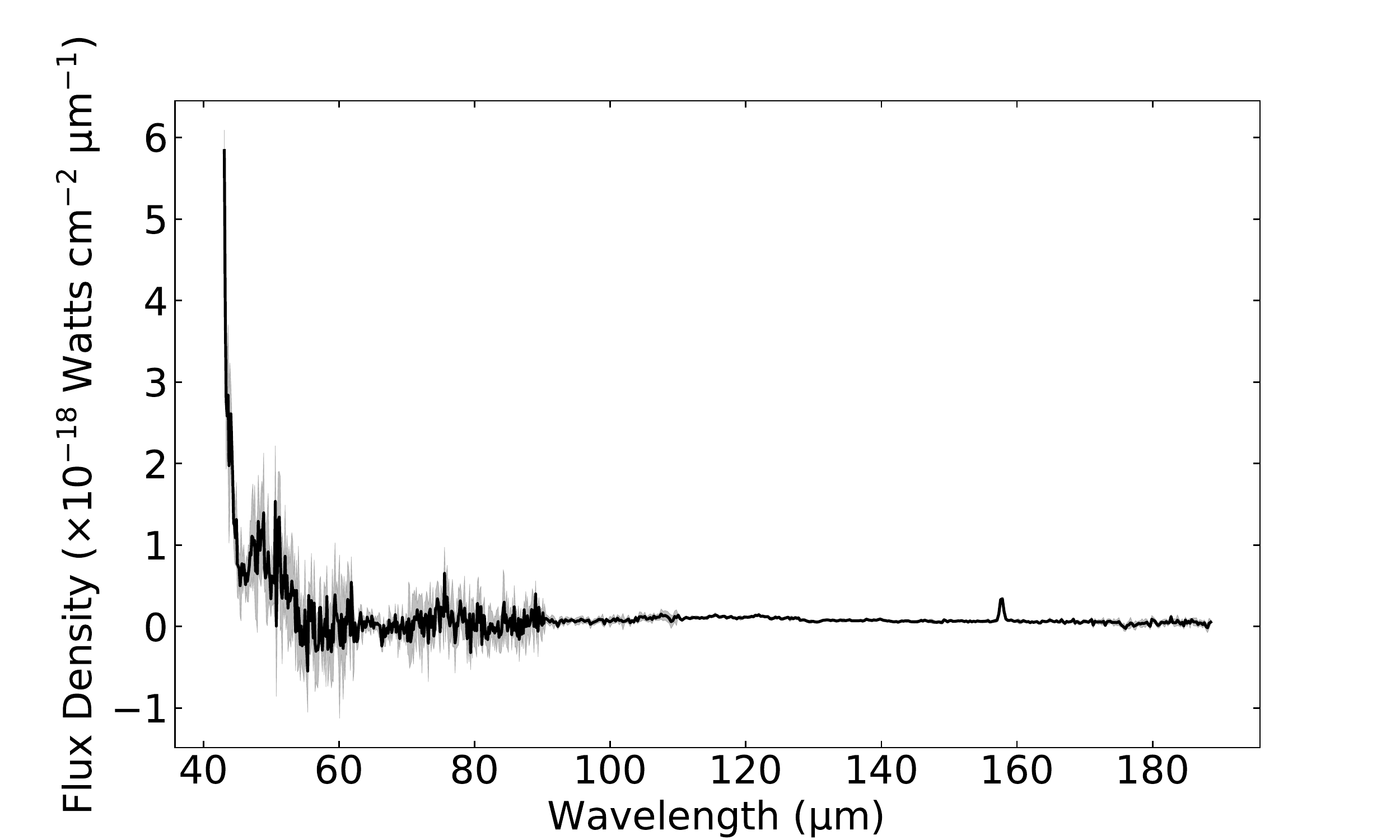}
 \put (12,55) {\Large \textbf{Tycho's SNR}}
\end{overpic}&
\includegraphics[width=5.2cm]{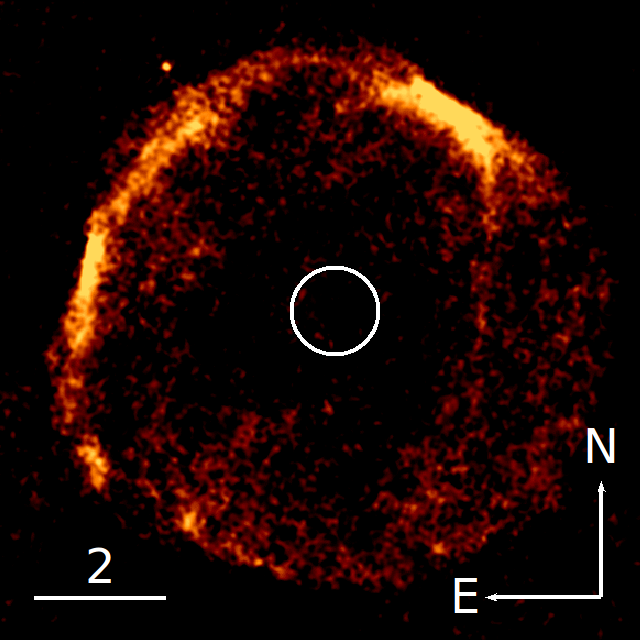} 
\end{tabular}
\caption{Left: The $ISO$ LWS spectrum of Tycho's SNR. Right: The {\it Herschel} PACS 70 \textmugreek{}m image of Tycho's SNR. The $ISO$ LWS pointing is shown with a white circle.
 } 
\label{fig:tych}
\end{figure*}

\begin{figure*}
\centering
\begin{tabular}{lc}
\begin{overpic}[width=8 cm,tics=10]{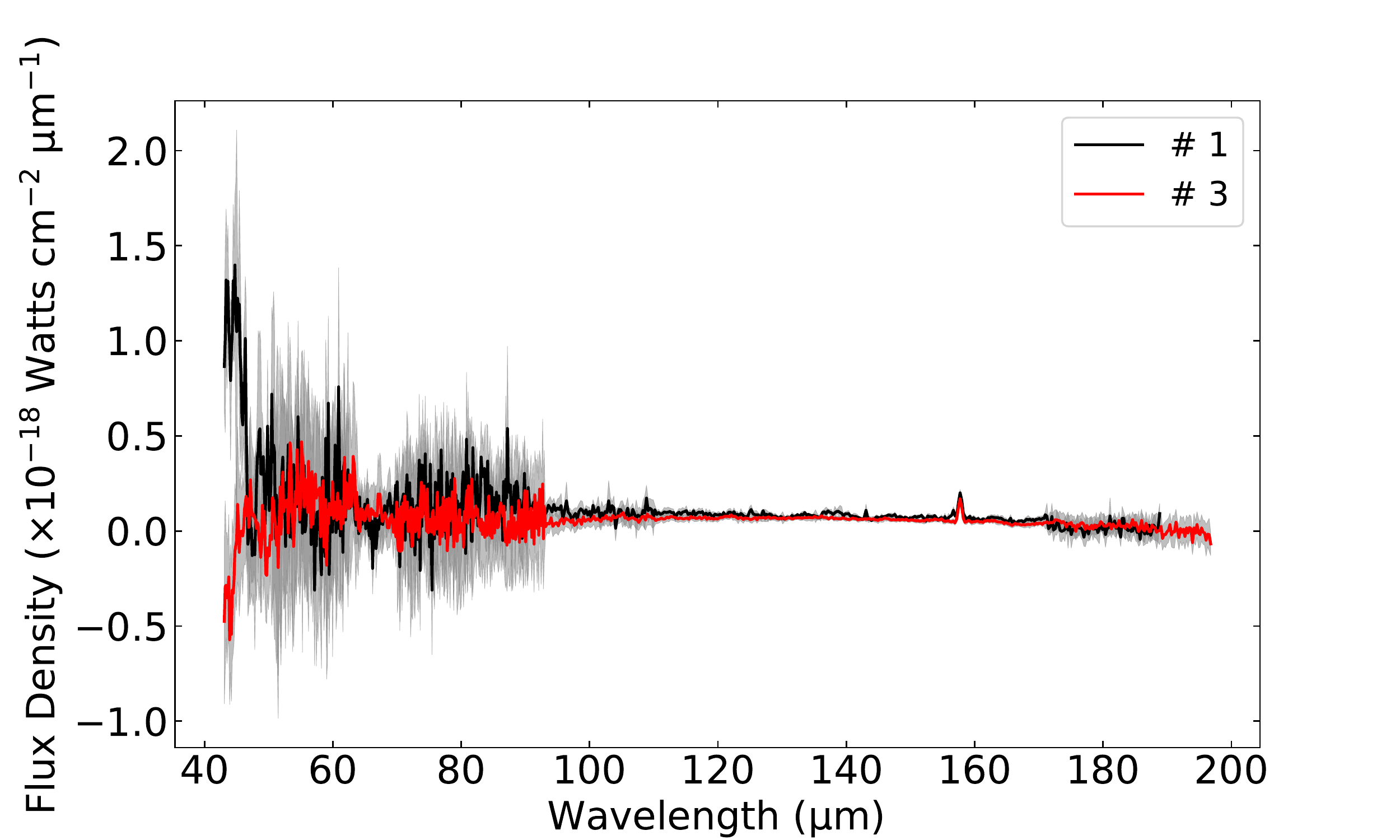}
 \put (12,55) {\Large \textbf{3C 58}}
\end{overpic}&
\includegraphics[width=5.2cm]{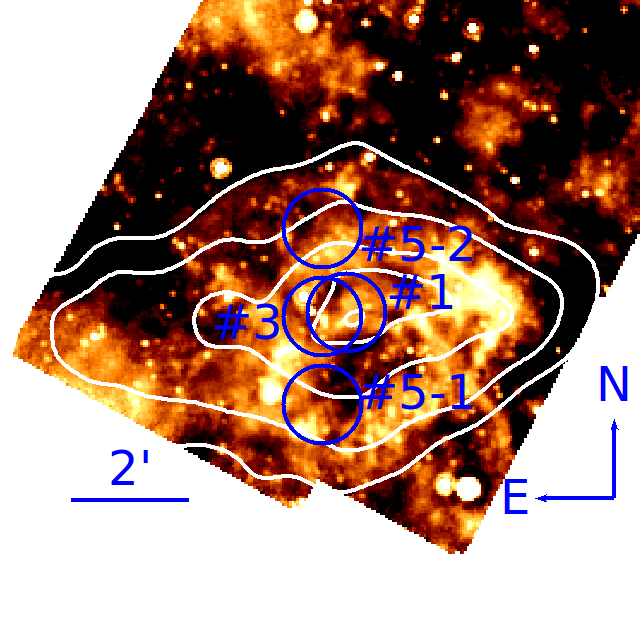}
\end{tabular}
\caption{Left: The $ISO$ LWS spectra of \hypertarget{crab-exact}{3C 58}. Right: The {\it Spitzer} MIPS 24 \textmugreek{}m image of 3C 58. The $ISO$ LWS pointings are shown with blue circles.
The white contours are from a VLA (1.48 GHz) image.  } 
\label{fig:3c58}
\end{figure*}

\begin{figure*}
\centering
\begin{tabular}{lc}
\begin{overpic}[width=8 cm,tics=10]{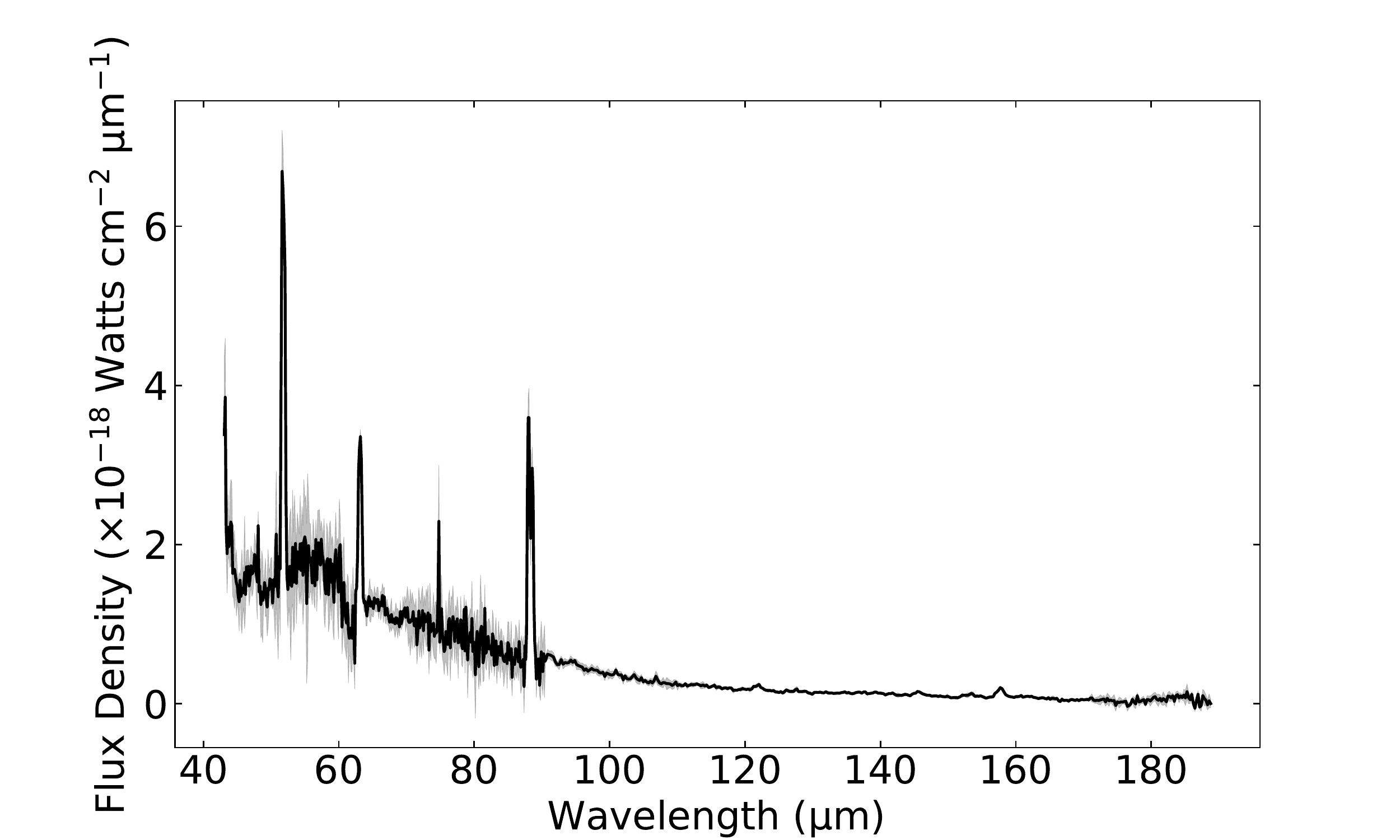}
 \put (12,55) {\Large \textbf{Crab Nebula}}
\end{overpic}&
\includegraphics[width=5.2cm]{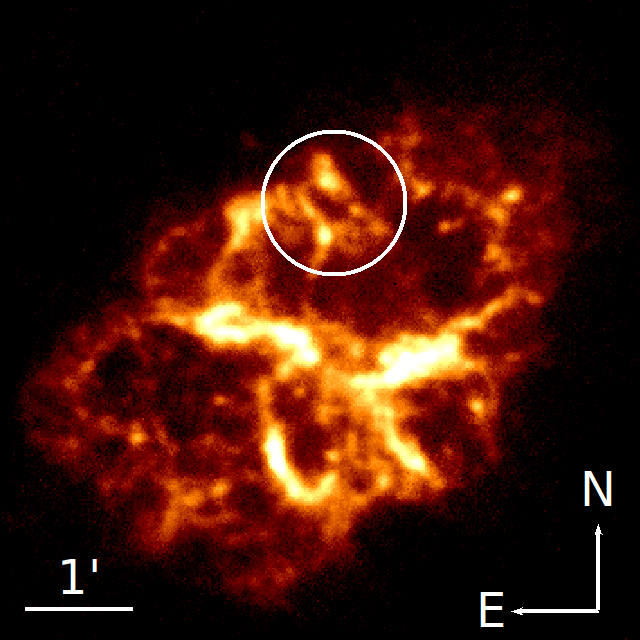}
\end{tabular}
\caption{Left: The $ISO$ LWS spectrum of \hypertarget{ic443-exact}{Crab Nebula}. Right: The {\it Herschel} PACS 70 \textmugreek{}m image of the Crab Nebula. The $ISO$ LWS pointing is shown with a white circle.} 
\label{fig:crab}
\end{figure*}

\begin{figure*}
\centering
\begin{tabular}{lc}
\begin{overpic}[width=8 cm,tics=10]{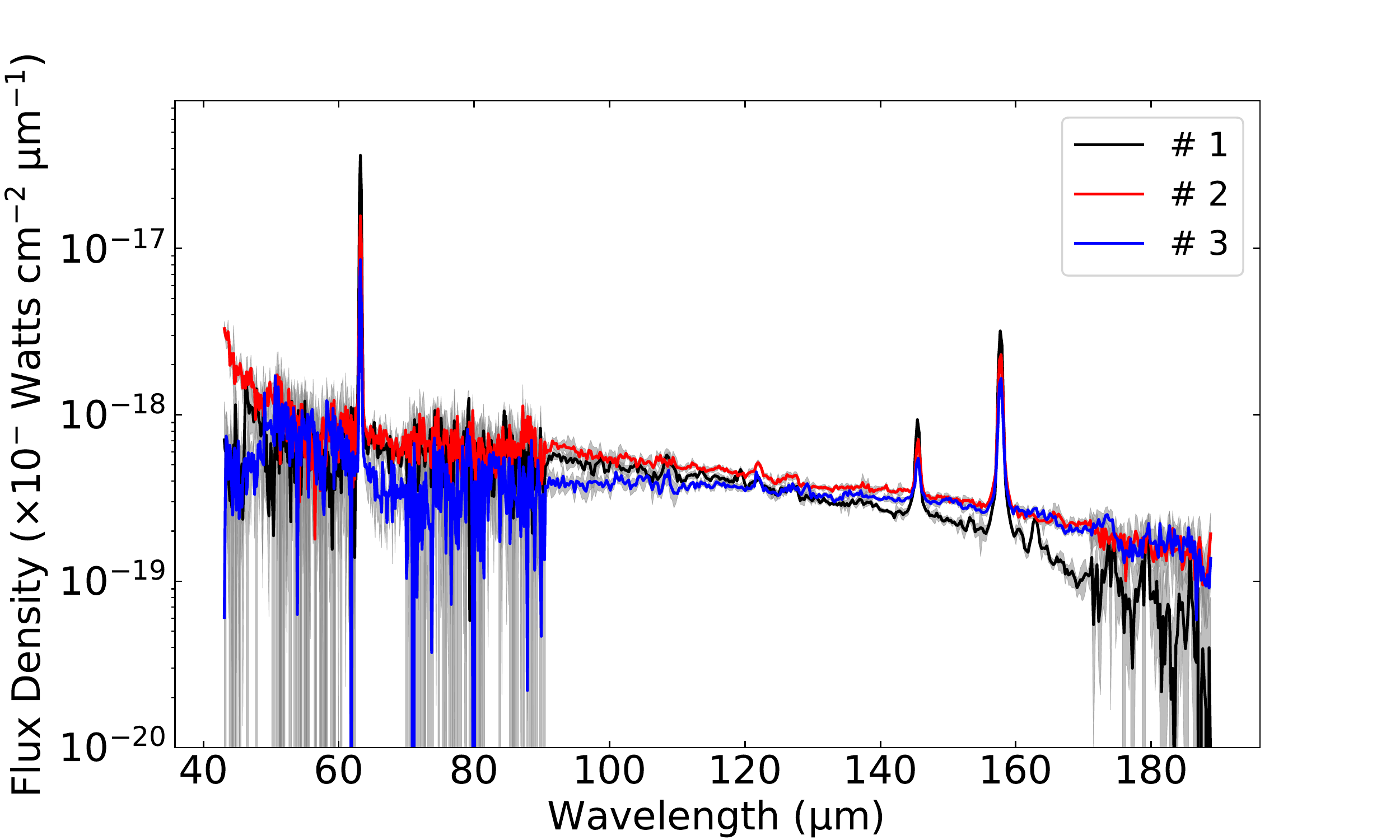}
 \put (12,55) {\Large \textbf{IC 443}}
\end{overpic}&
\includegraphics[width=5.2cm]{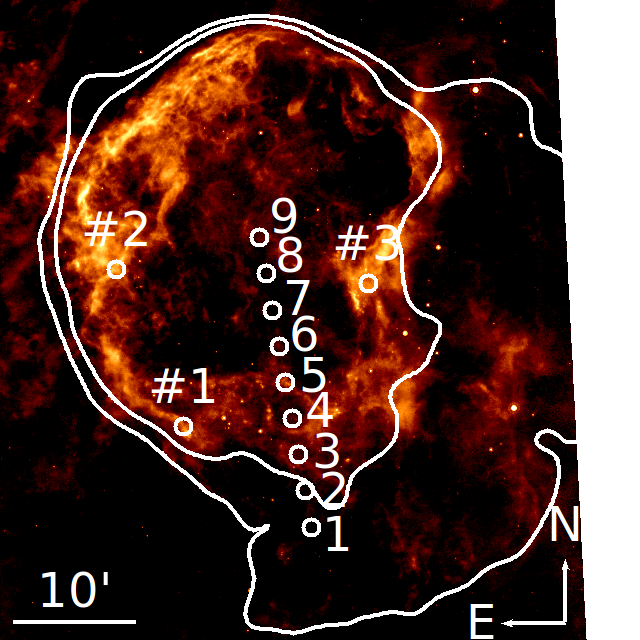}
\end{tabular}
\caption{
Left: The $ISO$ LWS spectra of \hypertarget{g292p0-exact}{IC 443}. Right: The {\it Spitzer} MIPS 24 \textmugreek{}m image of IC 443. The $ISO$ LWS pointings are shown with white circles. The raster line profiles are displayed in Figure \ref{fig:lineRasters}.  The white contours are from a VLA (330 MHz) image.}
\label{fig:ic443}
\end{figure*}

\begin{figure*}
\centering
\begin{tabular}{lc}
\begin{overpic}[width=8 cm,tics=10]{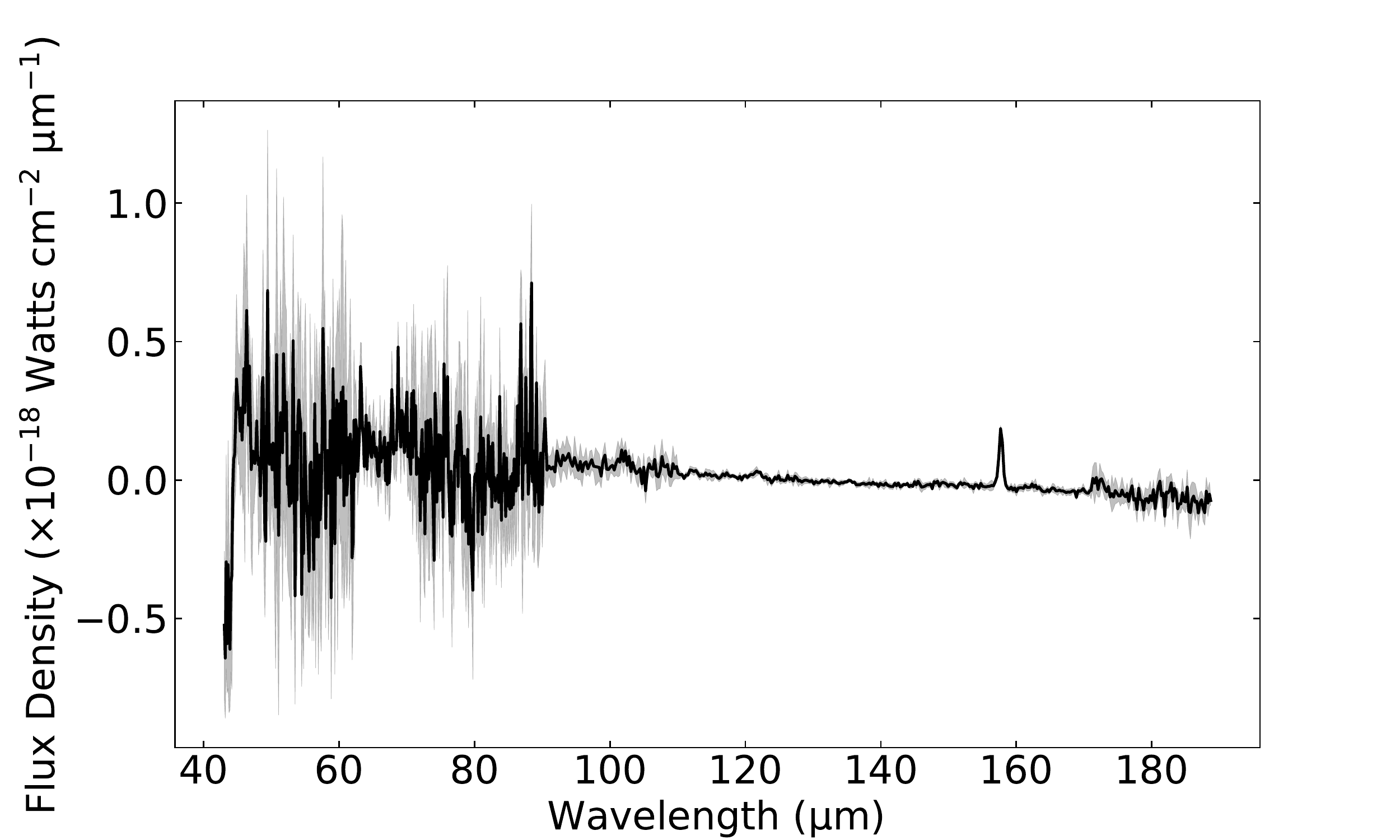}
 \put (12,55) {\Large \textbf{\hypertarget{bla}{G292.0+1.8}}}
\end{overpic}&
\includegraphics[width=5.2cm]{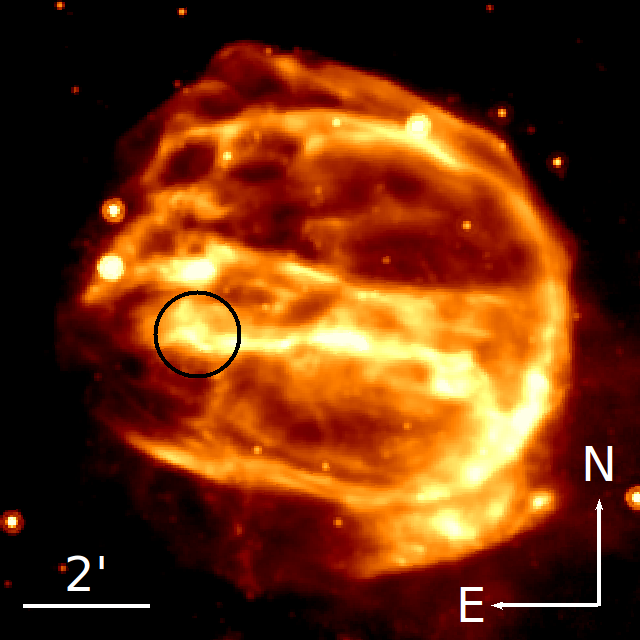}
\end{tabular}
\caption{Left: The $ISO$ LWS spectrum of \hypertarget{g292p0-exact}{G292.0+1.8}. Right: The {\it Spitzer} MIPS 24 \textmugreek{}m image of G292.0+1.8. The $ISO$ LWS pointing is shown with a black circle.} 
\label{fig:g292p0}
\end{figure*}

\begin{figure*}
\centering
\begin{tabular}{lc}
\begin{overpic}[width=8 cm,tics=10]{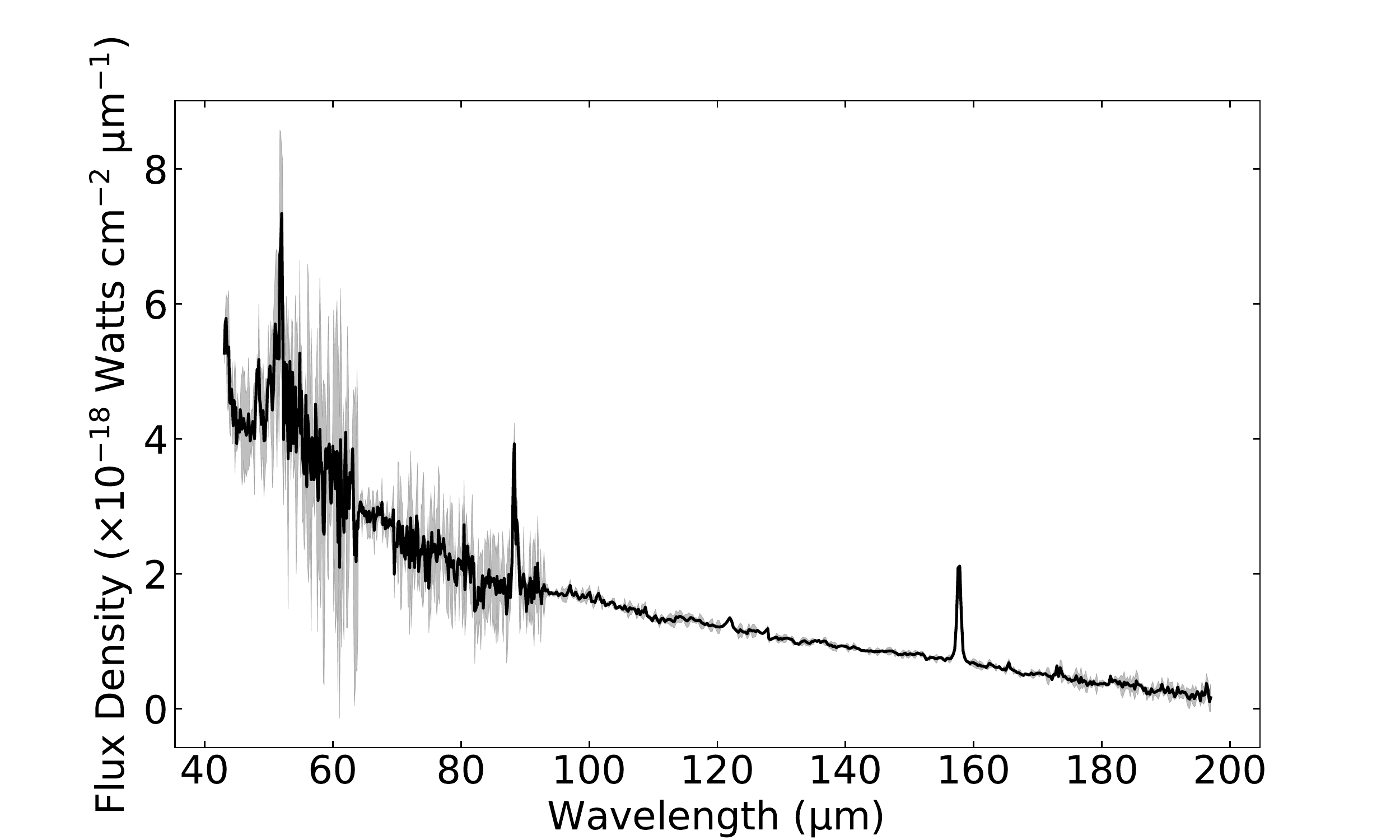}
 \put (12,55) {\Large \textbf{G320.4--1.2}}
\end{overpic}&
\includegraphics[width=5.2cm]{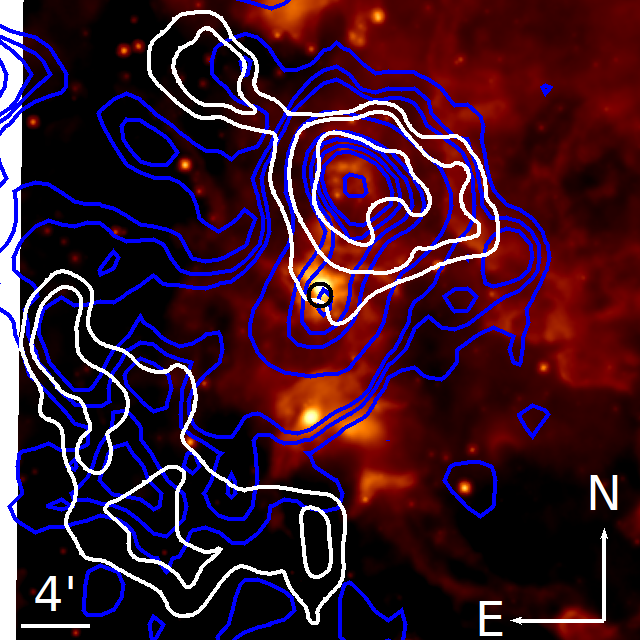}
\end{tabular}
\caption{
Left: The $ISO$ LWS spectrum of \hypertarget{rcw103-exact}{G320.4--1.2}. Right: The {\it WISE} 22 \textmugreek{}m image of G320.4--1.2. The $ISO$ LWS pointing is shown near the center with a black circle.
The white contours are from a MOST (843 MHz) image. The blue contours are from a broadband {\it ROSAT} image. }
\label{fig:g320p4}
\end{figure*}

\begin{figure*}
\centering
\begin{tabular}{lc}
\begin{overpic}[width=8 cm,tics=10]{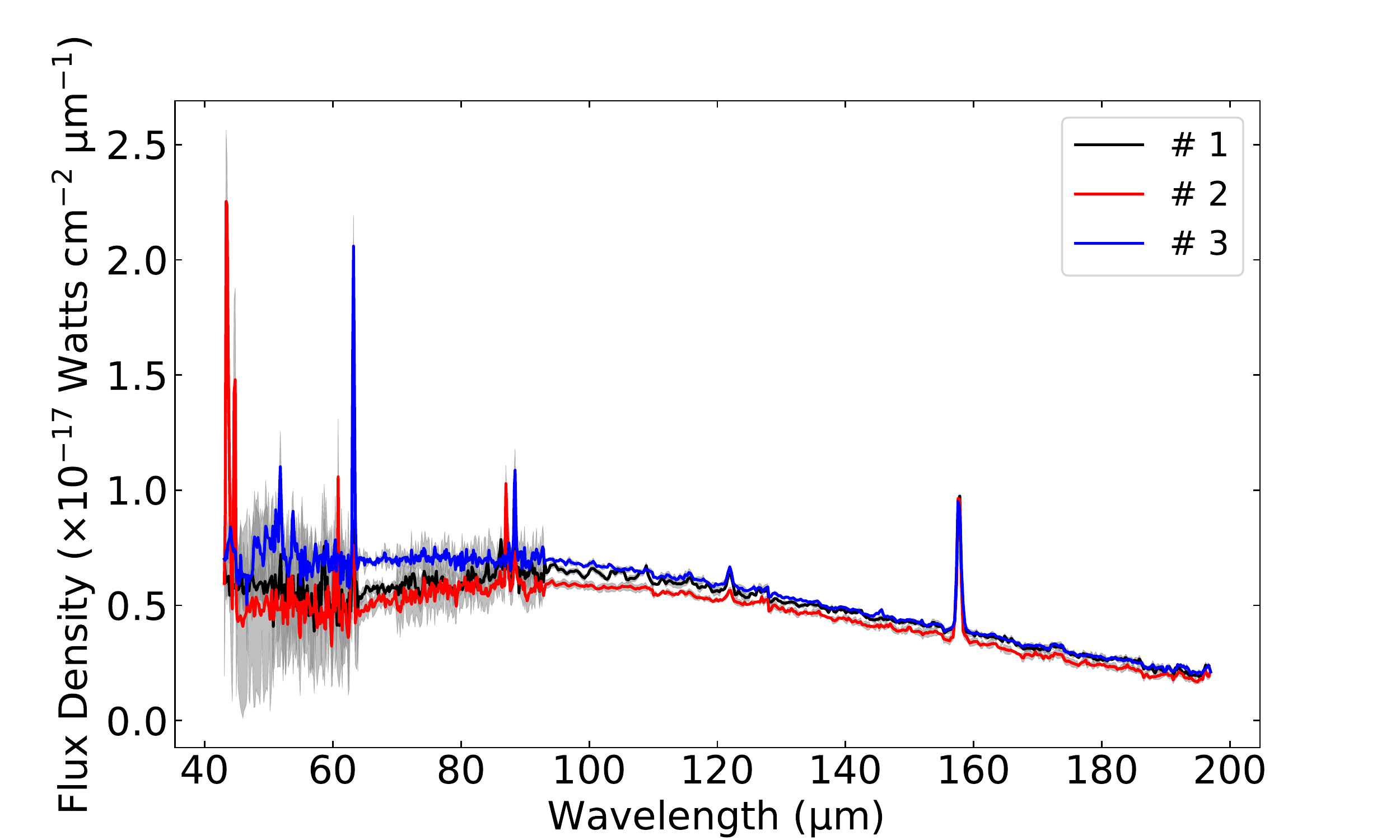}
 \put (12,55) {\Large \textbf{RCW 103}}
\end{overpic}&
\includegraphics[width=5.2cm]{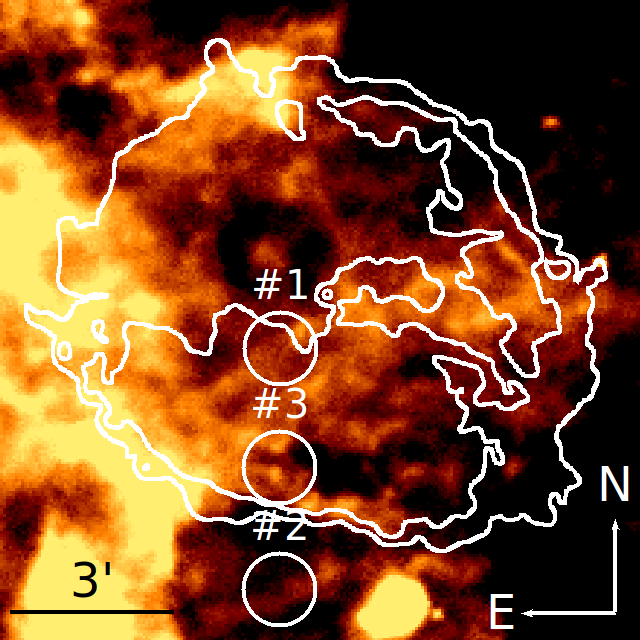} \\
\hspace{1.5cm}
\includegraphics[width=5.2cm]{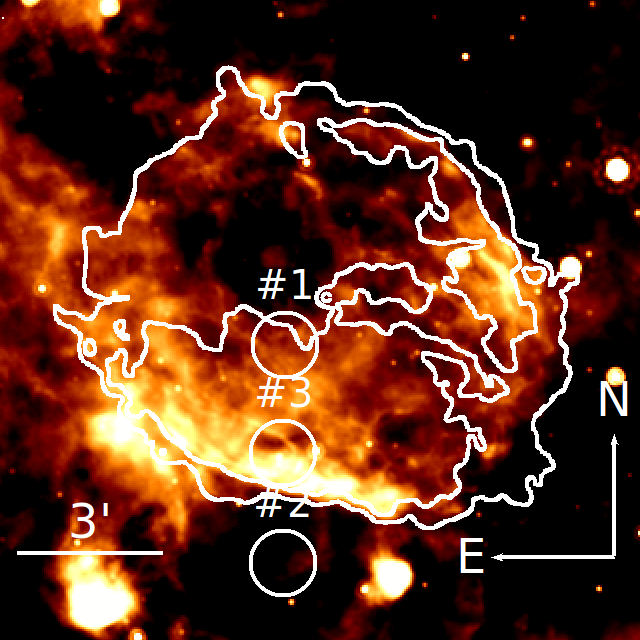}
&
\includegraphics[width=5.2cm]{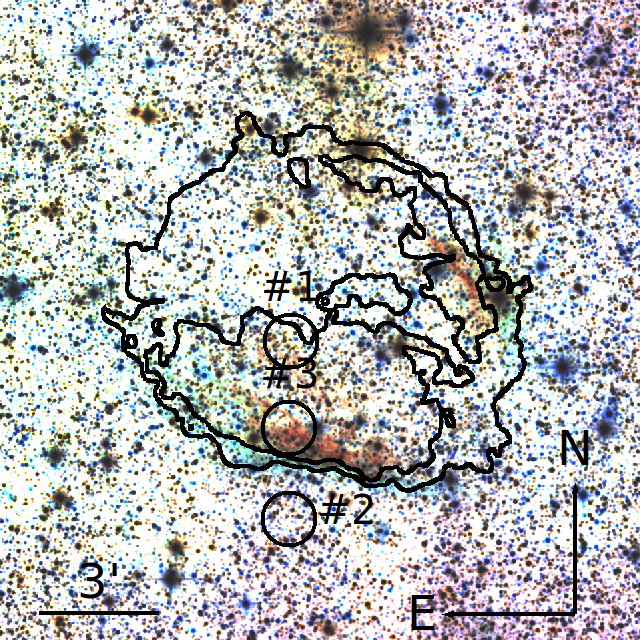}
\end{tabular}
\caption{Upper Left: The $ISO$ LWS spectra of \hypertarget{e0102-exact}{RCW 103}. Upper Right: The {\it Herschel} PACS 70 \textmugreek{}m image of RCW 103. Lower Left: The {\it Spitzer} MIPS 24 \textmugreek{}m image of RCW 103. Lower Right: The 3-color 2MASS image of RCW 103. The J, H, and K\textsubscript{s} bands are shown in red, green, and blue, respectively. In all three image panels, the $ISO$ LWS pointings are shown with white or black circles.
The contours are from a {\it Chandra} (0.5 -- 7.0 keV) image} 
\label{fig:rcw103}
\end{figure*}

\begin{figure*}
\centering
\begin{tabular}{lc}
\begin{overpic}[width=8 cm,tics=10]{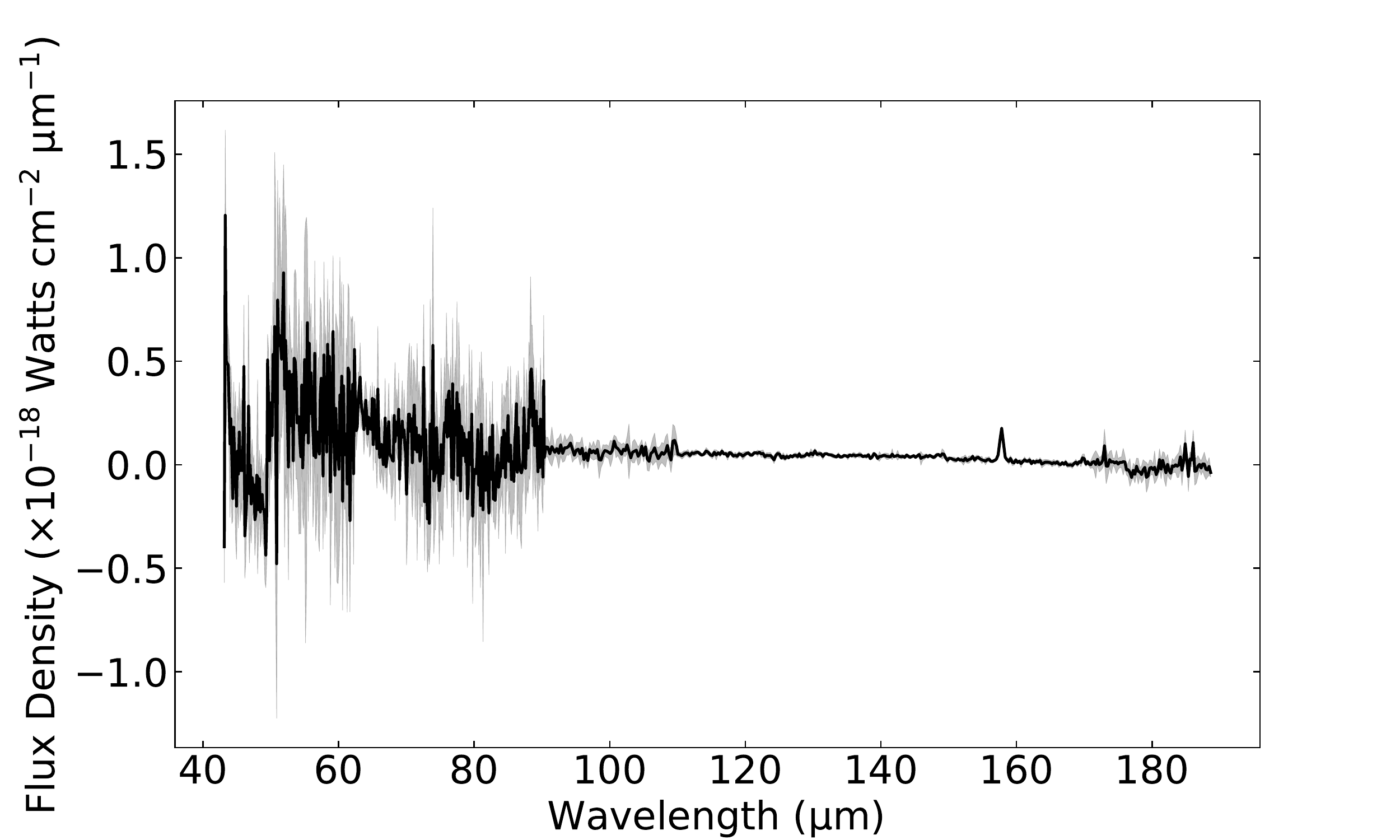}
 \put (12,55) {\Large \textbf{E0102.2--7219}}
\end{overpic}&
\includegraphics[width=5.2cm]{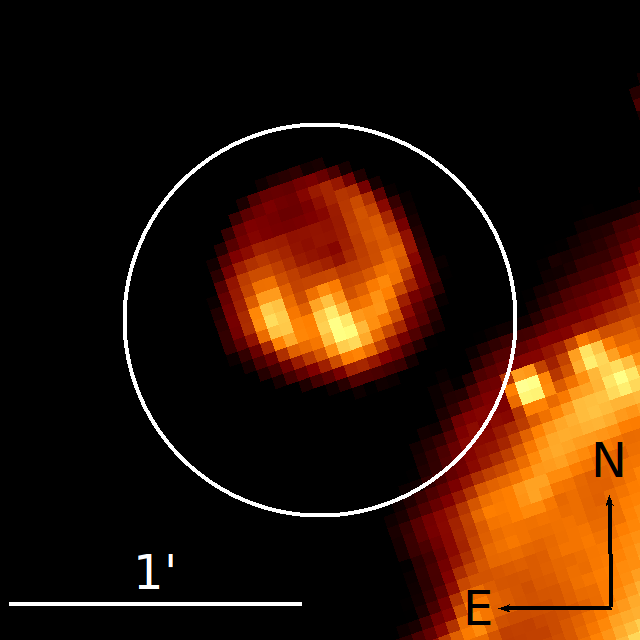}\\
\end{tabular}
\caption{Left: The $ISO$ LWS spectrum of \hypertarget{n132d-exact}{E0102.2--7219}. Right: The {\it Spitzer} MIPS 24 \textmugreek{}m image of E0102.2-7219. The $ISO$ LWS pointing is shown with a white circle.
} 
\label{fig:e0102}
\end{figure*}

\begin{figure*}
\centering
\begin{tabular}{lc}
\begin{overpic}[width=8 cm,tics=10]{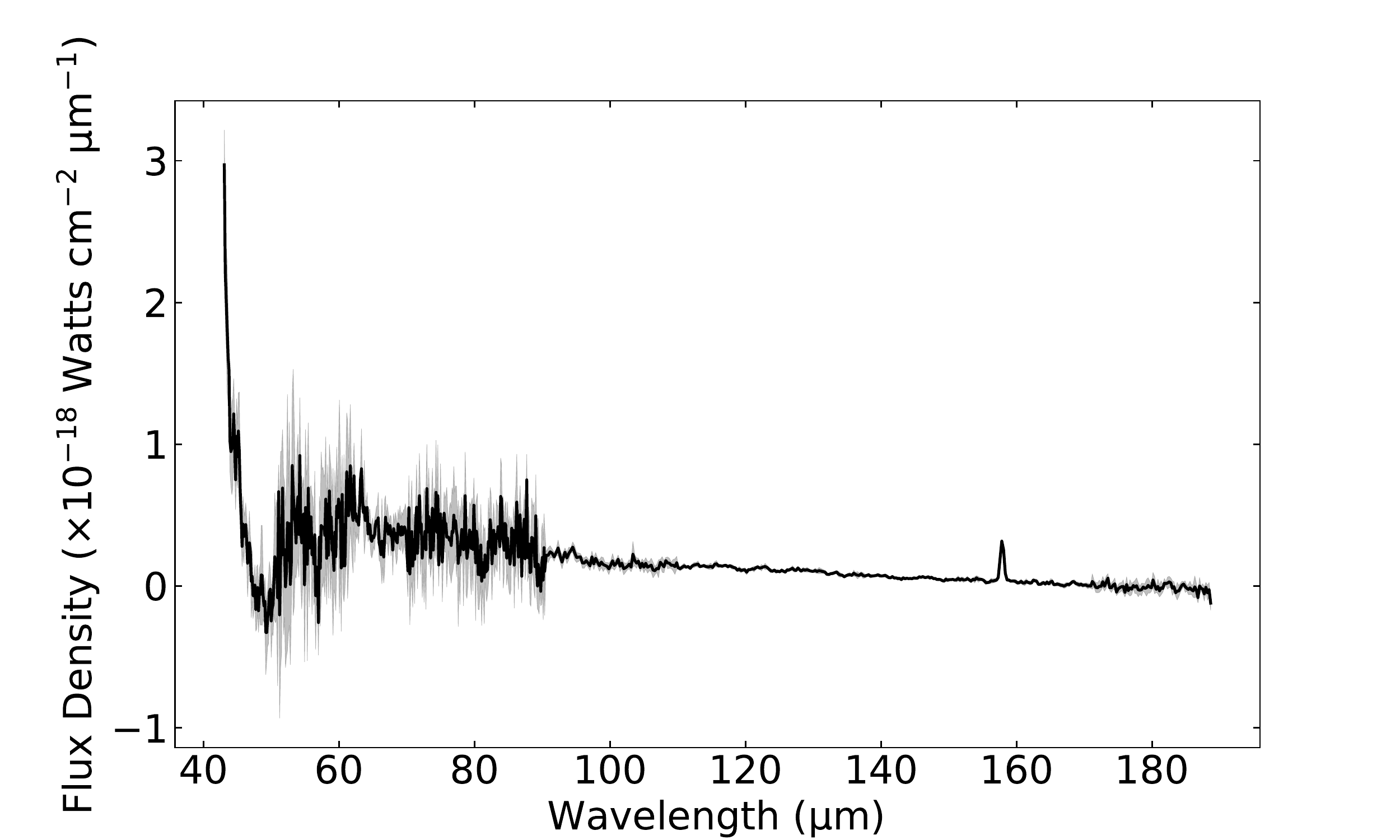}
 \put (12,55) {\Large \textbf{N132D}}
\end{overpic}&
\includegraphics[width=5.2cm]{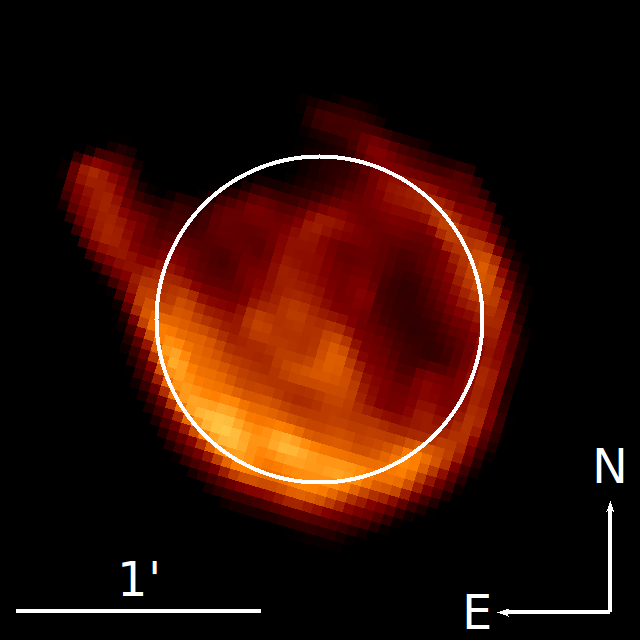}
\end{tabular}
\caption{Left: The $ISO$ LWS spectrum of N132D. Right: The {\it Spitzer} MIPS 24 \textmugreek{}m image of N132D. The $ISO$ LWS pointing is shown with a white circle.
\label{fig:n132d}} 
\end{figure*}

\begin{figure*}
\centering
\begin{tabular}{lc}
\begin{overpic}[width=8 cm,tics=10]{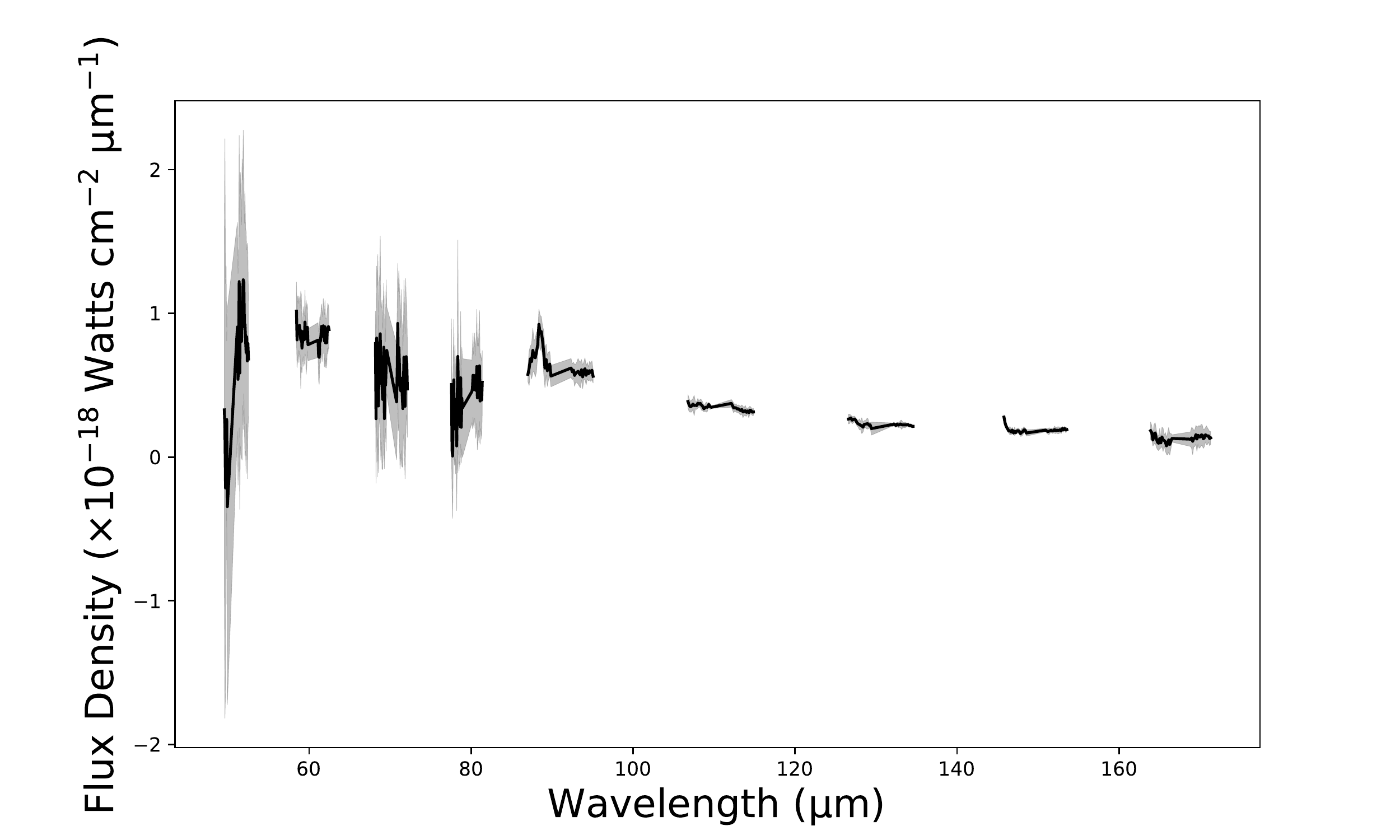}
 \put (12,55) {\Large \textbf{N49}}
\end{overpic}&
\includegraphics[width=5.2cm]{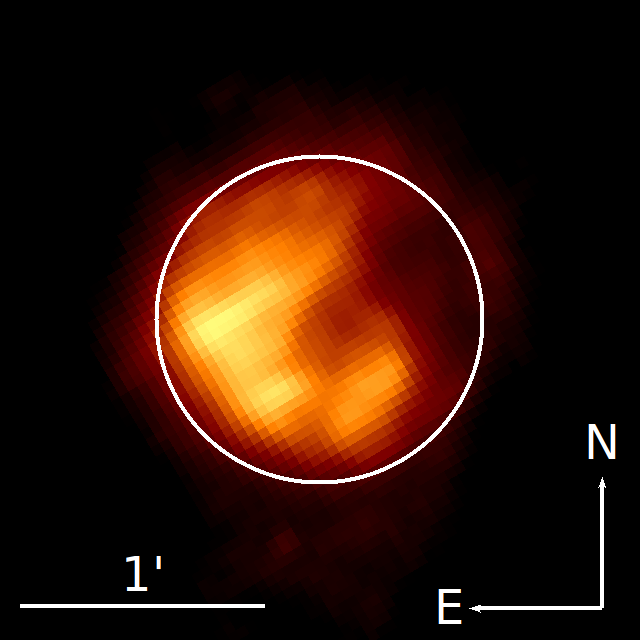}
\end{tabular}
\caption{Left: The $ISO$ LWS spectrum of \hypertarget{n63a-exact}{N49}. Right: The {\it Spitzer} MIPS 24 \textmugreek{}m image of N49. The $ISO$ LWS pointing is shown with a white circle.} 
\label{fig:n49}
\end{figure*}

\begin{figure*}
\centering
%
\begin{tabular}{lc}
\begin{overpic}[width=8 cm,tics=10]{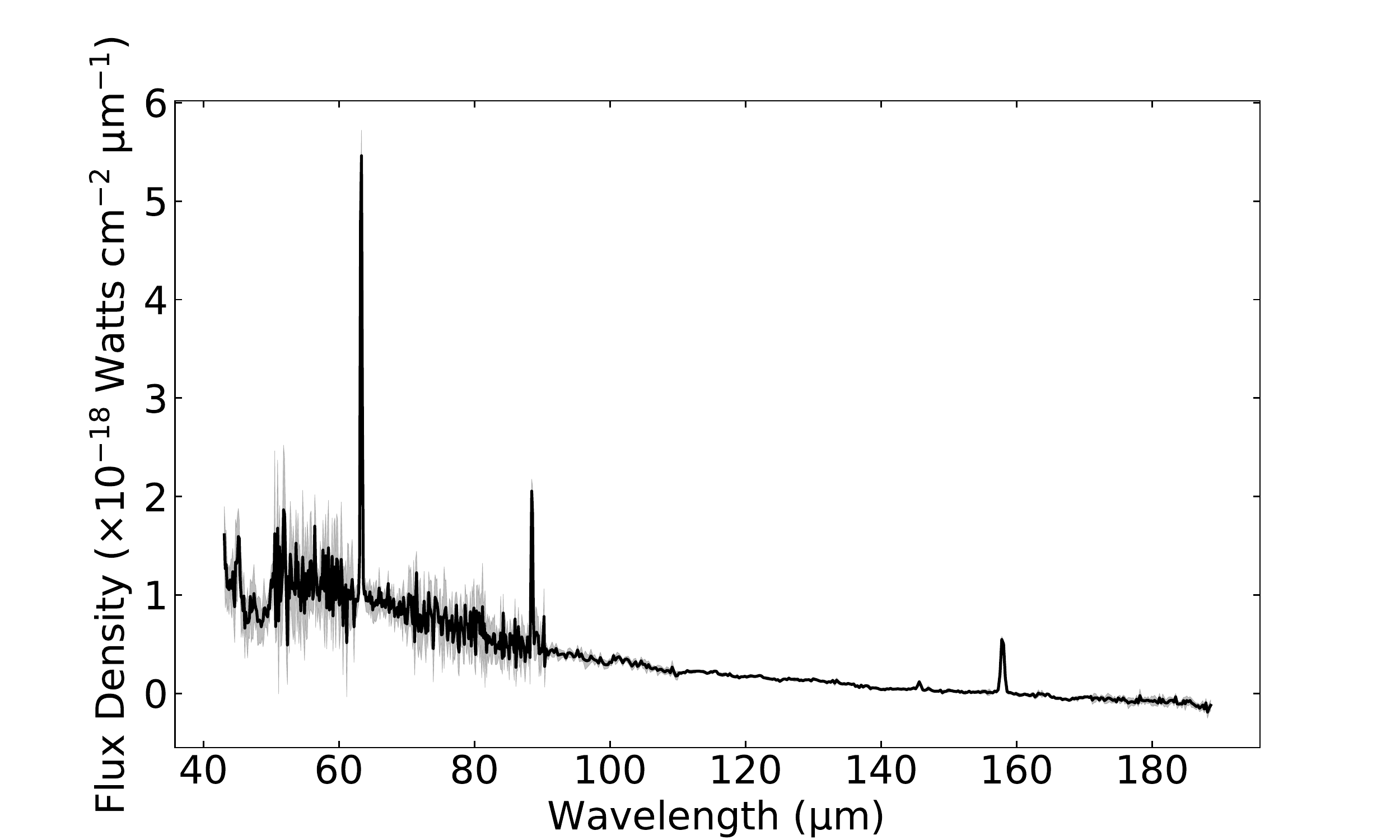}
 \put (12,55) {\Large \textbf{N63A}}
\end{overpic}&
\includegraphics[width=5.2cm]{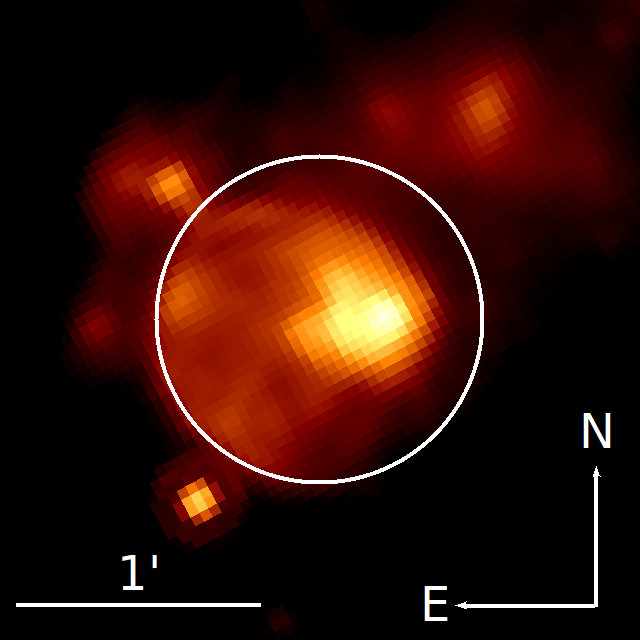}
\end{tabular}
\caption{Left: The $ISO$ LWS spectrum of \hypertarget{sn1987a-exact}{N63A}. Right: The {\it Spitzer} MIPS 24 \textmugreek{}m image of N63A. The $ISO$ LWS pointing is shown with a white circle.
} 
\label{fig:n63a}
\end{figure*}

\begin{figure*}
\centering
\begin{tabular}{lc}
\begin{overpic}[width=8 cm,tics=10]{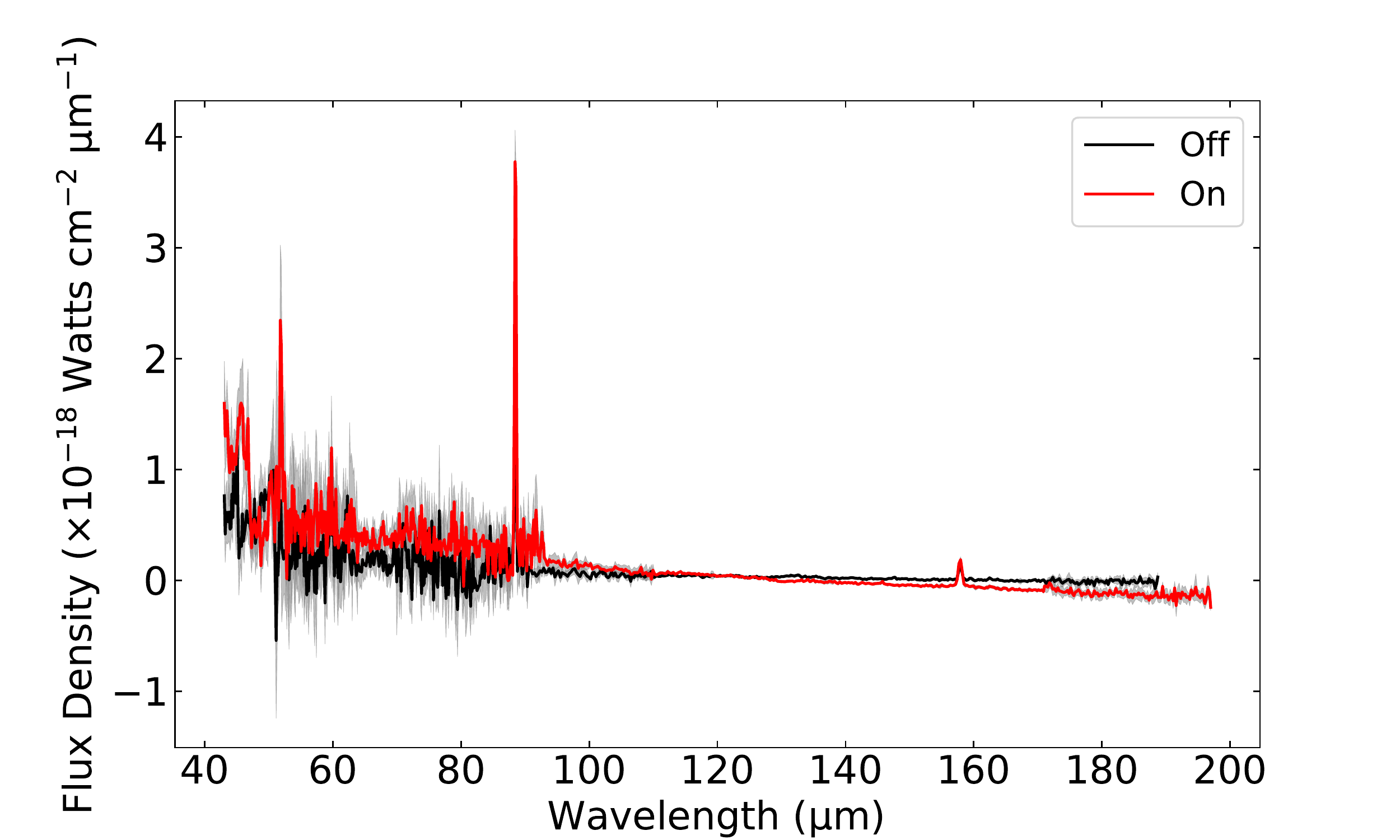}
 \put (12,55) {\Large \textbf{SN 1987A}}
\end{overpic}&
\includegraphics[width=5.2cm]{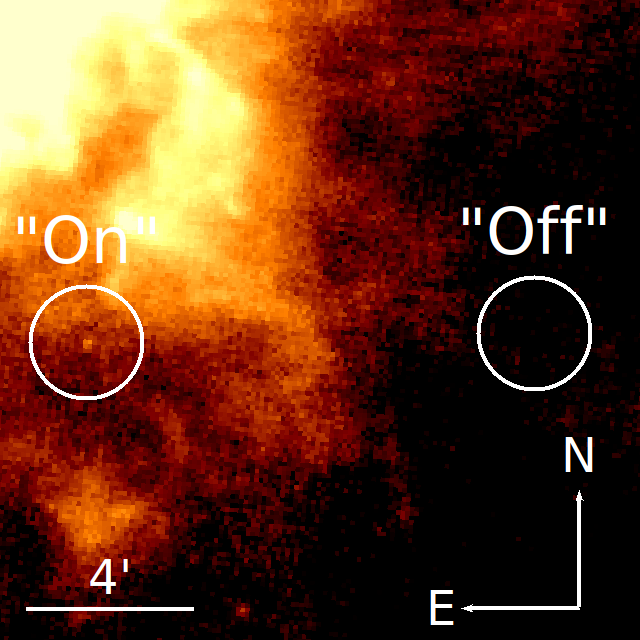}
\end{tabular}
\caption{Left: The $ISO$ LWS spectrum of \hypertarget{0540-69p3-exact}{SN 1987A}. Right: The {\it Herschel} PACS 100 \textmugreek{}m image of SN 1987A. The $ISO$ LWS pointings are shown with white circles.
} 
\label{fig:sn1987a}
\end{figure*}

\begin{figure*}
\centering
\begin{tabular}{lc}
\begin{overpic}[width=8 cm,tics=10]{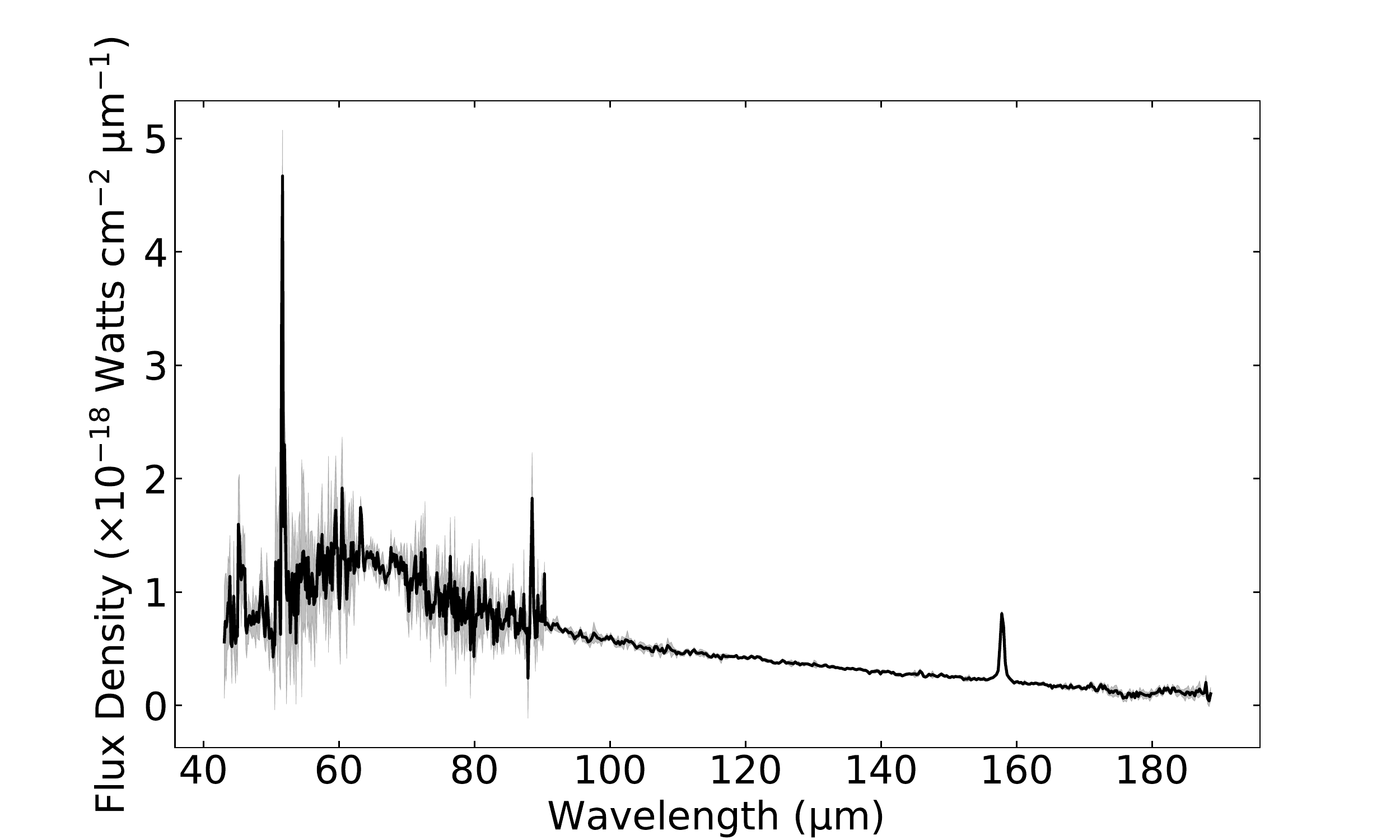}
 \put (12,55) {\Large \textbf{0540--69.3}}
\end{overpic}&
\includegraphics[width=5.2cm]{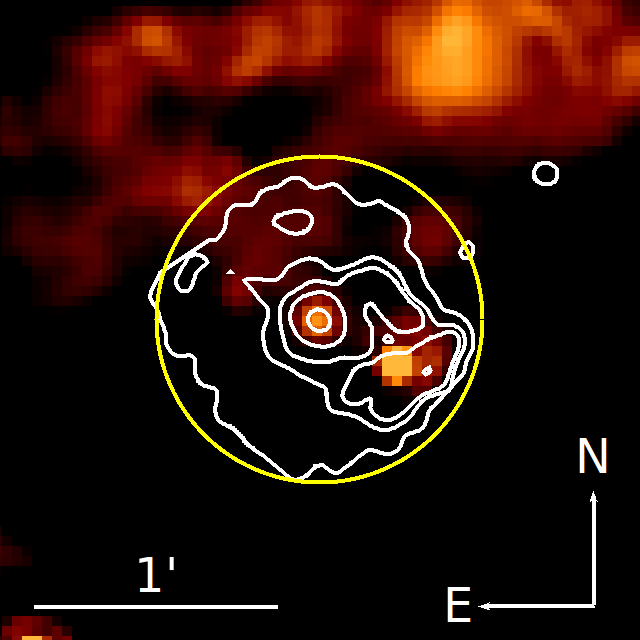}
\end{tabular}
\caption{Left: The $ISO$ LWS spectrum of 0540--69.3. Right: The {\it Spitzer} MIPS 24 \textmugreek{}m image of 0540-69.3. The $ISO$ LWS pointing is shown with a yellow circle. The white contours are from a Chandra (0.5 -- 7.0 keV) image.} 
\label{fig:0540-69p3}
\end{figure*}

\begin{figure*}
\centering
\begin{tabular}{lc}
\begin{overpic}[width=8 cm,tics=10]{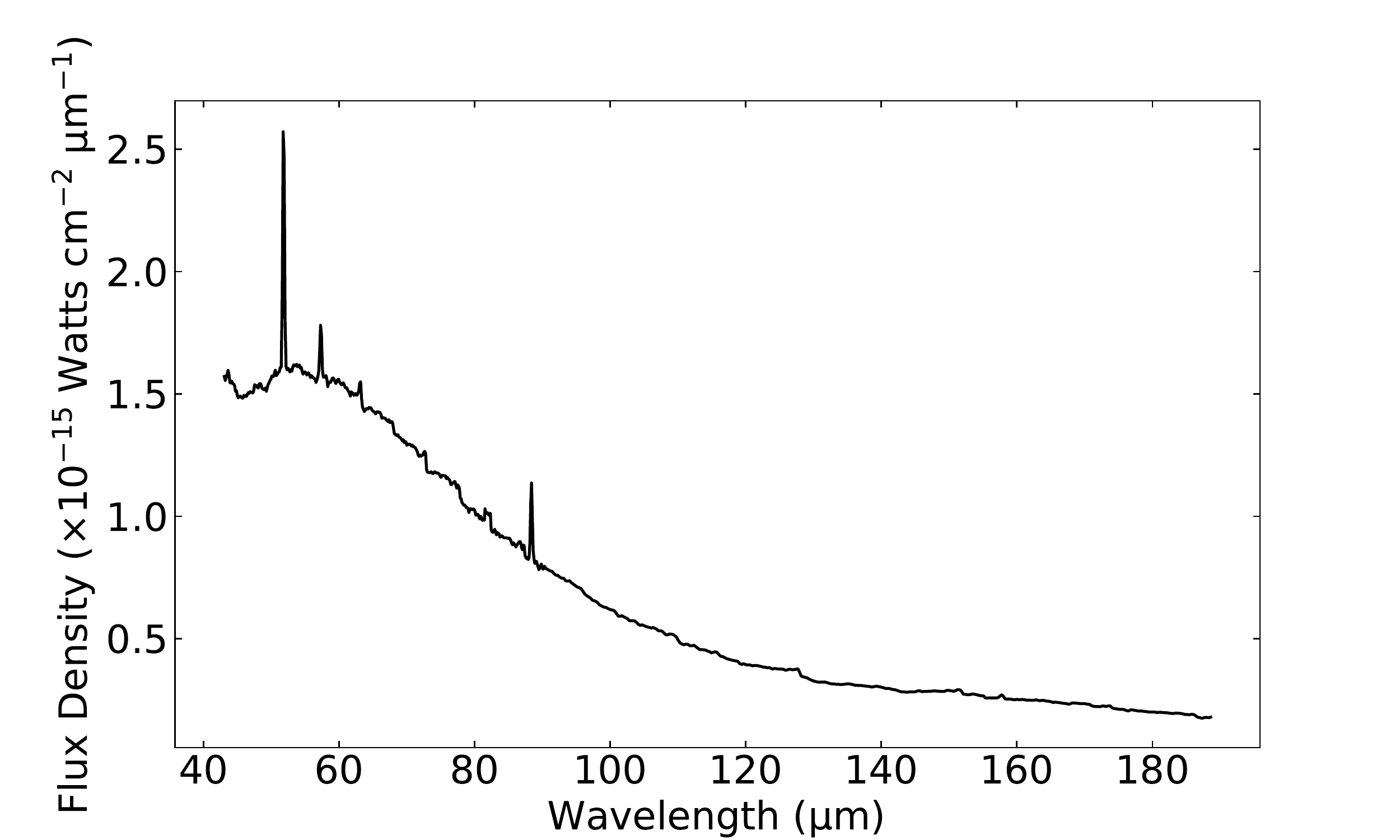}
 \put (12,55) {\Large \textbf{W51}}
\end{overpic}&
\includegraphics[width=5.2cm]{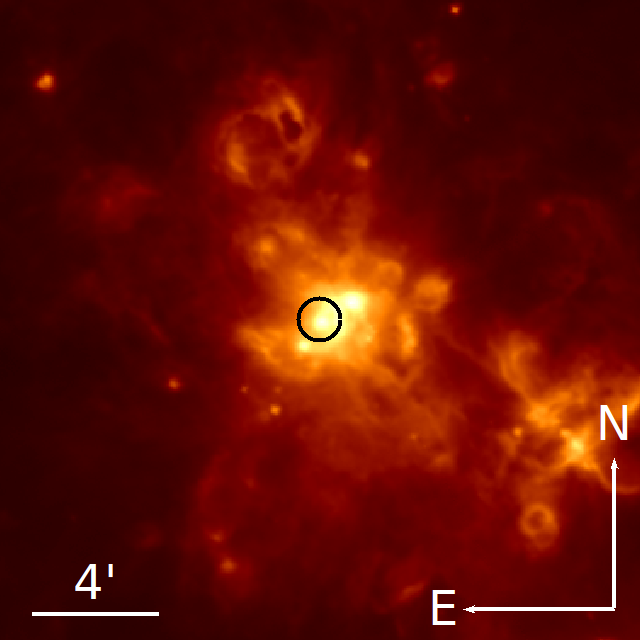}
\end{tabular}
\caption{ 
Left: The $ISO$ LWS spectrum of \hypertarget{g159p6-exact}{W51}. Right: The {\it Herschel} PACS 70 \textmugreek{}m image of W51. The $ISO$ LWS pointing is shown with a black circle.
} 
\label{fig:w51}
\end{figure*}

\begin{figure*}
\centering
\begin{tabular}{lc}
\begin{overpic}[width=8 cm,tics=10]{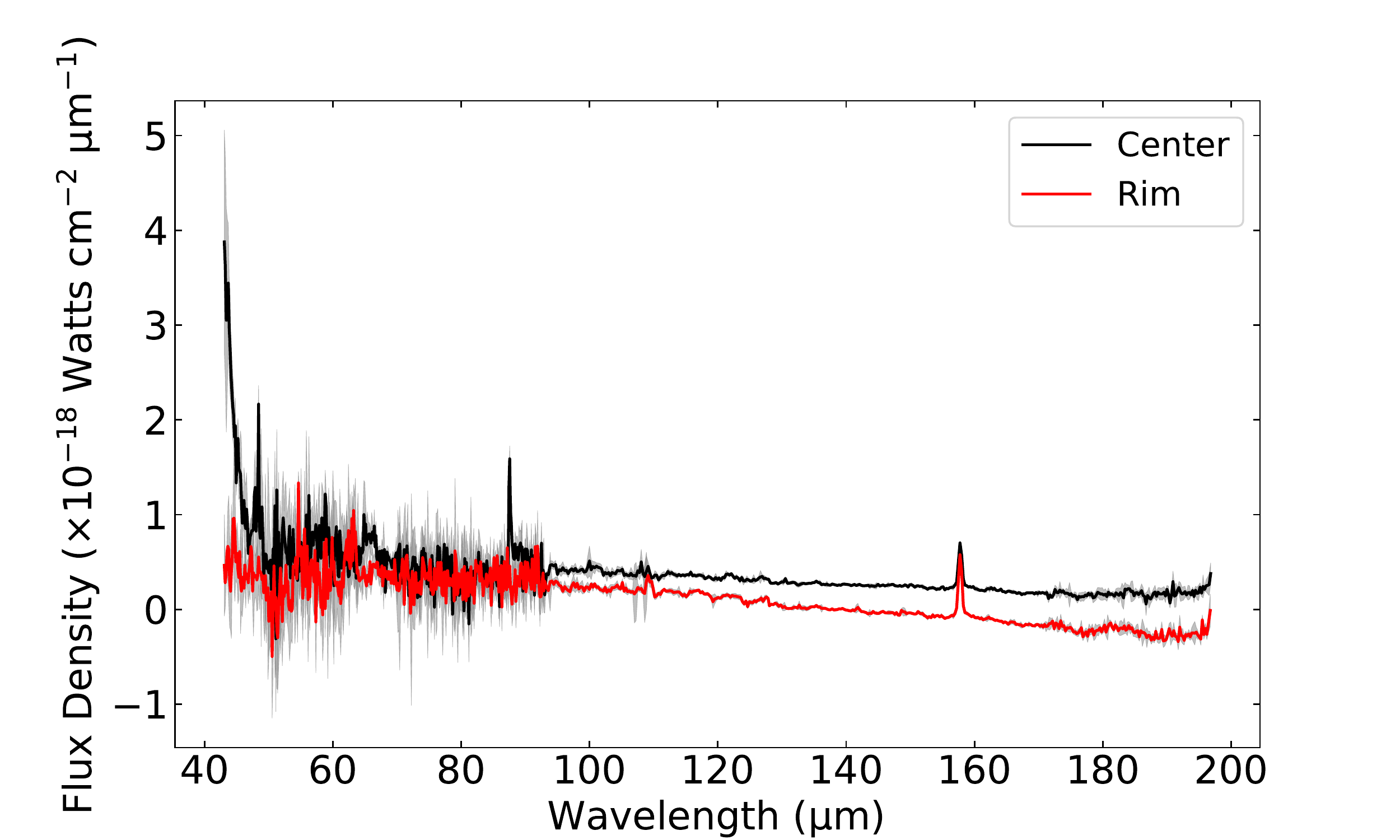}
 \put (12,55) {\Large \textbf{G159.6-18.5}}
\end{overpic}&
\includegraphics[width=5.2cm]{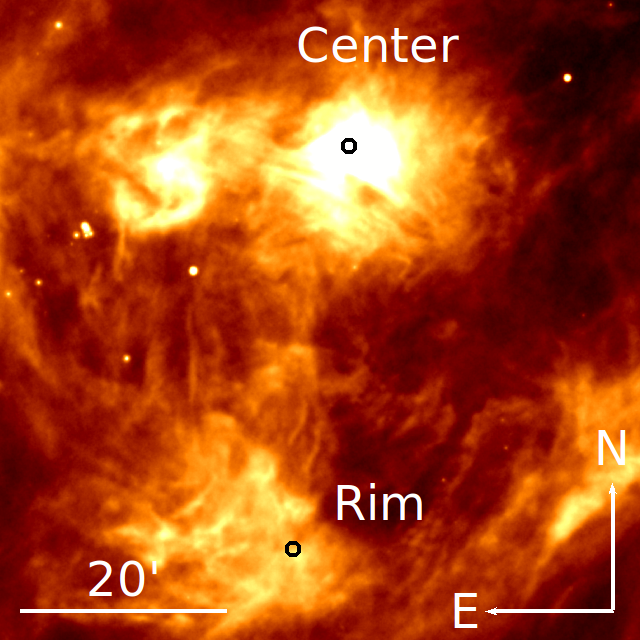}
\end{tabular}
\caption{
Left: The $ISO$ LWS spectra of \hypertarget{ngc6334-exact}{G159.6-18.5}. Right: The {\it WISE} 22 \textmugreek{}m image of G159.6-18.5. The $ISO$ LWS pointings are shown with black circles.
} 
\label{fig:g159p6}
\end{figure*}

\begin{figure*}[!ht]
\centering
\begin{tabular}{lc}
\begin{overpic}[width=8 cm,tics=10]{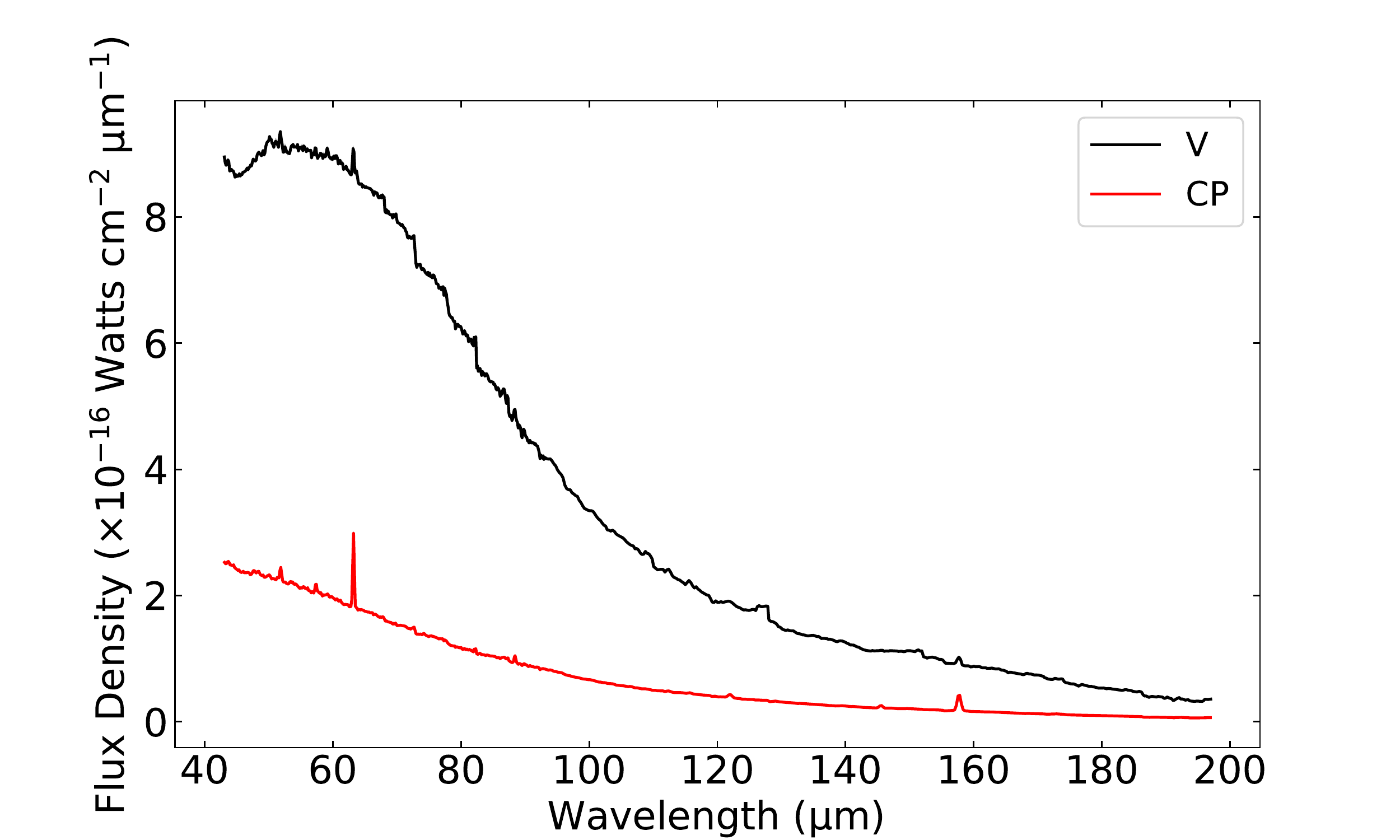}
 \put (12,55) {\Large \textbf{NGC 6334}}
\end{overpic}&
\includegraphics[width=5.2cm]{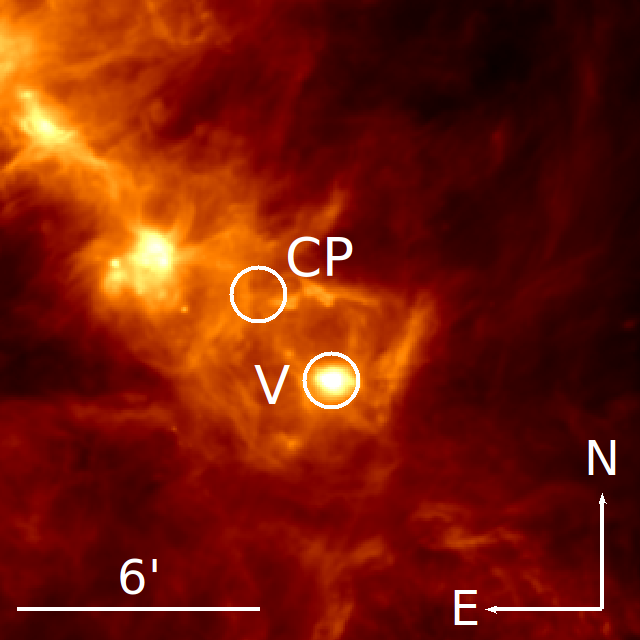}
\end{tabular}
\caption{
Left: The $ISO$ LWS spectra of NGC 6334. Right: The {\it Herschel} PACS 70 \textmugreek{}m image of NGC 6334. The $ISO$ LWS pointings are shown with white circles.
} 
\label{fig:ngc6334}
\end{figure*}

The LWS consists of 10 subdetectors (SW1 -- SW5 and LW1 -- LW5), each covering a fraction of the total LWS wavelength range, 43 -- 197 \textmugreek{}m. These spectra are often ``fractured'' -- the flux levels of some detectors do not match adjacent detectors by up to $\sim 10\%$.  To align the flux levels between adjacent detectors, we scaled each subdetector spectrum based on the average flux of the overlapping wavebands such that the full broadband spectrum becomes continuous without any jump in flux density. This scaling negligibly affects the emission line profiles. However, the scaled overall flux level of the continuum may vary by up to a factor of $\sim$ 2. Since SNRs in our sample are diffuse emission sources, we applied the effective area correction for the extended sources \citep{gry03} (Section \ref{sec:continuum}). Application of the extended source correction causes the overall flux level of the spectrum to be reduced by up to 40\%. In Figure \ref{fig:rcw103_before-after}, we show an example of the LWS spectrum before and after the flux scaling and application of extended source correction. Figures \ref{fig:kep} -- \ref{fig:ngc6334} show the resulting minimum broadband LWS spectrum of each SNR and HII region in our sample, along with the pointing position and aperture (80\arcsec{} diameter) overlaid on an infrared image of the SNR from $Herschel$, $Spitzer$, or $WISE$ observations. The line fluxes are listed in Table \ref{tab:lineflux} and the changes in the fluxes between the minimum and maximum broad-band spectra are less than 1\%.

We found that some flux values in the LWS spectra have zero uncertainties. The wavelengths at which these zero flux uncertainties appear vary depending on the observation. Their origin is unclear, although they may originate from glitches in the detector photocurrent.  In some cases, these fluxes were in regions of interest (at wavelengths corresponding to identified emission lines) in the LWS spectra.  Since these values were consistent with the fluxes that have non-zero uncertainties, instead of removing them, we assigned the largest uncertainty value in the subdetector spectrum to these fluxes.

\onecolumngrid

%
%
%
\clearpage

\onecolumngrid
%
\onecolumngrid
%
\clearpage

\vspace{-1cm}


%
%
%
%

\onecolumngrid

%
%

\startlongtable
\begin{deluxetable*}{lllllclllc}
\vspace{2.cm}
\tablecaption{$ISO$ LWS Observations of Our Sample of SNRs and HII Regions \label{tab:obsList}} 
\tablehead{\colhead{Name} & \colhead{Target Name*} & \colhead{RA, DEC (J2000)} & \colhead{Date} & \colhead{t$_{on}$ (s)} & \colhead{Publications} &\colhead{Figure}}
\startdata
\hline
\multicolumn{4}{l}{SNRs } \\
\hline
G4.5+6.8 & Kepler $\#$ 1 & 17\textsuperscript{h}30\textsuperscript{m}41\fs{}79 --21\degr{}29\arcmin{}29\farcs{}4 & 1996-02-16 & 814 &  This Work & \ref{fig:kep} \\
G4.5+6.8 & Kepler $\#$ 2  & 17\textsuperscript{h}30\textsuperscript{m}37\fs{}77 --21\degr{}28\arcmin{}19\farcs{}7 & 1996-02-16 & 812 &  This Work & \ref{fig:kep}\\
G6.4-0.1 & W 28 & 18\textsuperscript{h}00\textsuperscript{m}38\fs{}65 --23\degr{}25\arcmin{}03\farcs{}8 & 1996-02-24 & 814 &  This Work &  \ref{fig:w28}\\ 
G21.5--0.9 & G21.5--0.9 & 18\textsuperscript{h}33\textsuperscript{m}34\fs{}08 --10\degr{}34\arcmin{}04\farcs{}7 & 1996-10-11 & 1064 & This Work & \hyperlink{g21p5-exact}{\ref*{fig:g21p5}} \\
G29.7--0.3 & G29.7--0.2 & 18\textsuperscript{h}46\textsuperscript{m}25\fs{}19 --02\degr{}58\arcmin{}33\farcs{}9 & 1997-10-31 & 1126 & 3, This Work & \hyperlink{g29p7-exact}{\ref*{fig:g29p7}}\\ 
G33.6+0.1 & Kes 79 (L02) & 18\textsuperscript{h}52\textsuperscript{m}54\fs{}00 +00\degr{}41\arcmin{}23\farcs{}9 & 1996-10-04 & 1424 &  This Work & \ref{fig:kes79}\\ 
G54.1+0.3 & G54.1+0.3 & 19\textsuperscript{h}30\textsuperscript{m}30\fs{}69 +18\degr{}51\arcmin{}53\farcs{}1 & 1996-04-24  & 1052 & This Work & \hyperlink{g54p1-exact}{\ref*{fig:g54p1}}\\
\enddata
\tablecomments{{\scriptsize{ 1. \citep{snell05} 2. \citep{liseau06} 3. \citep{he05} 4. \citep{liu01}}}}
\tablenotetext{*}{The majority of targets were observed in the L01 mode, which covered the full range of the LWS detector. A few targets were observed in the L02 mode, which covered only certain wavelength bands specified by the observer.}
\end{deluxetable*}


\vspace{-3cm}


\startlongtable
\begin{deluxetable*}{lccccccll} 
\small
\centering
\tabletypesize{\small}
\tablecaption{Observed Spectral Lines from $ISO$ LWS Spectra of Supernova Remnants \label{tab:lineflux}}
\tablehead{
\colhead{Target} & \colhead{Wavelength}  &  \colhead{Line} & \colhead{FWHM (\%)$^c$}  & \colhead{Surface Brightness}  &  \colhead{Velocity$^a$} & \colhead{Shift$^{b,k}$}   \\
\colhead{} & \colhead{($\mu$m)}  & \colhead{} & \colhead{($\mu$m)} & \colhead{(10$^{-4}$ ergs s\textsuperscript{--1}cm\textsuperscript{--2}sr\textsuperscript{--1})} & \colhead{(km s\textsuperscript{--1})}  & \colhead{(km s\textsuperscript{--1})} 
}
\startdata
{\bf Kepler (\#1) } & 157.73$_{-0.01}^{+0.01}$ & [C II] & 0.55$_{-0.03}^{+0.04}$ &  0.06$_{-0.01}^{+0.02}$ & ... & ... \\
{\bf Kepler (\#2) } {\it (G4.5+6.8)} &$^h$157.74$_{-0.02}^{+0.02}$ & [C II] & 0.56$_{-0.06}^{+0.09}$ & 0.08$_{-0.02}^{+0.02}$ & ... & ... \\
{\bf W 28 }  &$^e$63.15$_{-0.01}^{+0.01}$ & [O I] & 0.31$_{-0.02}^{+0.02}$ (12$_{-7}^{+7}$\%)  & 1.3$_{-0.1}^{+0.2}$ & 700$_{-200}^{+200}$ & -160$_{-50}^{+60}$ \\
 {\it (G6.4--0.1)}              & 121.92$_{-0.02}^{+0.02}$ & [N II] & 0.76$_{-0.05}^{+0.06}$ (29$_{-8}^{+9}$\%)   & 0.8$_{-0.2}^{+0.2}$ & 1200$_{-200}^{+200}$  & ... \\ 
             &$^{e,i,g}$145.49$_{-0.04}^{+0.07}$ & [O I] & 0.60$_{-0.15}^{+0.09}$  & 0.14$_{-0.05}^{+0.09}$ & ... \\ 
                &$^g$157.73$_{-0.02}^{+0.02}$ & [C II] & 0.74$_{-0.03}^{+0.03}$ (27$_{-5}^{+5}$\%)   & 4.7$_{-0.4}^{+0.7}$ & 870$_{-90}^{+90}$ & ... \\
%
{\bf G21.5--0.9} & 63.21$_{-0.01}^{+0.01}$ & [O I] & 0.33$_{-0.01}^{+0.01}$ (20$_{-3}^{+3}$\%)   & 1.06$_{-0.08}^{+0.08}$ & 900$_{-80}^{+90}$ & 130$_{-20}^{+20}$ \\
              &$^g$121.83$_{-0.04}^{+0.03}$ & [N II] & 0.94$_{-0.09}^{+0.11}$ (61$_{-15}^{+18}$\%)   & 0.32$_{-0.04}^{+0.06}$ & 1800$_{-300}^{+400}$ & -180$_{-100}^{+80}$ \\
\enddata
\tablenotetext{a}{Estimated velocity dispersion after accounting for instrumental resolution.}
\tablenotetext{b}{Doppler shift from the line centroid.}
\tablenotetext{c}{Percent broader than instrument resolution.}
\tablenotetext{d}{Values are estimated from detector LW1.}
\tablenotetext{e}{Less than 5 $\sigma$ detection.}
\tablenotetext{f}{No line-broadening detected in adjacent subdetector.}
\tablenotetext{g}{High reduced chi-squared value ($\gtrsim$ 2).}
\tablenotetext{h}{Low reduced chi-squared value ($\lesssim$ 0.1).}
\tablenotetext{i}{Line not detected in adjacent subdetector.}
\tablenotetext{j}{L02 mode observation.}
\tablenotetext{\textsubscript{k}}{For remnants in the SMC or LMC, the Doppler-shift due to the motion of the galaxy has been subtracted.}
\tablenotetext{\textsubscript{l}}{The best--fit results for two--Gaussian + linear component model fits to each of the [O III] 52 and 88 \textmugreek{}m lines are shown.}
\tablenotetext{\textsubscript{...}}{ Value is approximately zero.}
%
\end{deluxetable*}


\begin{figure*}[!bp]
\centering
\begin{tabular}{llc}
\includegraphics[width=12.5cm]{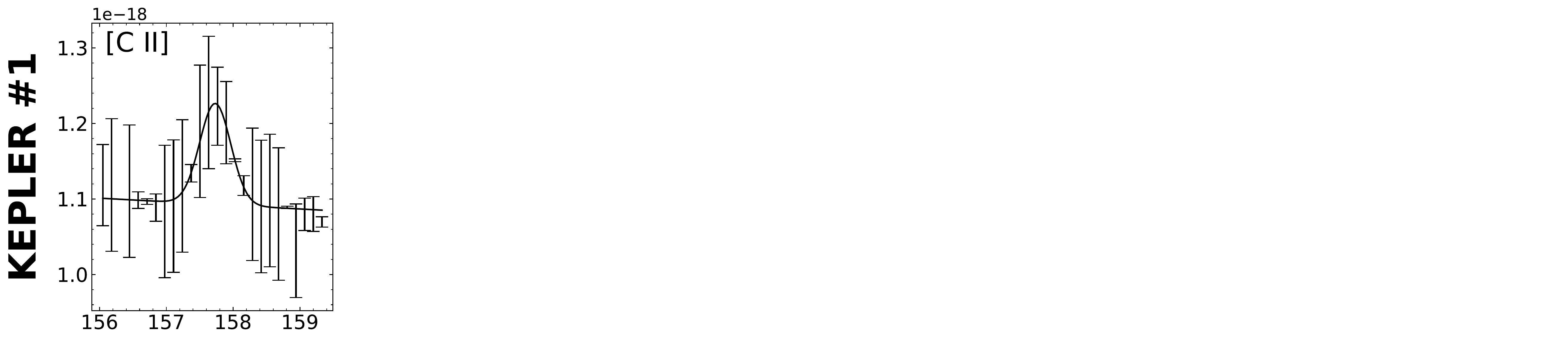} \hspace{-9.8cm} 
\includegraphics[width=12.5cm]{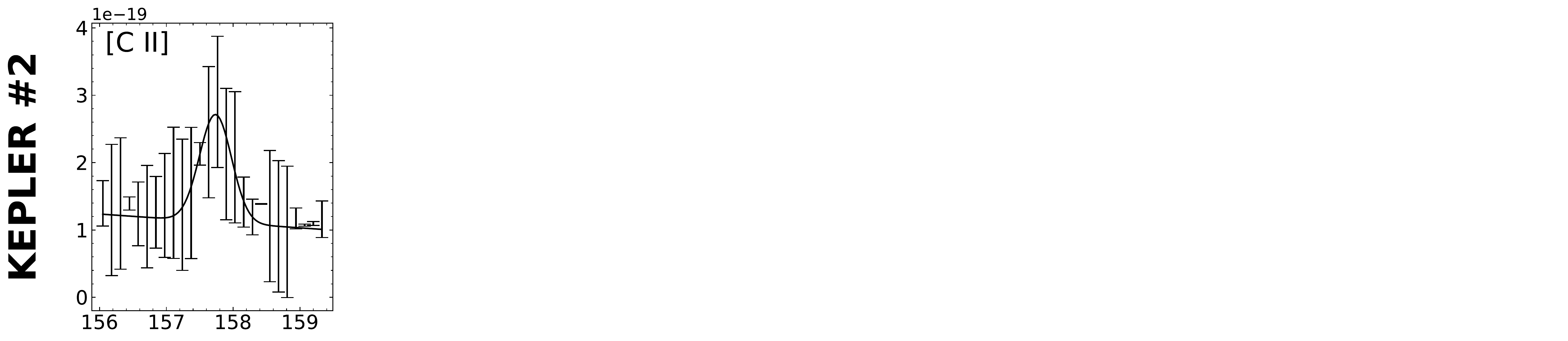}\hspace{-9.8cm} 
\includegraphics[width=20cm]{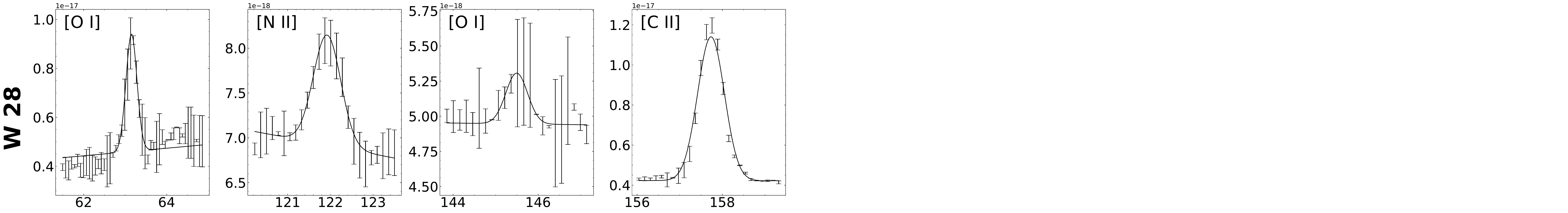}\\
\includegraphics[width=20cm]{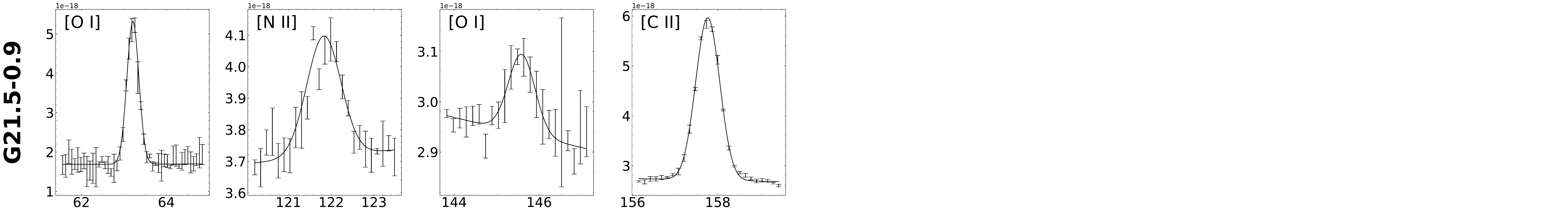}\\
\includegraphics[width=20cm]{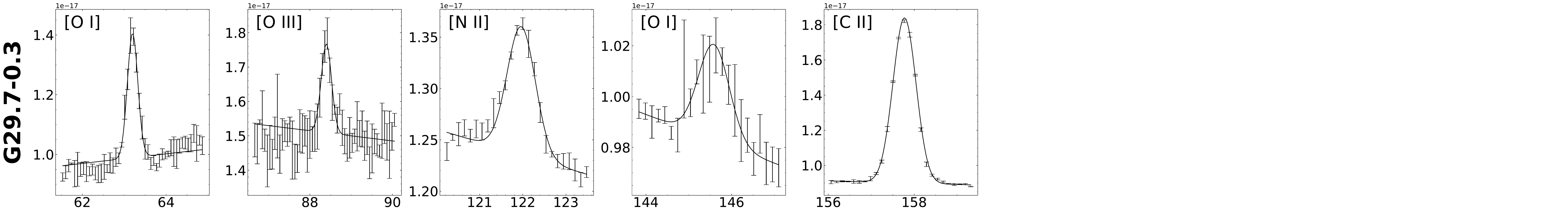}\\
\includegraphics[width=20cm]{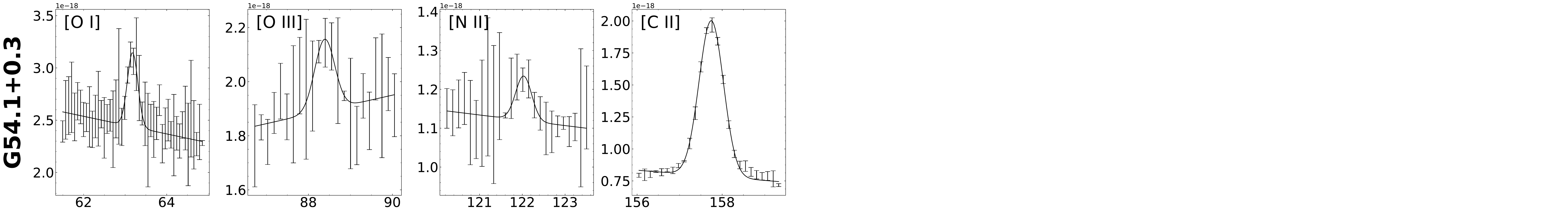}\\
\includegraphics[width=20cm]{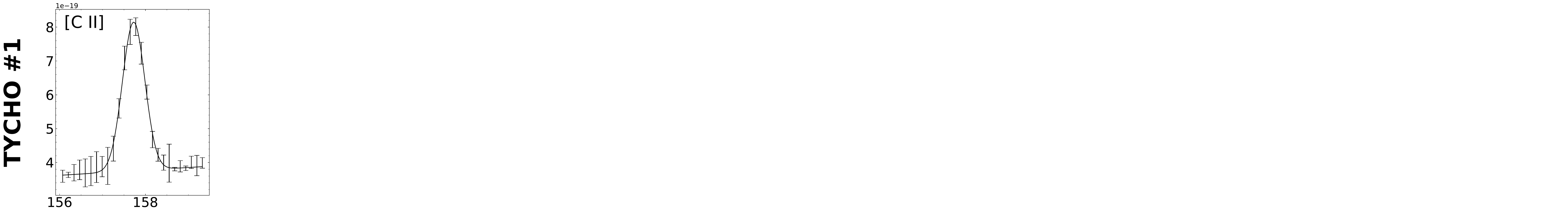}\hspace{-17.3cm}
\includegraphics[width=20cm]{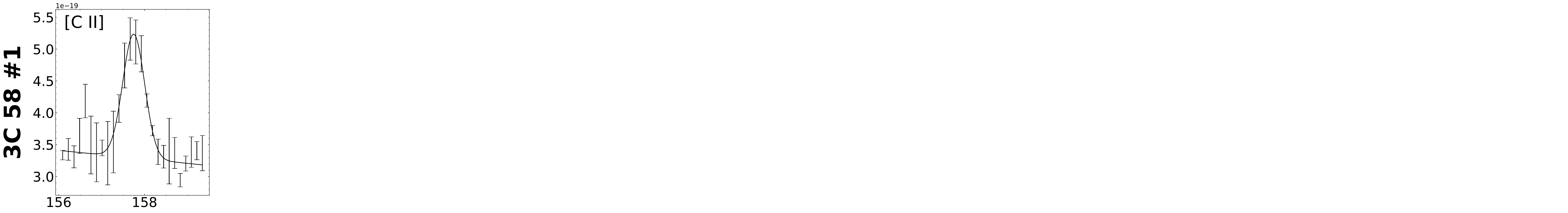} \hspace{-17.3cm} \includegraphics[width=20cm]{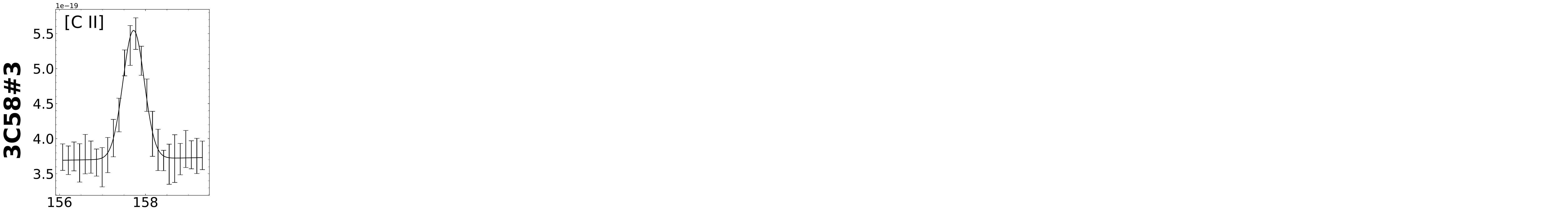} \hspace{-17.3cm}
\includegraphics[width=20cm]{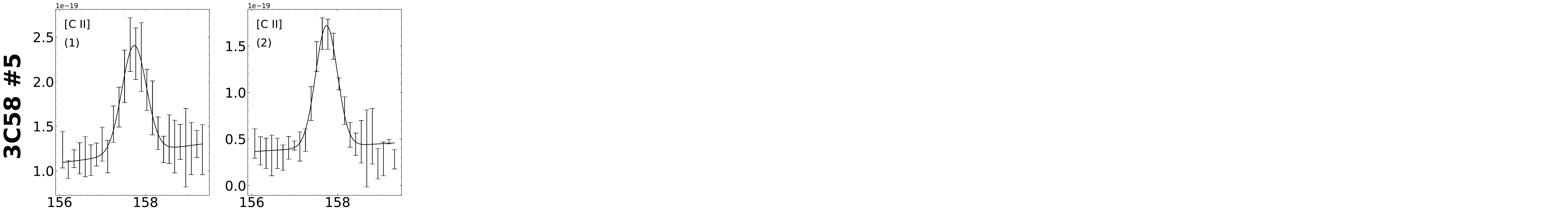} \\
\includegraphics[width=20cm]{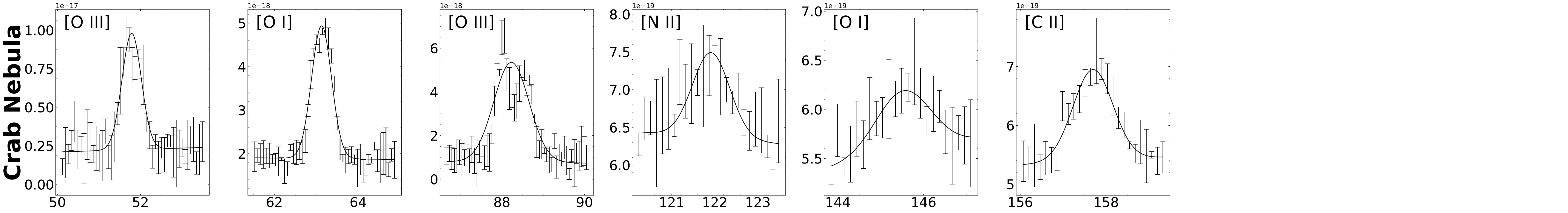}\\
\includegraphics[width=20cm]{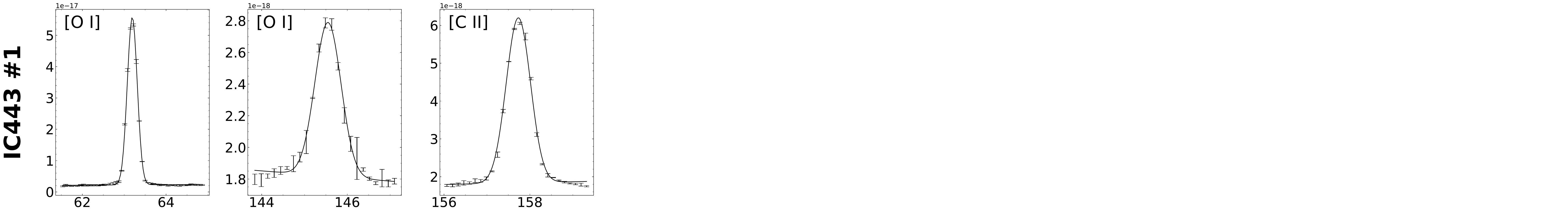}\hspace{-12.2cm}
\includegraphics[width=20cm]{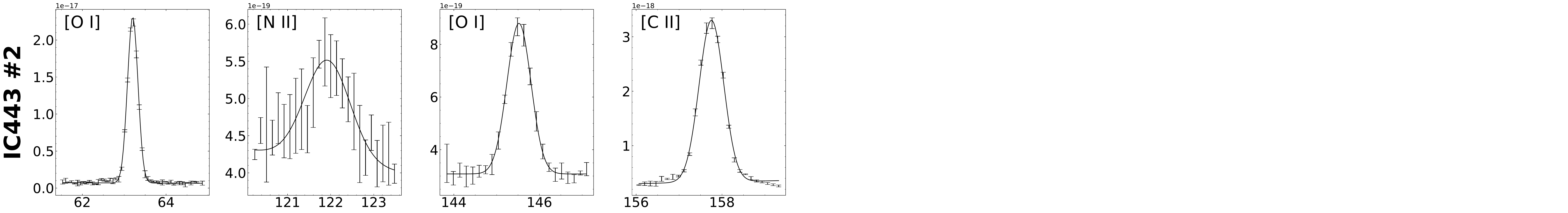}\\
\end{tabular}
    \caption{FIR Lines of SNRs. The best--fit Gaussian + linear model is overlaid. The y-axis is in units of flux density (erg s\textsuperscript{--1} cm\textsuperscript{--2} sr\textsuperscript{--1}) and x-axis in units of wavelength (\textmugreek{}m).   }
    \label{fig:lineFigs} 
\end{figure*}

\newpage
\onecolumngrid


\begin{figure*}[!htbp]
\centering
\begin{tabular}{lcc}
\includegraphics[width=20cm]{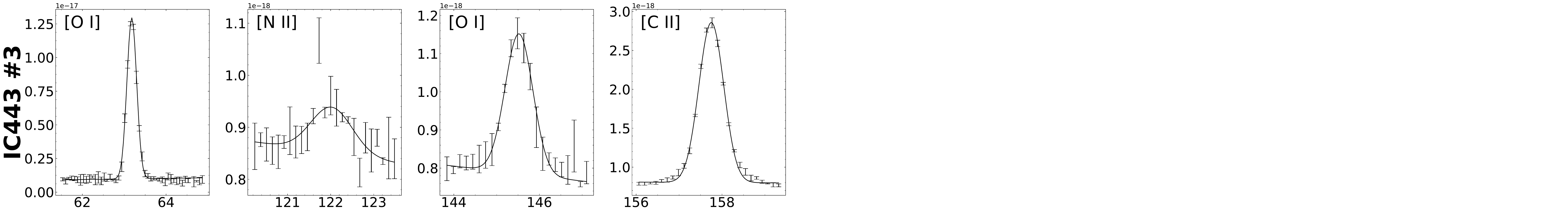} \hspace{-10cm}
\includegraphics[width=20cm]{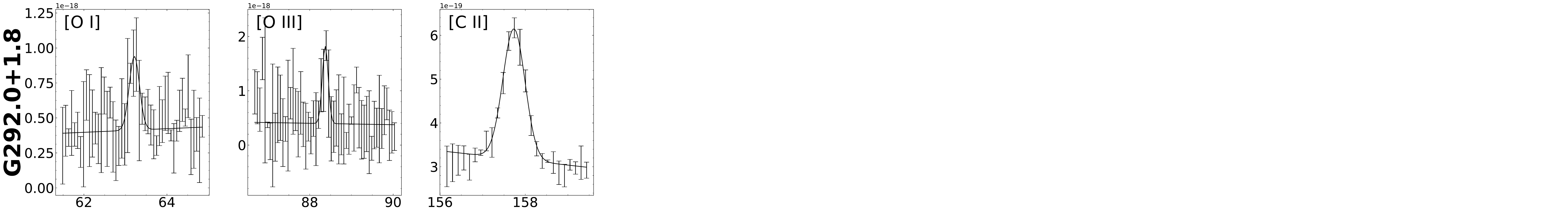}\\
\includegraphics[width=20cm]{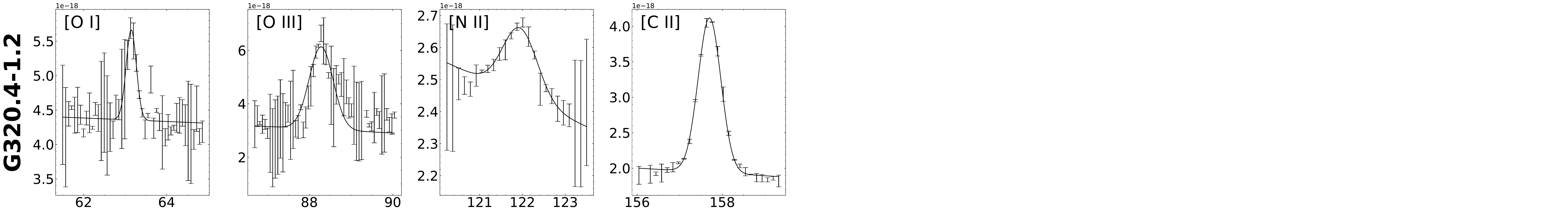}\\
\includegraphics[width=20cm]{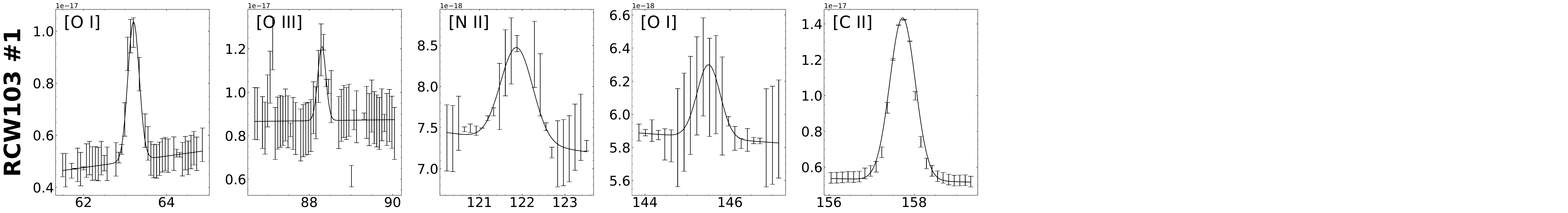}\\
\includegraphics[width=20cm]{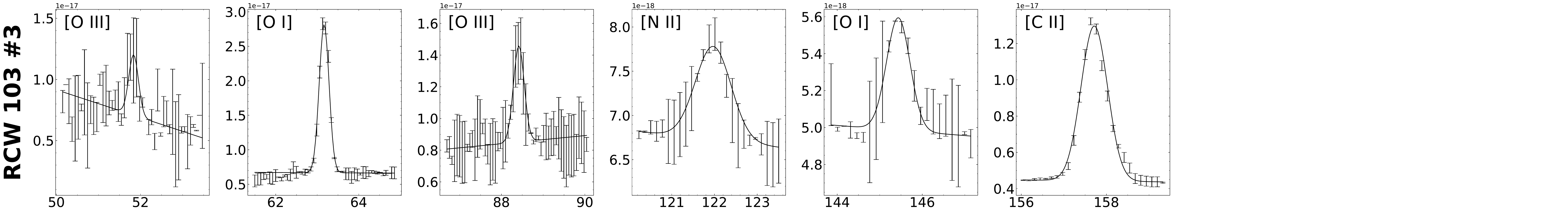}\\
\includegraphics[width=20cm]{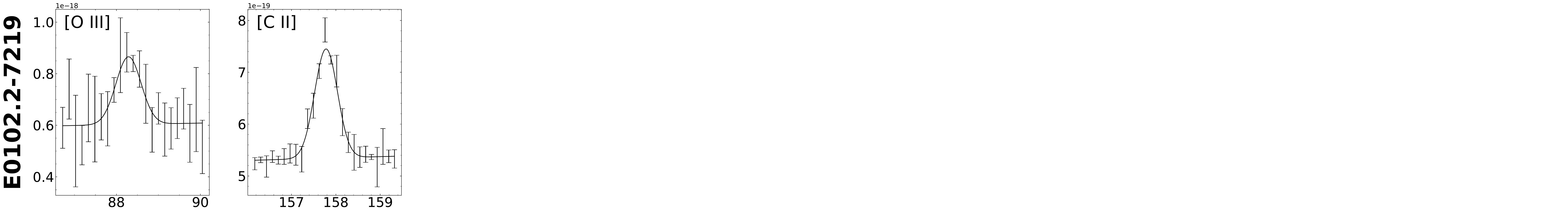} \hspace{-14.7cm}
\includegraphics[width=20cm]{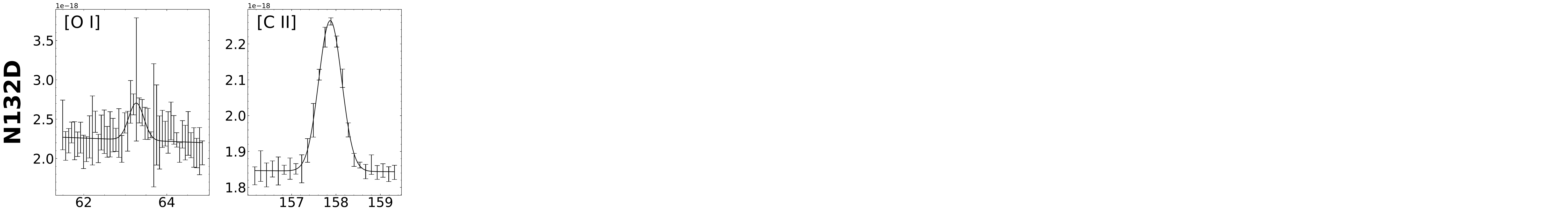}\\
\includegraphics[width=20cm]{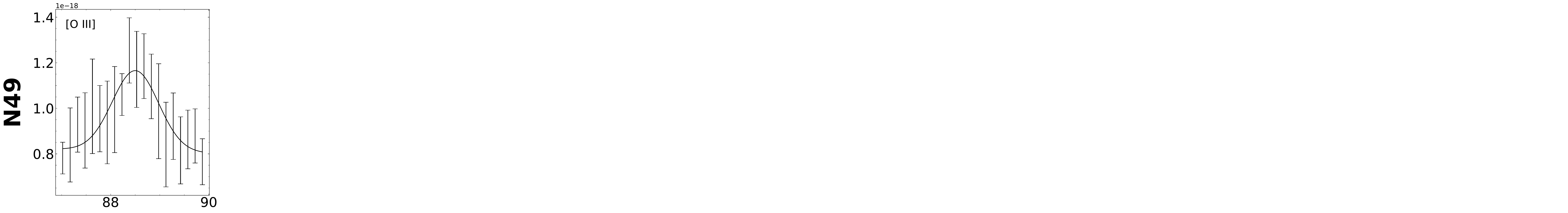}\hspace{-17cm}
\includegraphics[width=20cm]{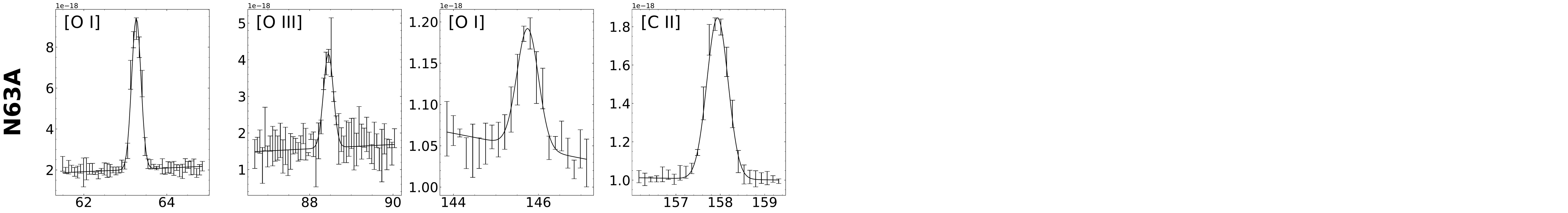}\\
\includegraphics[width=12.7cm]{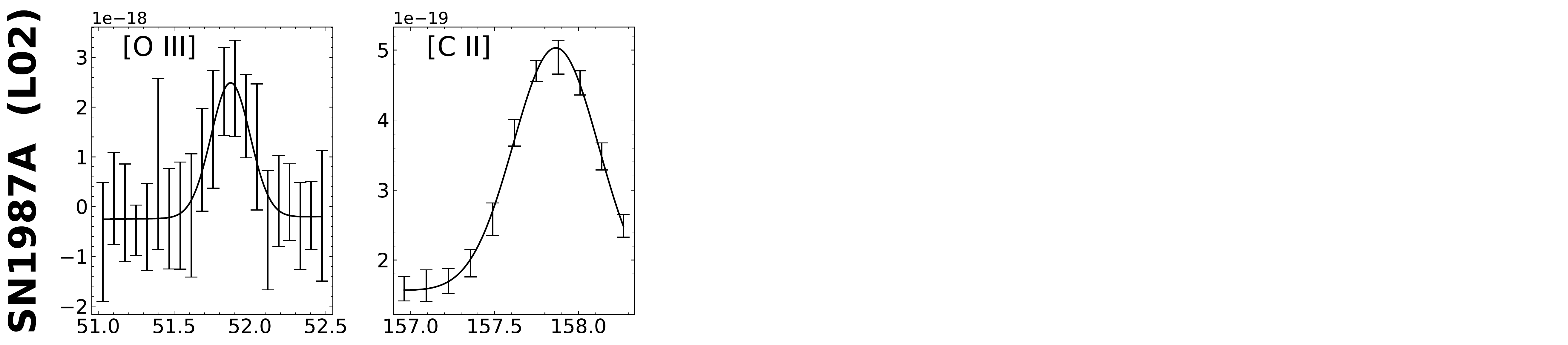} \hspace{-7.5cm}
\includegraphics[width=20cm]{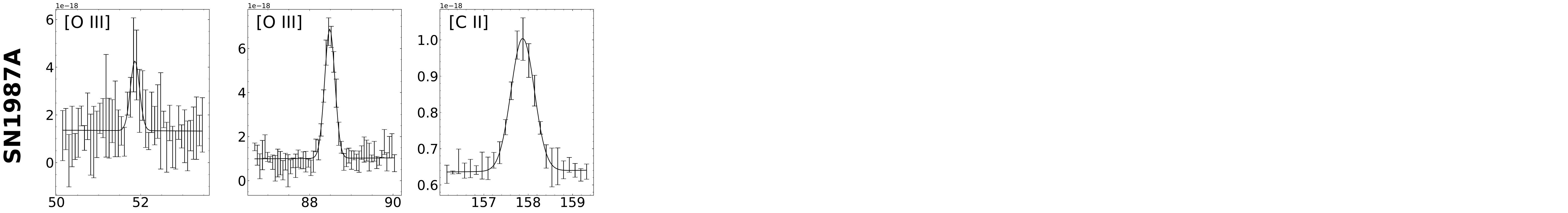}\\
\includegraphics[width=20cm]{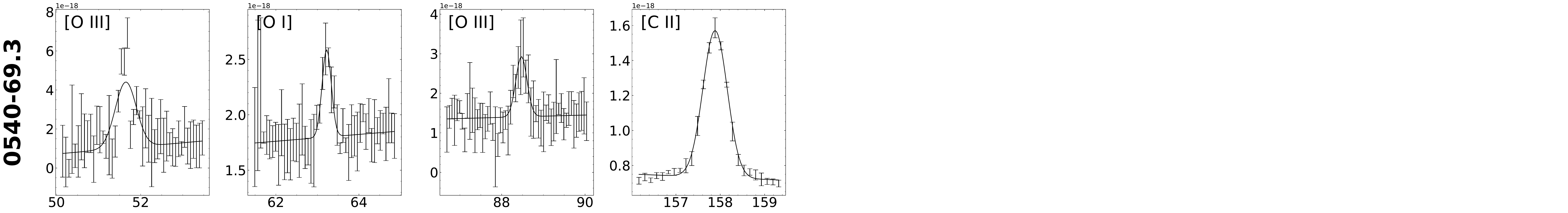}\\
\end{tabular}
    \caption{FIR Lines of SNRs (continued) }
    \label{fig:lineFigs} 
\end{figure*}

\clearpage

\onecolumngrid


\startlongtable
\begin{deluxetable*}{lccccccc}
\small
\centering
\tabletypesize{\footnotesize}
\tablecaption{Observed Spectral Lines from $ISO$ LWS Spectra of HII Regions and PN NGC 6720  \label{tab:hiispec}}
\tablehead{
\colhead{Target} & \colhead{Wavelength}  &  \colhead{Line} & \colhead{FWHM (\%)$^c$}   & \colhead{Surface Brightness} \\ 
\colhead{} &\colhead{($\mu$m)} & \colhead{} & \colhead{($\mu$m)}   & \colhead{(10$^{-4}$ ergs s\textsuperscript{--1}cm\textsuperscript{--2}sr\textsuperscript{--1})} 
}
\startdata
%
%
{\bf W51} & 51.82$_{-0.01}^{+0.01}$ & [O III] & 0.283$_{-0.004}^{+0.004}$  & 289$_{-8}^{+8}$ \\ 
                                & 63.17$_{-0.01}^{+0.01}$ & [O I] & 0.283$_{-0.008}^{+0.009}$  & 41$_{-2}^{+2}$ \\ 
                                & 88.39$_{-0.01}^{+0.01}$ & [O III] & 0.287$_{-0.005}^{+0.005}$  & 124$_{-4}^{+4}$ \\ 
                                &145.54$_{-0.01}^{+0.01}$ & [O I] & 0.56$_{-0.03}^{+0.03}$  & 4.9$_{-0.4}^{+0.5}$ \\
                                & 157.74$_{-0.01}^{+0.01}$ & [C II] & 0.58$_{-0.03}^{+0.03}$  & 12$_{-1}^{+1}$ \\ 
{\bf G159 (Center)} {\it (G159.6-18.5)} & 157.76$_{-0.01}^{+0.01}$ & [C II] & 0.62$_{-0.01}^{+0.01}$  & 0.37$_{-0.02}^{+0.02}$ \\ 
{\bf G159 (Rim)} & 157.76$_{-0.01}^{+0.01}$ & [C II] & 0.57$_{-0.01}^{+0.01}$  & 0.49$_{-0.02}^{+0.02}$ \\ 
{\bf NGC6334 (V)} &51.81$_{-0.03}^{+0.03}$ & [O III] & 0.38$_{-0.08}^{+0.09}$    & 12$_{-4}^{+4}$ \\
\enddata
\tablecomments{Notation the same as in Table \ref{tab:lineflux}}
\end{deluxetable*}

%





%
%

\startlongtable
\begin{deluxetable*}{cccccccc}
%
%
\small
\centering
\tablecaption{Kes 79 Raster Map 63 \textmugreek{}m [O I]  Observations \label{tab:kes79raster}}
\tablehead{
\colhead{Raster} & \colhead{R.A.} & \colhead{Decl.} & \colhead{FWHM (\%)$^c$}  & \colhead{Surface Brightness}  &  \colhead{Velocity$^a$} &  \colhead{Shift$^b$}\\
 \colhead{} &  \colhead{(J2000)} &  \colhead{(J2000)} & \colhead{($\mu$m)}  & \colhead{(10$^{-4}$ ergs s\textsuperscript{--1}cm\textsuperscript{--2}sr\textsuperscript{--1})} & \colhead{km s\textsuperscript{--1}} & \colhead{(km s\textsuperscript{--1})}
}
\startdata
1 & 18\textsuperscript{h}52\textsuperscript{m}42\fs{}33 & +00\degr{}39\arcmin{}46\farcs{}89 & 0.27$_{-0.01}^{+0.01}$  & 0.68$_{-0.04}^{+0.04}$ & ... & ...\\
2 & 18\textsuperscript{h}52\textsuperscript{m}45\fs{}25 & +00\degr{}40\arcmin{}11\farcs{}13  & 0.29$_{-0.01}^{+0.01}$  & 0.69$_{-0.05}^{+0.05}$ & ... & ...\\
3& 18\textsuperscript{h}52\textsuperscript{m}48\fs{}17 & +00\degr{}40\arcmin{}35\farcs{}37 & 0.291$_{-0.01}^{+0.01}$  & 0.65$_{-0.04}^{+0.04}$ & ... & ...\\
4& 18\textsuperscript{h}52\textsuperscript{m}51\fs{}08 & +00\degr{}40\arcmin{}59\farcs{}62  & 0.29$_{-0.01}^{+0.02}$  & 0.59$_{-0.06}^{+0.06}$ & ... & ...\\
5& 18\textsuperscript{h}52\textsuperscript{m}54\fs{}00 & +00\degr{}41\arcmin{}23\farcs{}86  & 0.28$_{-0.01}^{+0.01}$  & 0.52$_{-0.02}^{+0.02}$ & ... & 100$_{-10}^{+10}$ \\
6& 18\textsuperscript{h}52\textsuperscript{m}56\fs{}92 & +00\degr{}41\arcmin{}48\farcs{}09  & 0.28$_{-0.01}^{+0.01}$  & 0.62$_{-0.03}^{+0.03}$ & ... & 80$_{-20}^{+20}$ \\
7& 18\textsuperscript{h}52\textsuperscript{m}59\fs{}83 & +00\degr{}42\arcmin{}12\farcs{}34  & 0.28$_{-0.01}^{+0.01}$ & 0.63$_{-0.02}^{+0.02}$ & ... & 70$_{-10}^{+10}$ \\
8& 18\textsuperscript{h}53\textsuperscript{m}02\fs{}75 & +00\degr{}42\arcmin{}36\farcs{}58  & 0.31$_{-0.01}^{+0.01}$  & 0.71$_{-0.03}^{+0.03}$ & ... & ...\\
\enddata
\tablecomments{Notation the same as in Table \ref{tab:lineflux}}
\end{deluxetable*}

\onecolumngrid

\begin{figure*}[!b]
    \centering
\begin{tabular}{ccc}
\includegraphics[width=20cm]{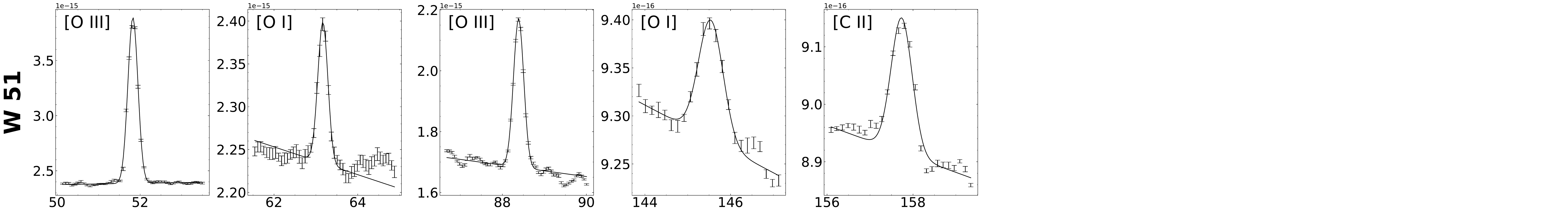}\\
\includegraphics[width=20cm]{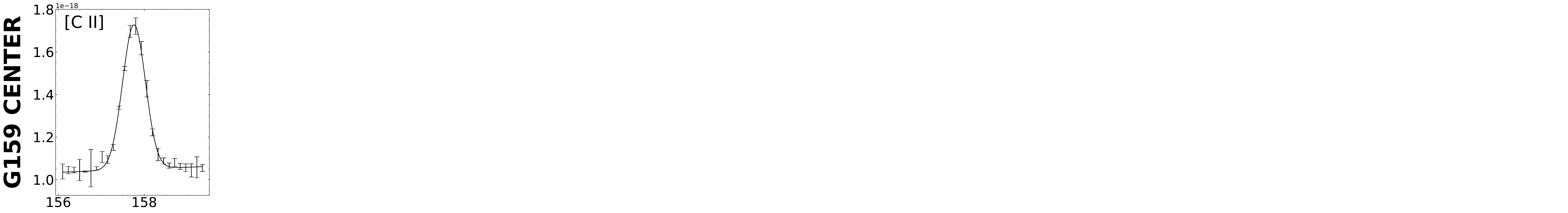} & \hspace{-17.5cm} \includegraphics[width=20cm]{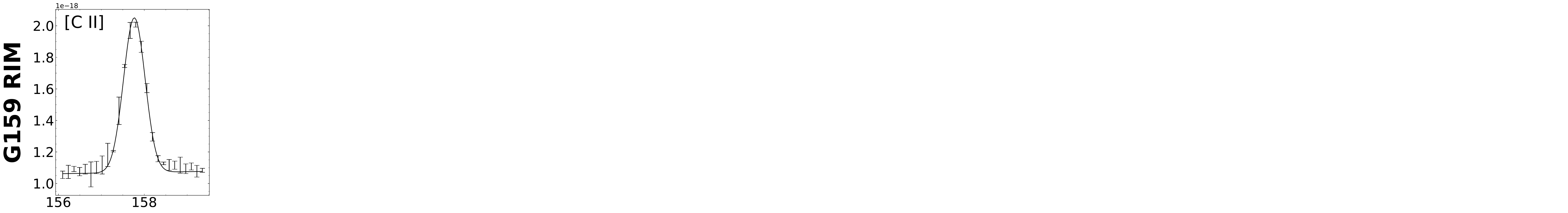} \\
\includegraphics[width=20cm]{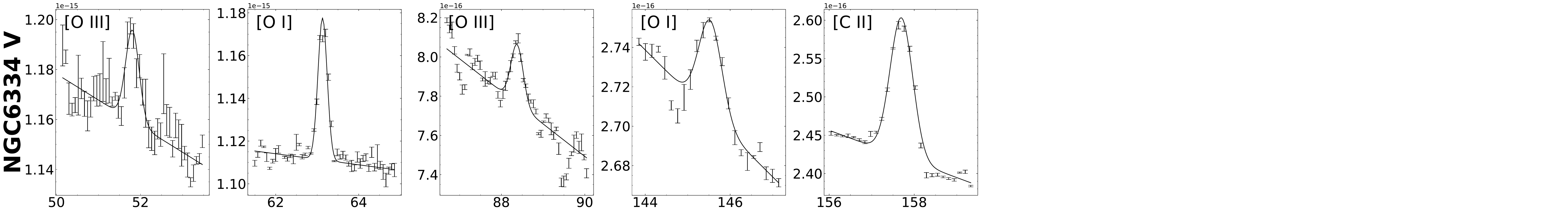}\\
\includegraphics[width=20cm]{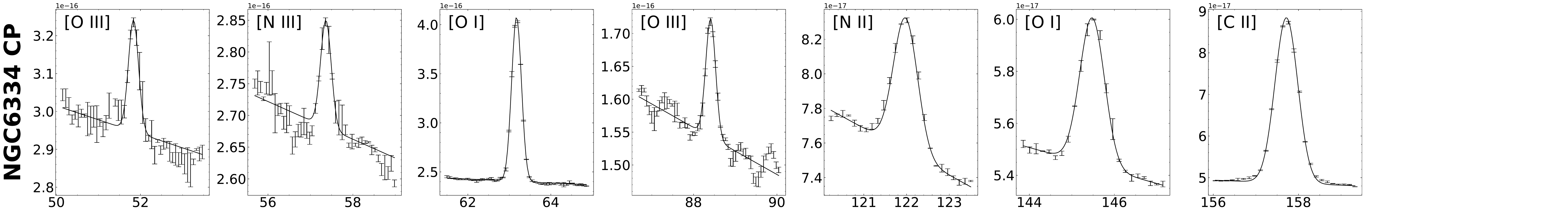}\\
\includegraphics[width=20cm]{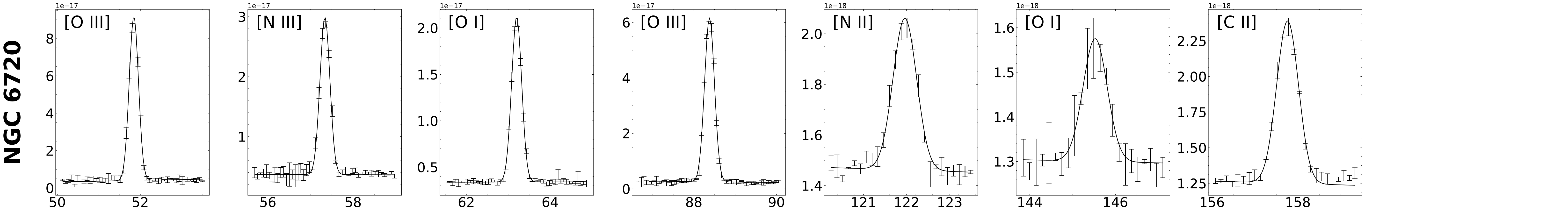}\\
\end{tabular}
    \caption{FIR Lines of HII Regions. Presentation is the same as Figure \ref{fig:lineFigs}.}
    \label{fig:lineHIIFigs} 
\end{figure*}


\startlongtable
\begin{deluxetable*}{cccccccc}
%
%
\small
\centering
\tablecaption{CTB 109 Raster Map 63 \textmugreek{}m [O I]  Observations \label{tab:ctb109raster}}
\tablehead{
\colhead{Raster} & \colhead{R.A.} & \colhead{Decl.} & \colhead{FWHM (\%)$^c$}   & \colhead{Surface Brightness}  &  \colhead{Velocity$^a$} &  \colhead{Shift$^b$}\\
 \colhead{} &  \colhead{(J2000)} &  \colhead{(J2000)} & \colhead{($\mu$m)} & \colhead{(10$^{-4}$ ergs s\textsuperscript{--1}cm\textsuperscript{--2}sr\textsuperscript{--1})} & \colhead{(km s\textsuperscript{--1})} & \colhead{(km s\textsuperscript{--1})}
}
\startdata
1 & 22\textsuperscript{h}59\textsuperscript{m}29\fs{}61 & +58\degr{}50\arcmin{}11\farcs{}35  & 0.28$_{-0.03}^{+0.02}$  & 0.20$_{-0.04}^{+0.03}$ & ... & ...\\
2 & 22\textsuperscript{h}59\textsuperscript{m}46\fs{}21 & +58\degr{}52\arcmin{}16\farcs{}99  & 0.29$_{-0.01}^{+0.01}$ & 0.26$_{-0.02}^{+0.02}$ & ... & -50$_{-10}^{+10}$ \\
3 & 23\textsuperscript{h}00\textsuperscript{m}02\fs{}87 & +58\degr{}54\arcmin{}22\farcs{}51 & 0.39$_{-0.05}^{+0.06}$ ($38_{-18}^{+21}$\%)   & 0.22$_{-0.05}^{+0.05}$ & 1300$_{-400}^{+400}$ & -150$_{-90}^{+100}$ \\
4 & 23\textsuperscript{h}00\textsuperscript{m}19\fs{}55 & +58\degr{}56\arcmin{}27\farcs{}89 & 0.27$_{-0.03}^{+0.03}$  & 0.14$_{-0.02}^{+0.02}$ & ... & ... \\
5 & 23\textsuperscript{h}00\textsuperscript{m}36\fs{}26 & +58\degr{}58\arcmin{}33\farcs{}14  & 0.24$_{-0.03}^{+0.03}$  & 0.09$_{-0.02}^{+0.02}$ & ... & ... \\
6 & 23\textsuperscript{h}00\textsuperscript{m}53\fs{}01 & +59\degr{}00\arcmin{}38\farcs{}25& 0.27$_{-0.07}^{+0.06}$  & 0.12$_{-0.05}^{+0.04}$ & ... & ... \\
7 & 23\textsuperscript{h}01\textsuperscript{m}09\fs{}79 & +59\degr{}02\arcmin{}43\farcs{}22  & 0.24$_{-0.05}^{+0.08}$ & 0.12$_{-0.07}^{+0.08}$ & ... & -200$_{-100}^{+200}$ \\
8 & 23\textsuperscript{h}01\textsuperscript{m}26\fs{}61 & +59\degr{}04\arcmin{}48\farcs{}07 & 0.35$_{-0.04}^{+0.03}$ (24$_{-14}^{+11}$\%)  & 0.21$_{-0.04}^{+0.04}$ & 1000$_{-300}^{+300}$ & ... \\
\enddata
\tablecomments{Notation the same as in Table \ref{tab:lineflux}}
\end{deluxetable*}

\onecolumngrid


\startlongtable
\begin{deluxetable*}{cccccccc}
\small
\centering
\tablecaption{IC 443 Raster Map Observations \label{tab:ic443raster}}\
\tablehead{
\colhead{Raster} & \colhead{R.A.} & \colhead{Decl.} & \colhead{FWHM (\%)$^c$}   & \colhead{Surface Brightness}  &  \colhead{Velocity$^a$} &  \colhead{Shift$^b$}\\
 \colhead{} &  \colhead{(J2000)} &  \colhead{(J2000)} & \colhead{($\mu$m)} & \colhead{(10$^{-4}$ ergs s\textsuperscript{--1}cm\textsuperscript{--2}sr\textsuperscript{--1})} & \colhead{(km s\textsuperscript{--1})} & \colhead{(km s\textsuperscript{--1})}
}
\startdata
\hline
63 \textmugreek{}m [O I] \\
\hline
1 & 06\textsuperscript{h}16\textsuperscript{m}58\fs{}52 & +22\degr{}13\arcmin{}45\farcs{}22  & 0.27$_{-0.02}^{+0.02}$  & 0.35$_{-0.07}^{+0.07}$ & ... & ... \\
2 & 06\textsuperscript{h}17\textsuperscript{m}00\fs{}78 & +22\degr{}16\arcmin{}42\farcs{}47 & 0.28$_{-0.02}^{+0.02}$  & 0.62$_{-0.06}^{+0.06}$ & ... & 120$_{-40}^{+40}$ \\
3 & 06\textsuperscript{h}17\textsuperscript{m}03\fs{}05 & +22\degr{}19\arcmin{}39\farcs{}71  & 0.30$_{-0.01}^{+0.01}$  & 1.97$_{-0.05}^{+0.05}$ & ... & ... \\
4 & 06\textsuperscript{h}17\textsuperscript{m}05\fs{}32 & +22\degr{}22\arcmin{}36\farcs{}95  & 0.31$_{-0.01}^{+0.01}$  & 0.77$_{-0.04}^{+0.04}$ & ... & ...\\
5 & 06\textsuperscript{h}17\textsuperscript{m}07\fs{}59 & +22\degr{}25\arcmin{}34\farcs{}19 & 0.26$_{-0.01}^{+0.01}$   & 9.7$_{-0.4}^{+0.4}$ & ... & 60$_{-10}^{+10}$ \\
6 & 06\textsuperscript{h}17\textsuperscript{m}09\fs{}86 & +22\degr{}28\arcmin{}31\farcs{}43 & 0.32$_{-0.01}^{+0.02}$  & 0.81$_{-0.06}^{+0.06}$ & ... & ...\\
7 & 06\textsuperscript{h}17\textsuperscript{m}12\fs{}14 & +22\degr{}31\arcmin{}28\farcs{}66 & 0.29$_{-0.02}^{+0.02}$  & 0.20$_{-0.08}^{+0.08}$ & ... & 290$_{-20}^{+20}$ \\
\enddata
%
\tablecomments{Notation the same as in Table \ref{tab:lineflux} }
\end{deluxetable*}

\begin{figure*}[!b]
    \centering
\begin{tabular}{cccccc}
\includegraphics[height=13.5cm]{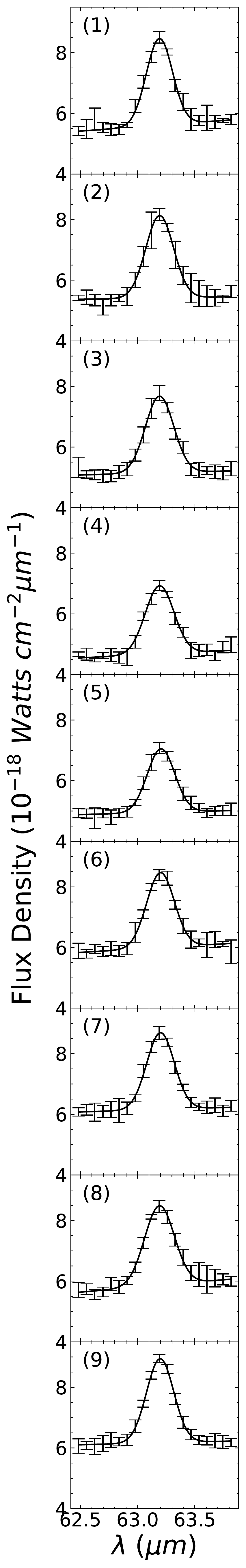} & \includegraphics[height=13.5cm]{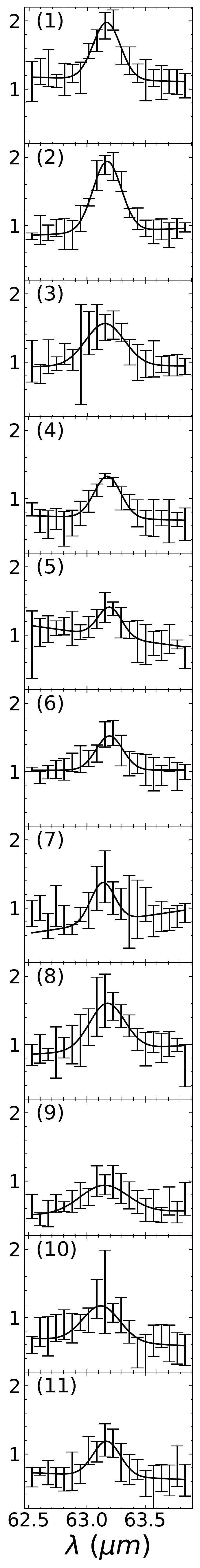} &  \includegraphics[height=13.5cm]{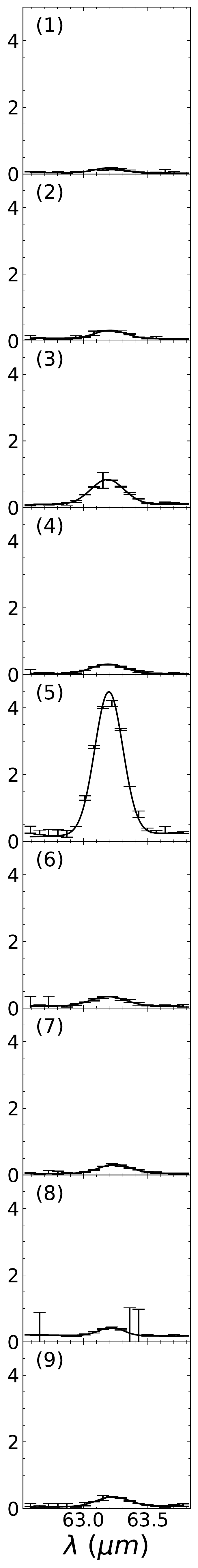} &
\includegraphics[height=13.5cm]{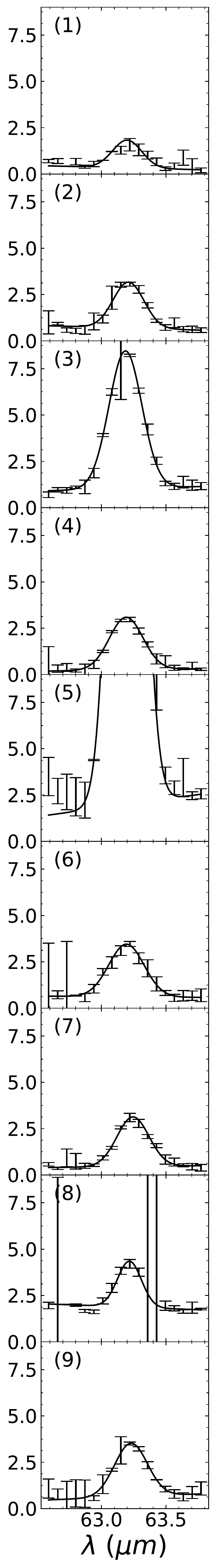}
 &
\includegraphics[height=13.5cm]{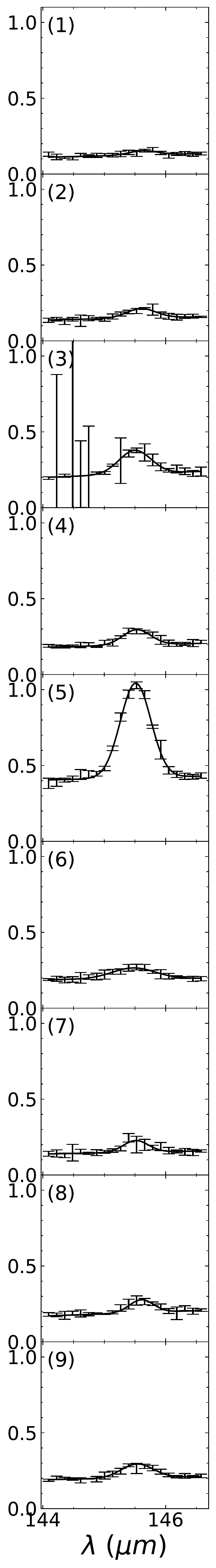} & \includegraphics[height=13.5cm]{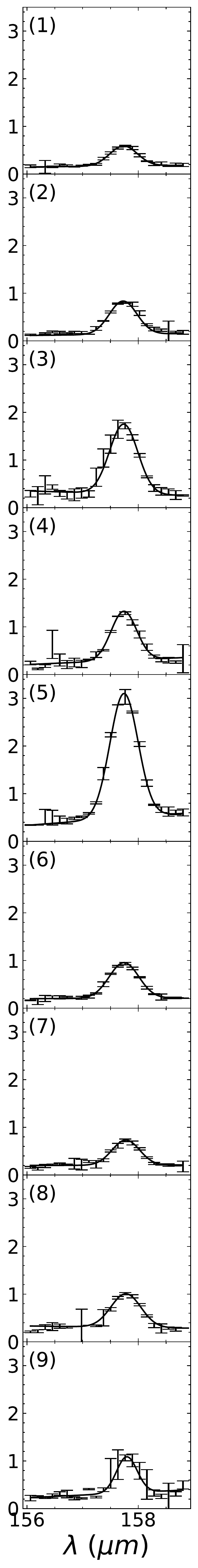} \\ 
\multicolumn{1}{c}{(a) Kes 79: 63 \textmugreek{}m } & \multicolumn{1}{c}{(b) CTB 109: 63 \textmugreek{}m} & \multicolumn{1}{c}{(c) IC 443: 63 \textmugreek{}m} &
\multicolumn{1}{c}{(d) Same as (c) Zoomed } &
\multicolumn{1}{c}{(e) IC 443: 145 \textmugreek{}m} & \multicolumn{1}{c}{(f) IC 443: 158 \textmugreek{}m}  \\
\end{tabular}
    \caption{FIR Lines of LWS Observations with multiple raster pointings.}
    \label{fig:lineRasters} 
\end{figure*}


\begin{table*}[!h]
\caption{Spectral Resolution of $ISO$ LWS} 
\begin{center}
\begin{tabular}{llccccccc}
\hline \hline
Detector & Line & \multicolumn{2}{c}{Detector FWHM\textsuperscript{a}}  & \multicolumn{2}{c}{NGC 6720 FWHM}  
& \\
 &  & (\textmugreek{}m) & (km s$^{-1}$) &  (\textmugreek{}m) & (km s$^{-1}$) \\
\hline
SW2 (49.5--64 \textmugreek{}m) &52 $\mu$m ([O III]) & 0.283$\pm$0.009 & 1640$\pm$50 & 0.262$\pm$0.006 & 1510$\pm$40 \\
SW3 (57--70 \textmugreek{}m)  &57 $\mu$m ([N III])  & 0.283$\pm$0.009 & 1480$\pm$50 & 0.277$\pm$0.007 & 1450$\pm$40 \\
SW3 &63 $\mu$m ([O I])  & 0.283$\pm$0.009  & 1400$\pm$40 & 0.283$\pm$0.003 & 1400$\pm$20 \\
SW5 (76--93 \textmugreek{}m) &88 $\mu$m ([O III])  & 0.283$\pm$0.009 & 1340$\pm$40 & 0.283$\pm$0.005 & 1340$\pm$20 \\
LW1 (84--110 \textmugreek{}m) &88 $\mu$m ([O III]) & 0.584$\pm$0.015 & 1980$\pm$50 & 0.618$\pm$0.013 & 2100$\pm$40 \\
LW2 (103--128 \textmugreek{}m) &122 $\mu$m ([N II])  & 0.584$\pm$0.015  & 1440$\pm$40 & 0.634$\pm$0.024 & 1560$\pm$60 \\
LW4 (142--171 \textmugreek{}m)&145 $\mu$m ([O I])  & 0.584$\pm$0.015  & 1200$\pm$30 & 0.680$\pm$0.044 & 1400$\pm$90\\
LW4  &158 $\mu$m ([C II])  & 0.584$\pm$0.015 & 1100$\pm$30 & 0.599$\pm$0.016 & 1140$\pm$30\\
\hline \hline
\end{tabular}
\end{center}
\label{tab:specRes}
\tablenotetext{a}{$ISO$ LWS Handbook \citep{gry03}}
\end{table*}

\twocolumngrid

\section{Detected Line Profiles}\label{sec:lines}
The candidate atomic fine-structure emission lines  are approximately Gaussian in shape and the underlying continuum is roughly linear over small wavelength ranges. Thus, we fit the narrow-band spectrum of each line with a Gaussian + linear component model (Levenberg--Marquardt algorithm) using the Python package ``LMFIT'' \citep{newville14}.  These narrow-band line spectra and best--fit models are plotted in Figure \ref{fig:lineFigs}, and our results are summarized in Table \ref{tab:lineflux}. We clearly detect atomic fine-structure emission lines from [C II] at 158 \textmugreek{}m, [N II] at 122 \textmugreek{}m, [O I] at 63 and 145 \textmugreek{}m, and [O III] at 52 and 88 \textmugreek{}m in several remnants. On several occasions, the [O I] 63  and 145 \textmugreek{}m and [O III] 88 \textmugreek{}m lines are detected in adjacent subdetectors. The SW3 subdetector always covers the [O I] 63 \textmugreek{}m band, while its adjacent subdetector (SW2) covers this band only in a few observations. For the [O I] 145 \textmugreek{}m line, while both the LW3 and LW4 subdetectors always cover the wavelength band of interest, LW4 consistently provides better signal-to-noise (S/N). The [O III] 88 \textmugreek{}m line is detected with both the SW5 and LW1 detectors. The SW5 detector provides roughly twice the resolving power of LW1 (Table \ref{tab:specRes}). Therefore, we report the [O I] 63  and 145 \textmugreek{}m and [O III] 88 \textmugreek{}m line measurements detected with the SW3, LW4, and SW5 subdetectors, respectively. In a few cases, the data quality is poor in these detectors, so we instead report measurements using the adjacent subdetector. We note these exceptions in Table \ref{tab:lineflux}.

A clear signature of emission lines emanating from fast-moving ejecta gas from a SN is their significant Doppler broadening, which may be detected in the $ISO$ LWS spectra of SNRs. To calculate the velocity broadening ($B$) of each emission line, we remove the effect of the instrumental broadening through the relation, $B_{Line} = \sqrt{B_{Obs}^2 - B_{LWS}^2}$, where $B_{Obs}$ is the observed full width at half maximum (FWHM) of the emission line, $B_{LWS}$ is the instrument resolution of the LWS detector (Table \ref{tab:specRes}), and $B_{Line}$ is the true FWHM of the line.
We measure an apparent Doppler broadening of $B_{Line} \gtrsim$ 1000 km s\textsuperscript{--1} for several emission lines, mostly from the 63 \textmugreek{}m [O I] and 88 \textmugreek{}m [O III]  lines (see Figure \ref{fig:velocity} and Table \ref{tab:snrinfo}). 
The detection of [C II] lines requires caution because [C II] 158 \textmugreek{}m emission is the primary coolant in interstellar gas. Galactic plane surveys of the [C II] 158 \textmugreek{}m line revealed ambient intensities of $\sim 10^{-5} - 10^{-4}$ erg s\textsuperscript{--1} cm\textsuperscript{--2} sr\textsuperscript{--1} \citep{bennett94,nakagawa98,pineda13}.  Thus, our measured line profiles may be contaminated by emissions originating from along the line--of--sight, not associated with the SNR. Unfortunately, in most cases, no accompanying background observation is available to distinguish between the line--of--sight and SNR emission. Future on--off observations are required to unambiguously determine if the [C II] emission belongs to the SNR. Nonetheless, a clear case of Doppler-broadening of the [C II] 158 \textmugreek{}m line is in the Crab Nebula ($B_{Line} \sim$ 1800 km s\textsuperscript{--1}). We also note that the $Herschel$ spectroscopic observation of Kes 75 (G29.7-0.3)
resolves the [C II] line into two components of narrow and broad (1300 -- 1500 \kms) \citep{temim19}, while the $ISO$ spectra indicate 8\% broadening of this [C II] line. The $Herschel$ Photodetector Array Camera and Spectrometer (PACS) resolution ($\sim$240 \kms) at 158 \textmugreek{}m is a factor of $\sim$5 superior to that of the $ISO$ LWS ($\sim$1100 \kms). The marginally detected broadening of the [C II] line in the LWS spectra may need to be verified (or disputed) by future follow--up observations.

%

\section{High velocity SN Ejecta}\label{sec:ejecta}
We compared our measured FWHM values for the detected emission lines with the instrumental FWHM reported in the $ISO$ LWS Handbook, $\sim$ 0.283 \textmugreek{}m and $\sim$ 0.584 \textmugreek{}m for lines detected with the SW and LW subdetectors (see Table \ref{tab:specRes}, respectively \citep{gry03}. We note that these instrumental resolutions are consistent (within statistical uncertainties) with the line width measurements for our HII region sample and the planetary nebula NGC 6720 (see Table \ref{tab:hiispec}). For each emission line, we take the instrumental FWHM ($B_{LWS}$) to be the smaller value of either the FWHM reported in the $ISO$ LWS Handbook or our measured FWHM of the emission line in NGC 6720 (Table \ref{tab:specRes}). Note that the handbook provides only one number for each of SW and LW and the FWHM is expected to be a function of wavelength because the LWS is a grating instrument. We confirm that using either the $ISO$ Handbook or our measured NGC 6720 values for the detector resolutions does not significantly affect our scientific conclusions.

\begin{figure*}
\centering
\includegraphics[angle=0,width=6.5truecm,height=8truecm]{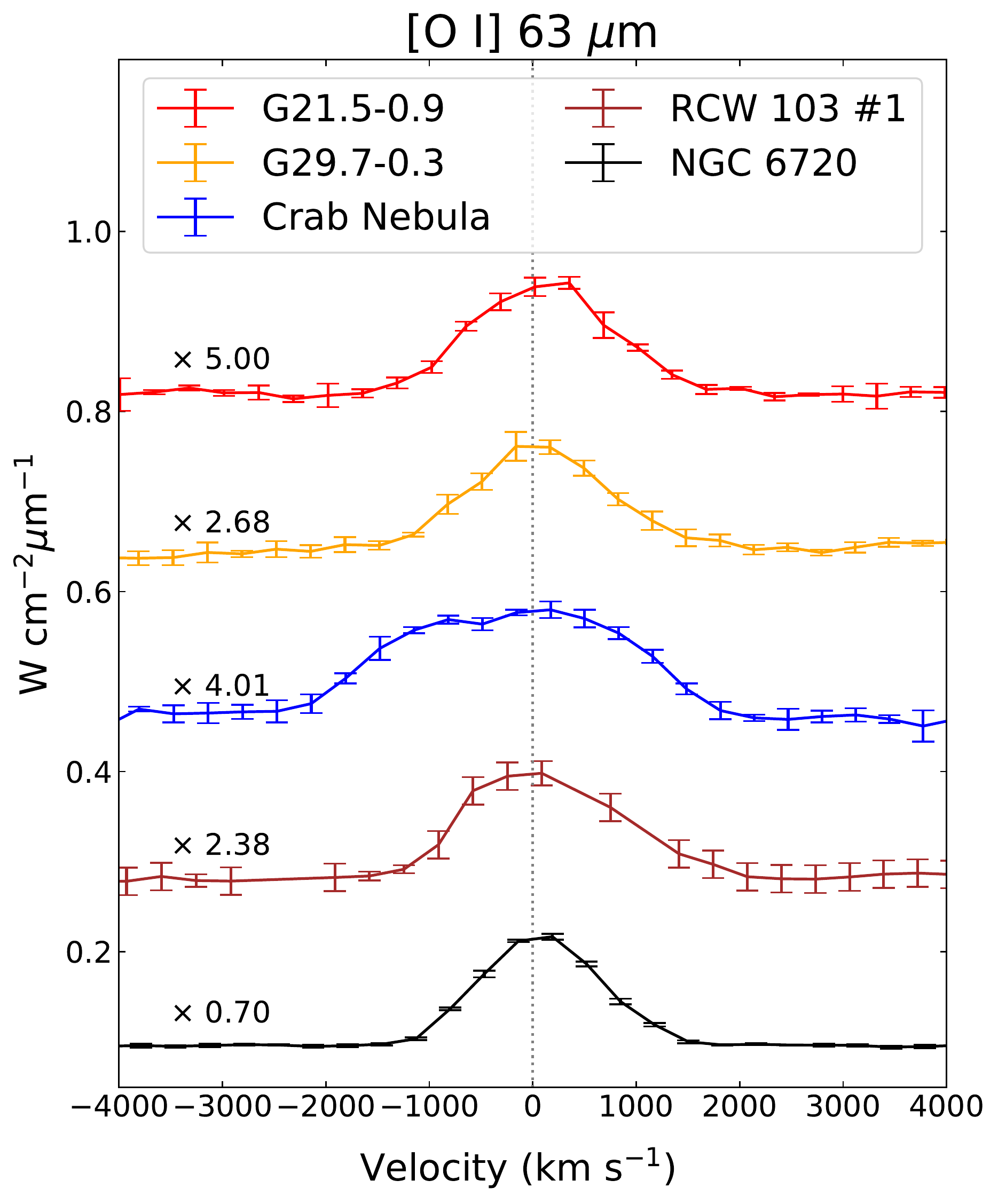} 
\includegraphics[angle=0,width=6.5truecm,height=8truecm]{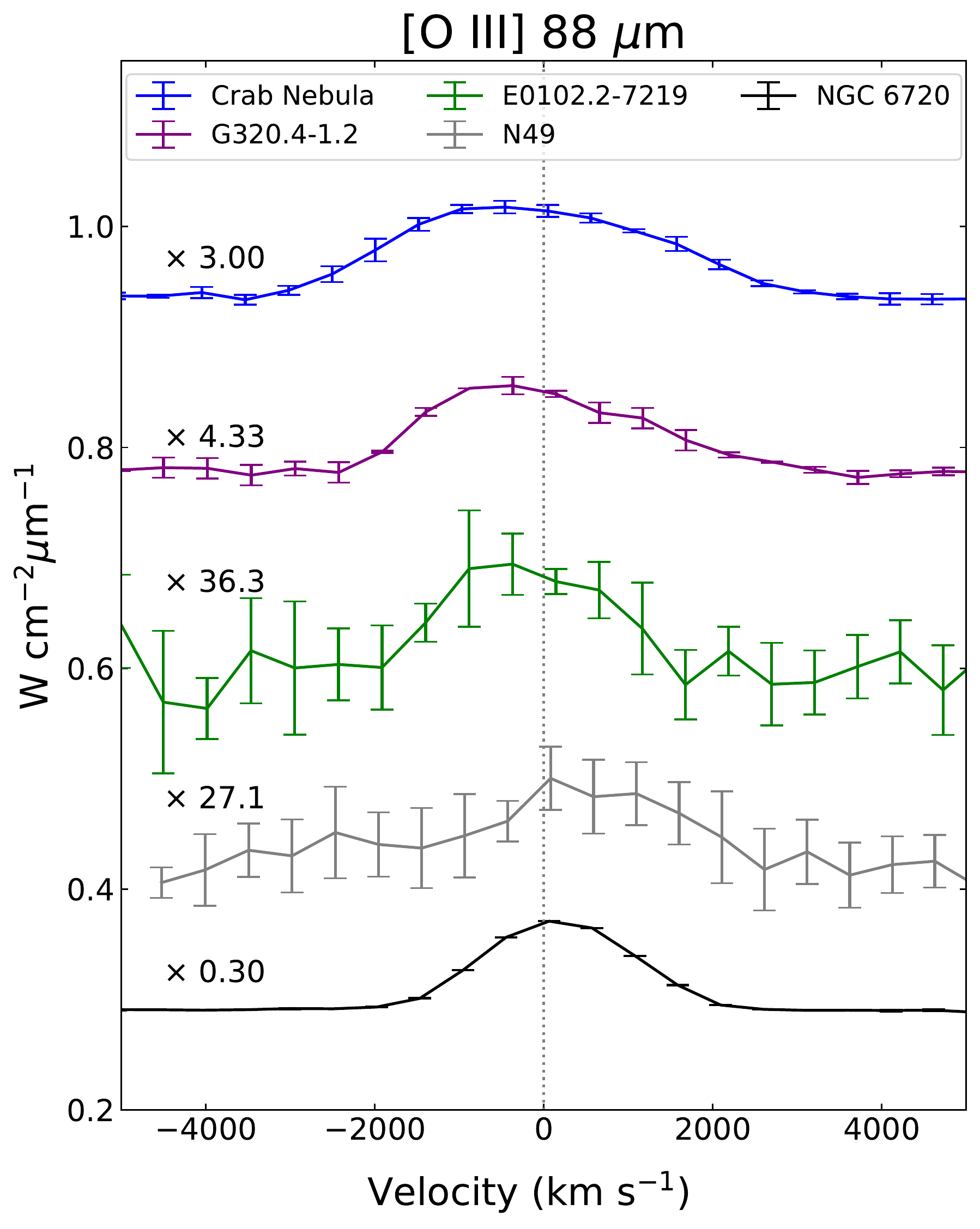}
\caption{$ISO$ LWS velocity profiles of apparently broad 63 \textmugreek{}m [O I] and 88 \textmugreek{}m [O III] SNR emission lines. The profiles are scaled by the value listed above each line, and shifted for comparison. The solid black line shows the profile of planetary nebula NGC 6720, which represents the velocity resolution of the detector.}
\label{fig:velocity}
\end{figure*}

We detect broad lines (up to $\sim$ 3000 km s\textsuperscript{--1}) in several SNRs including G21.5--0.9, G29.7--0.3, the Crab Nebula, and G320.4--1.2. The velocity profiles of a few representative cases are shown in Figure \ref{fig:velocity}. We classify these SNRs into three groups based on the statistical significance of the measured intensity and broadening of at least one emission line in each SNR spectrum: i.e.,  -- Group A: both a firm line detection with S/N $>$ 5 and a clear 5 \textsigma{} broadening detection, Group B: the line is firmly detected (S/N $>$ 5), but the broadening is marginal ($<$ 5 \textsigma{}), and Group C: the lines appear to be broad, but the line detection is marginal (3 $<$ S/N $<$ 5), or show a marginal broadening only in the [C II] 158 \textmugreek{}m line. In Table \ref{tab:snrinfo}, we summarize these results.

The spectrum of the Crab Nebula (see Figure \ref{fig:velocity}) shows clear evidence of broad emission lines from high-speed ejecta. Its spectrum exhibits double--peaked line features (Figure \ref{fig:lineFigs}, [O III] 88 \textmugreek{}m), indicating both blue- and red-shifted emission components expanding along the line of sight. In fact, all detected emission lines  (i.e.,  52 and 88 \mic\ [O\,III], 63 and 145 \mic\ [O\,I], 122 \mic\ [N\,II], and 158 \mic\ [C\,II]) from its spectra are broad, with line widths of  $B_{Line} \sim$ 2000--3000 km s\textsuperscript{--1}. The 88 \textmugreek{}m [O III] line in G320.4--1.2 shows an asymmetrical broadening, which may also suggest double--peaks with differential blue and red shifts (Figure \ref{fig:velocity}). Below, we discuss individual SNRs with evidence of ejecta emission in detail.

\textbf{\textit{ G21.5--0.9}} -- Observations of the PWN with the $Herschel$ PACS detector revealed broad [O I] 63 \textmugreek{}m and [C II] 158 \textmugreek{}m lines with measured FWHMs of 850 km s\textsuperscript{--1} and 1000 km s\textsuperscript{--1}, respectively, indicating the presence of PWN-shocked ejecta \citep{hattori20,cha21}.  The LWS pointing encompasses almost all of the IR emission of the PWN (see Figure \ref{fig:g21p5}). The FIR lines at 63 \mic\ \oif\, 122 \mic\ [N II], and 145 \mic\ [O I] show broadening (Table \ref{tab:lineflux}, Figure \ref{fig:lineFigs}). The broadening at 63 \textmugreek{}m is firmly detected, thus we classify the detection into Group A, indicating that the emission likely originates from fast-moving ejecta. Our estimates of the  [O I] 63 \textmugreek{}m and 145 \textmugreek{}m line FWHMs of $\sim$ 900 km s\textsuperscript{--1}, are consistent with the measured values from the $Herschel$ PACS observations.

\textbf{\textit{ G29.7--0.3}} --  $Herschel$ PACS observations of the PWN showed narrow and broad components for each of the [O I] 63 \textmugreek{}m, [O III] 88 \textmugreek{}m, and [C II] 158 \textmugreek{}m emission lines \citep{temim19}. \citet{temim19} attributed the narrow component to background emission, and estimated FWHMs of 1270 -- 1570 km s\textsuperscript{--1} for the broad components, suggesting that the PWN shock is interacting with ejecta in the interior of the remnant.

The $ISO$ LWS pointing covers the entirety of the PWN as seen in the FIR (Figure \ref{fig:g29p7}). We detect broad [N II] 122 \textmugreek{}m and [O I] 145 \textmugreek{}m lines. The broadening is firmly detected for the [N II] 122 \textmugreek{}m (1200 km s\textsuperscript{--1}) and [O I] 145 \textmugreek{}m (1300 km s\textsuperscript{--1}) lines (Group A). This is the first detection of [O I] 145 and [N II] 122  \textmugreek{}m line broadening in this SNR. Our measured line widths are consistent with those detected at [O I] 63 and [O III] 88 \textmugreek{}m based on the $Herschel$ PACS data \citet{temim19}.


\textbf{\textit{G54.1+0.3}} -- \citet{temim10} suggested that the IR  morphology is due to the PWN driving shocks into the expanding SN ejecta. Using $Spitzer$ archival IRS, IRAC, and MIPS data, \citet{rho18} found dust emission spatially coincident with the ejecta line emission as traced by [Ar II]. In addition to Ar, they also detected atomic fine-structure emission from Ne, Cl, Si, and S. These lines are broadened, corresponding to velocities up to several 10\textsuperscript{2} km s\textsuperscript{--1}. The [C II] line at 158 \textmugreek{}m shows a slight, but statistically significant ($>$ 5 \textsigma{}) broadening of $B_{Line} = 640$ km s\textsuperscript{--1}. However, the [C II] emission contribution from the remnant is unclear due to contamination from line--of--sight emission. Thus, we classify this SNR as Group C.

\textbf{\textit{Crab Nebula}} -- \citet{gomez12} showed that broad atomic lines observed with the $ISO$ LWS correspond with resolved redshifted and blueshifted emission peaks in the $Herschel$ PACS spectra from a bright filament at the eastern side of the remnant. With the higher-resolution PACS data, they estimate that the redshifted and blueshifted emission peaks are in the velocity range of 1290 -- 1750 km s\textsuperscript{--1}.

Here, our results are based on a pointing toward the northern part of the remnant (Figure \ref{fig:crab}). The [O III] 52 and 88 \textmugreek{}m, [N III] 57 \textmugreek{}m, [O I] 63 and 145 \textmugreek{}m, [N II] 122 \textmugreek{}m, and [C II] 158 \textmugreek{}m lines are broad ($\sim$ 1800 -- 3300 km s\textsuperscript{--1}, Table \ref{tab:lineflux}). The double--peaked structure due to red-- and blue--shifted components is resolved in the LWS data for the [O III] 88 \textmugreek{}m line (Figure \ref{fig:lineFigs}). Applying a two--Gaussian + linear component model fit to this line gives line centroid values corresponding to $\pm$ 1000 km s\textsuperscript{--1}. The Crab Nebula exhibits the clearest case in our sample of broad line emission originating from high-speed ejecta.  

%

\startlongtable
\begin{deluxetable*}{llllllllll}
\tabletypesize{\tiny}
\tablecaption{Properties of Young SNRs with Broad FIR Lines \label{tab:snrinfo}}
\tablehead{
\colhead{SNR} & \colhead{Distance} & \colhead{Age} &  \colhead{ M$_{dust}$} & \colhead{Progenitor} & \colhead{Broad Lines\textsuperscript{a}} & \colhead{Measured\textsuperscript{a}} & \colhead{Group} & \colhead{52/88 \textmugreek{}m\textsuperscript{a}} & \colhead{63/145 \textmugreek{}m\textsuperscript{a}} \\  
\colhead{}& \colhead{(kpc)} &  \colhead{(yr)} & \colhead{{\it Herschel} (M$_{\odot}$)} & \colhead{Mass (M$_{\odot}$)} & \colhead{(\textmugreek{}m)} & \colhead{FWHM(km s\textsuperscript{--1})} & \colhead{}&
} 
\startdata
G21.5--0.9        & 4.8 \textsuperscript{1}  & $<$1000 \textsuperscript{2,3}    & 0.29 \textsuperscript{4}& --  & 63,122,145 & 900--1800 & A & -- & 11\\ 
G29.7--0.3     & 5.1--10.6 \textsuperscript{5,6}     & 723 \textsuperscript{7} & 0.51 \textsuperscript{4} & 8--12 \textsuperscript{8} & 122,145 & 1200--1300 & A & -- & 1.3\\
G54.1+0.3    & 6.2 \textsuperscript{16} & 1800-2400 \textsuperscript{17}  & 0.08--0.9 \textsuperscript{18} & 15--20 \textsuperscript{19} & 158 & 640 & C & -- & --\\
Crab Nebula    &2 \textsuperscript{9}& 966    &0.016--0.6 \textsuperscript{10-13} & 8--10 \textsuperscript{14} & 51,63,88,122,145,158 & 1800-3300 & A& 1.4, 1.0\tablenotemark{b} & 17\\
G320.4--1.2      & 5.2 \textsuperscript{15}   &1700 \textsuperscript{15}  & \textsuperscript{e}0.009 & -- \textsuperscript{} & 88,122 & 1800-2100 & A& -- & --\\ 
RCW 103 & 3.1 \textsuperscript{20} & 2000 \textsuperscript{21} & 0.18 \textsuperscript{22}, 1.1 \textsuperscript{23} & 18--20 \textsuperscript{24} & 63,122 & 800-2000 & C & -- & 6.1, 15\tablenotemark{c}\\
E0102.2--7219 & 61  & 2050 \textsuperscript{25}  & \textsuperscript{e}0.014\textsuperscript{26} &  25 -- 40 \textsuperscript{27,28,39} &  88 & 1300 & C & -- & --\\
N132D        & 50 & 2500 \textsuperscript{29} & \textsuperscript{d}0.013 \textsuperscript{30} & 15 $\pm$ 5 \textsuperscript{31} & 63 & 1500 & B& -- & -- \\
N49    & 50 & 4800 \textsuperscript{32}     &  $\gtrsim$ 0.1 \textsuperscript{33}& 13 -- 17 \textsuperscript{34}  & 88 & 3000 & C & -- & --\\
0540-69.3  & 50 & 800 -- 1100 \textsuperscript{35} & \textsuperscript{d,}\textsuperscript{e}1 -- 3$\times$10\textsuperscript{-3} \textsuperscript{36,37} &  20 -- 25 \textsuperscript{38}  & 52 & 2900 & C & 4.5 & --\\
\enddata
%
%
%
\tablecomments{1. \citet{tian08} 2. \citet{guest19} 3. \citet{bietenholz08} 4. \citet{chawner19} 5. \citet{leahy08b} 6. \citet{su09} 7. \citet{gotthelf00} 8. \citet{temim19} 9. \citet{trimble68} 10. \citet{delooze19} 11. \citet{gomez12} 12. \citet{temim13} 13. \citet{owen15} 14. \citet{smith13} 15. \citet{gaensler99} 16. \citet{leahy08} 17. \citet{bocchino10} 18. \citet{rho18} 19. \citet{gelfand15} 20. \citet{reynoso04} 21. \citet{carter97} 22. \citet{pinheiroconcalves11} 23. \citet{andersen11} 24. \citet{frank15} 25. \citet{finkelstein06} 26. \citet{rho09} 27. \citet{blair00} 28. \citet{flanagan04} 29. \citet{law20} 30. \citet{seok08} 31. \citet{sharda20} 32. \citet{park12} 33. \citet{otsuka10} 34. \citet{zhou19} 35. \citet{reynolds85} 36. \citet{williams08} 37. \citet{lundqvist20} 38. \citep{lundqvist11} 39. \citet{alan19}}
\tablenotemark{a}{This work ($ISO$ LWS).}
\tablenotemark{b}{Blue, red-shifted line}
\tablenotemark{c}{Center, rim} 
\tablenotemark{d}{$AKARI$ estimate}
\tablenotemark{e}{$Spitzer$ estimate}\\
\tablenotemark{}{Group A: Line flux and line broadening are both significant to $\geq 5 \sigma$; Group B: Significance of broad line emission is $> 5 \sigma$, line broadening significance is $< 5  \sigma$; Group C: Significance of broad line flux $< 5 \sigma$. }
\end{deluxetable*}

\textbf{\textit{G320.4--1.2}} -- \citet{koo11} reported a clump of gas near the PWN emitting a [Ne II] 12.81 \textmugreek{}m line with a line-of-sight velocity of $\sim$ +1000 km s\textsuperscript{--1}, indicating SN ejecta. A recent X-ray proper motion study of compact ejecta knots located to the north of the pulsar (in the RCW 89 nebula) indicates high velocities up to 5000 \kms\ \citep{borkowski20}.

The LWS pointing is centered on the PWN, which shows a complicated structure of filaments in the $Herschel$ 70 \textmugreek{}m image (Figure \ref{fig:G320.4-1.2_images}). We detect [O I] 63 \textmugreek{}m, [O III] 88 \textmugreek{}m,  [N II] 122 \textmugreek{}m,  and [C II] 158 \textmugreek{}m emission lines. The 88 \textmugreek{}m [O III] is broadened by $\sim$ 2000 km s\textsuperscript{--1} (Group A).   Our detection of broad lines is consistent with similar high-speed ejecta emission reported in \citet{koo11} and \citet{borkowski20}. G320.4--1.2 shows promising evidence for emission from both ejecta and dust from the same region (see Section \ref{sec:continuum} for a detailed discussion of the continuum emission).

\textbf{\textit{RCW 103}} --  A previous $ISO$ study found that the post-shock gas has low densities ($n_e \sim$ 10\textsuperscript{3} cm\textsuperscript{-3}) and abundances close to solar values, and thus no clear indication of ejecta emission \citep{oliva99}. A $Chandra$ study found sparse ejecta emission throughout SNR, with abundance values slightly above solar values \citep{frank15}.

We examined the $ISO$ LWS spectra of two sub-regions within RCW 103, one near the SNR's center and the other at the bright southern shell (regions \# 1 and \# 3, respectively, see Figure \ref{fig:rcw103}).  We extracted the background spectrum from another LWS data pointed at a region just outside of the southern shell of RCW 103 (region \# 2, Figure \ref{fig:rcw103}). We find that the region \# 3 spectrum clearly shows enhanced [O I] line emission at 63 \textmugreek{}m above the background spectrum (see Table \ref{tab:lineflux}), which may be related to shock interactions with a molecular cloud (see Section \ref{sec:rasters}).  The central region (\# 1) shows marginal evidence for line--broadening, however the line strengths are similar to (or even weaker than) those measured in the background region. Thus, the line--broadening is not compelling, and we classify this SNR as Group C.

 \textbf{\textit{E0102.2-7219}} --  Optical observations suggest the presence of fast-moving ejecta emitting [S II], [S III], [Ar III], and H$\alpha$ and H$\beta$ lines at speeds up to $\sim$ 1800 km s\textsuperscript{--1} \citep{seitenzahl18}.  \citet{rho09} measured several MIR emission lines from ejecta, including broad \neiiif\ and \nevf\ lines which suggest velocity dispersion ranges of $\sim$ 2000 -- 4000 km s\textsuperscript{--1}. 
 
 We measure a candidate broad [O III] 88 \textmugreek{}m line, suggesting a velocity up to $\sim$ 1300 km s\textsuperscript{--1}. However, the broadening is statistically insignificant ($\sim$ 1 \textsigma{}), and the line detection is marginal ($\sim$ 4 \textsigma{}, Table \ref{tab:lineflux}). Thus, we classify this evidence into Group C.

\textbf{\textit{N132D}} -- Based on $Spitzer$ IRS, IRAC, and MIPS data, \citet{tappe12} detected [Ne II] and [O IV] MIR lines from a fast-moving ejecta knot position, while the southeastern shell is dominated by ISM \citep{tappe06}. We detect the [O I] 63 \textmugreek{}m and 145 \textmugreek{}m lines. The [O I] 63 \textmugreek{}m line appears to be broadened (a 3 \textsigma{} detection, Group B). The implied velocity dispersion is $\sim$ 1500 km s\textsuperscript{--1}. If our suggested line broadening would be confirmed with follow--up observations, this velocity would be in plausible agreement with kinematic studies of the optical [O III] ejecta which estimate expansion velocities up to  $\sim$ 3000 km s\textsuperscript{--1} \citep{morse95,law20}.

\textbf{\textit{N49}} -- \citet{bilikova07} estimated regional expansion velocities up to $\sim$ 500 km s\textsuperscript{--1} based on echelle spectra of H$\alpha$ and [N II] emission lines, while X-ray observations indicate that N49 is enriched in Si and S \citep{park03,park12}, as well as O and Ne \citep{zhou19}.
We find marginal evidence for a broad [O III] 88 \textmugreek{}m line (a 3 \textsigma{} detection, Group C, Table \ref{tab:lineflux}). If confirmed, this broad [O III] line may be associated with the O--rich ejecta in this SNR.

\textbf{\textit{0540-69.3}} -- Optical [O III] emission lines with velocity dispersions of $\sim$ 3000 km s\textsuperscript{--1} have been detected in 0540-69.3 \citep{mathewson80,kirshner89}. X-ray measurements hint at metal-rich ejecta in the southern part of the SNR \citep{park10}. Our best--fit to the [O III] 52 \textmugreek{}m line suggests potential broadening ($\sim$ 2900 km s\textsuperscript{--1}). However, the line intensity and broadening are both measured with only 3 \textsigma{} confidence (Group C). The line-center is also shifted by +1300 km s\textsuperscript{--1}, however, this shift is not detected in the [O III] 88 \textmugreek{}m line.

Follow--up observations with higher spectral resolution (e.g., the FIFI-LS on board $SOFIA$, or the proposed $Origins$ Space Telescope \citep{leisawitz21}) are required to firmly detect the candidate broad lines in G54.1+0.3, RCW 103, E0102.2--7219, N132D, N49, and 0540--69.3. Furthermore, due to the large aperture (diameter = 80\arcsec{}) of the LWS, constructing a detailed spatial distribution of ejecta emission in SNRs of our sample is not feasible. Establishing the spatial associations between ejecta and dust emission features based on high--resolution imaging is necessary to test the dust formation in the SN ejecta.


\section{SNRs Interacting with Molecular Clouds}\label{sec:rasters}

\subsection{Diagnostic Line Results}

Progenitors of CC-SN burn through their fuel in relatively short time periods, and therefore their SNRs may interact with the molecular clouds (MCs) where they were born. The high expansion speed of the SNR drives a shock into the interacting MC. The radiative cooling of the shocked gas produces a number of molecular and atomic emission lines in IR bands.  

The [O I] 63 \textmugreek{}m and 145 \textmugreek{}m, and [C II] 158 \textmugreek{}m line strengths are useful diagnostics for the pre-shock density of the cloud and the speed of the shock that it encounters \citep{hollenbach89}. In general, two types of shock interactions are considered, i.e.,  a continuous (C)-type, or jump (J)-type shock. In a C-type shock, the shock front moves at a speed slower than the magnetosonic speed in the medium, and thus the gas conditions (e.g., temperature and density) change gradually. In a J-type shock, the shock front moves faster than the magnetosonic speed, and there is a sharp change in the gas temperature and density as the shock front moves through the medium. The type of shock affects the post-shock chemistry and overall dust destruction. Here, we aim to identify shock types from our measurements of FIR [O I] and [C II] lines in these MC-interacting SNRs.

\begin{figure}
\centering
\includegraphics[angle=0,width=8.truecm]{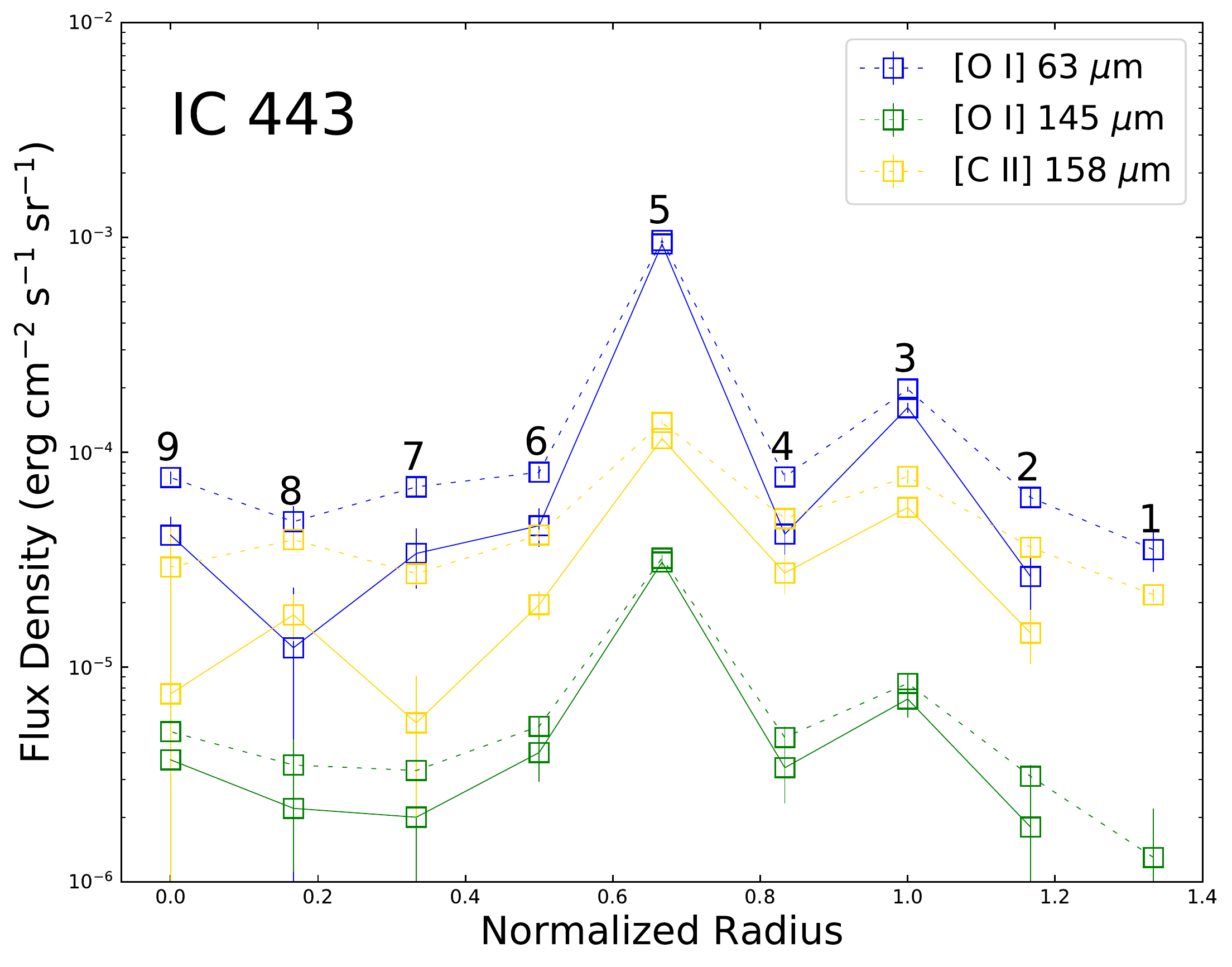}
\includegraphics[angle=0,width=8.truecm]{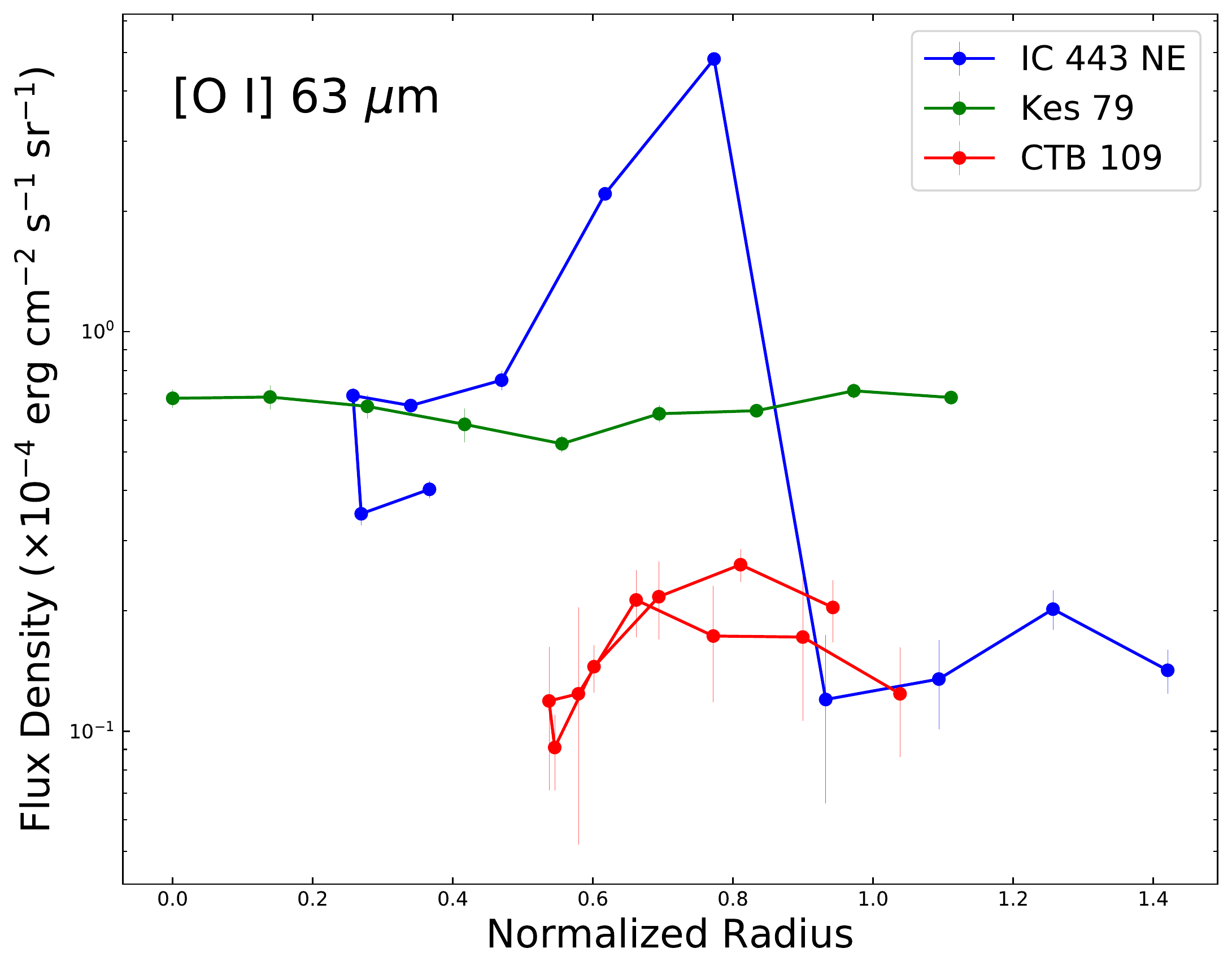}
\caption{Upper: Distribution of [O I] and [C II] line intensity in the southern region of IC 443 (see Figure \ref{fig:ic443}) as a function of distance from the center of the SNR. The total flux density in each region is represented by dashed lines, while solid lines indicate the background--subtracted values. The emission peaks at the double shell (position 5 at $r$=0.7 and 3 at $r$=1.0). Lower: Total [O I] 63 \textmugreek{}m line intensity vs. radius for raster observations of Kes 79, CTB 109, and the northeast region of IC 443 from \citet{rho01} are compared with those of the southern region of IC 443 from the upper panel.} 
\label{fig:flux_v_radius}
\end{figure}

Our sample includes SNRs W 28, IC 443, Kes 79, CTB 109, and RCW 103, for which the shock-MC interaction has been reported in literature \citep{tatematsu90,dickman92,rho97,arikawa99,stanimirovic03,paron06,kilpatrick16,zhou16b}. For SNRs Kes 79, CTB 109, and IC 443, the [O I] 63 \textmugreek{}m  band was observed with a raster map across the face of remnant. The raster maps allow for measurements of spatial variations in line strength across the remnant, whereby the locations of various shocks can be identified. These raster map observations focus on narrow wavelength ranges of select emission lines, not covering the full range of the LWS. The raster maps of IC 443 also include spectra of the [O I] 145 \textmugreek{}m and [C II] 158 \textmugreek{}m bands. The raster map narrow-band line spectra and our spectral model fits are shown in Figure \ref{fig:lineRasters}, while Tables \ref{tab:kes79raster}, \ref{tab:ctb109raster}, and \ref{tab:ic443raster} show the line profiles and best--fit results for Kes 79, CTB 109, and IC 443, respectively.   Figure \ref{fig:flux_v_radius} shows the radial distribution of line intensity for each of the raster observations of the molecular-cloud interacting SNRs in our sample. In Figure \ref{fig:flux_v_radius}, we also overlay the published radial distribution of the [O I] 63 um line intensity across the northeastern shell of IC 443 \citep{rho01}.  

The raster map of Kes 79 consists of 9 observations, each separated by 1 arcmin, which extend across the eastern boundary (see Figure {\ref{fig:kes79}}) 
of the remnant. The [O I] 63 \textmugreek{}m line intensities are consistent with the mean, $I_{63\mu{}m} =$ 6.4 $\times$ 10$^{-5}$ \ergsb), to within 3 \textsigma{} uncertainties. Thus, there is no clear emission enhancement between regions inside the boundary of the remnant versus outside of it.

The CTB 109 raster map (see Figure~\ref{fig:ctb109}) consists of 11 observations positioned across the northwest quadrant of the remnant, with 3 arcmin spacing between each pointing. We note the lowest line fluxes in regions 5 –- 7  located inside the boundary of the remnant, and highest near of the western perimeter.  The dimmer regions roughly coincide with the location of a CO arm that reaches across the remnant \citep{kothes02}, which could indicate that the emission is being absorbed by a foreground MC.

The raster map of IC 443 consists of 9 observations spaced 3 arcmin apart, from the geometric center of the remnant to outside the southern shell (see Figures \ref{fig:ic443} and \ref{fig:flux_v_radius}:Upper). The [O I] 63 \textmugreek{}m, 145 \textmugreek{}m and [C II] 158 \textmugreek{}m lines are strongest at a bright IR ridge in the southern shell (position 5), $I_{63\mu{}m}$ = $9.7\pm0.4 \times $10$^{-4}$ \ergsb, and weakest at the position exterior to the boundary of the remnant (position 1), $I_{63\mu{}m}$= $0.35\pm0.07 \times $10$^{-4}$ cm\textsuperscript{--2} s\textsuperscript{--1} sr\textsuperscript{--1}. The \oif\ emission peaks at the southern shell (more than an order of magnitude higher than the background intensity), indicating significant cooling from the SNR shocks encountering dense MCs. Similar sharp increases in line intensity are present in LWS observations of W44 and 3C391 \citep{reach96}. IC 443 was also observed using single (non--raster) pointings directed at the southeastern (\# 1), eastern (\# 2), and western (\# 3) boundaries of the remnant (see Table \ref{tab:lineflux}). The strongest [O I] and [C II] emission is observed at the southeast position. We note that [N\,II] lines from IC 443 show somewhat high velocity (2000 -- 3000 \kms), but the significance of the detection is less than 5 \textsigma{}. Higher resolution spectroscopy is required to resolve if the [N\,II] lines are truly broad.

RCW 103 and W 28 were observed with non--raster LWS pointings.   RCW 103 is close to the Galactic plane ($b = -0.4\degr{}$).  After background subtraction using a separate background pointing (\# 2), nearly all of the [C II] is removed from the central (\# 1) and southern border (\# 3) spectra (see Figure \ref{fig:rcw103}). At the central region, the [O I] line strength is also reduced significantly.  However, at the southern ridge  the [O I] flux remains high, $I_{63\mu{}m}$ = $3.7_{-0.5}^{+0.5}\times$10$^{-4}$ \ergsb. We note that the background emission around RCW 103 is non-uniform with a complex morphology, which may imply a significant spatial variation in the background intensity. Thus, our estimated background--subtracted flux of the [O I] line should be considered with caution. For W 28, there is no background observation. Since it is located close to the Galactic plane ($b = -0.1\degr{}$), the observed line intensity may be mixed with a significant amount of line-of-sight emission.

%

\subsection{SNR--MC Interactions and Shock Models}

\begin{figure*}
\centering
\begin{tabular}{ccc}
\includegraphics[angle=0,height=4.5truecm]{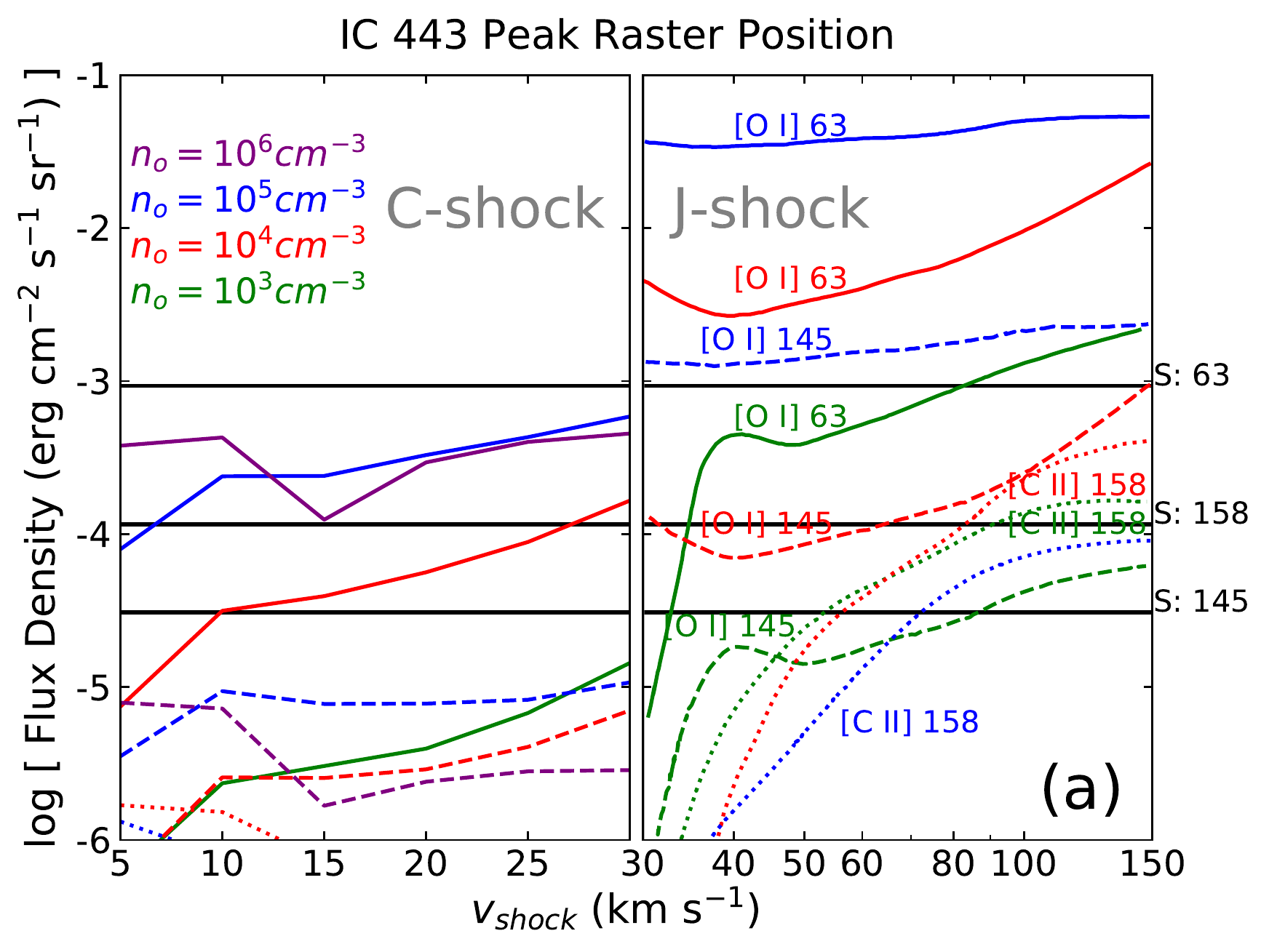} & 
\includegraphics[angle=0,height=4.5truecm]{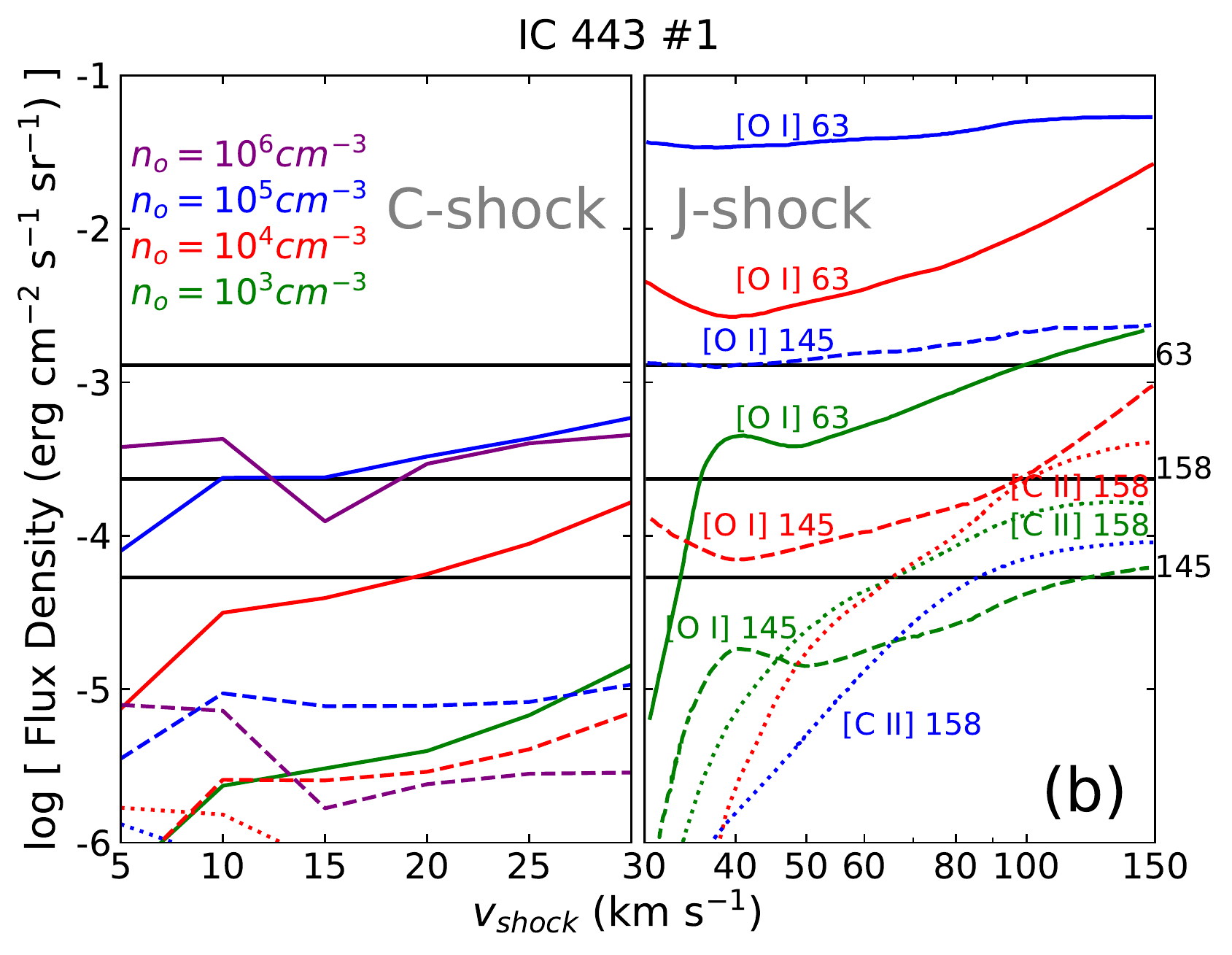}&
\includegraphics[angle=0,height=4.5truecm]{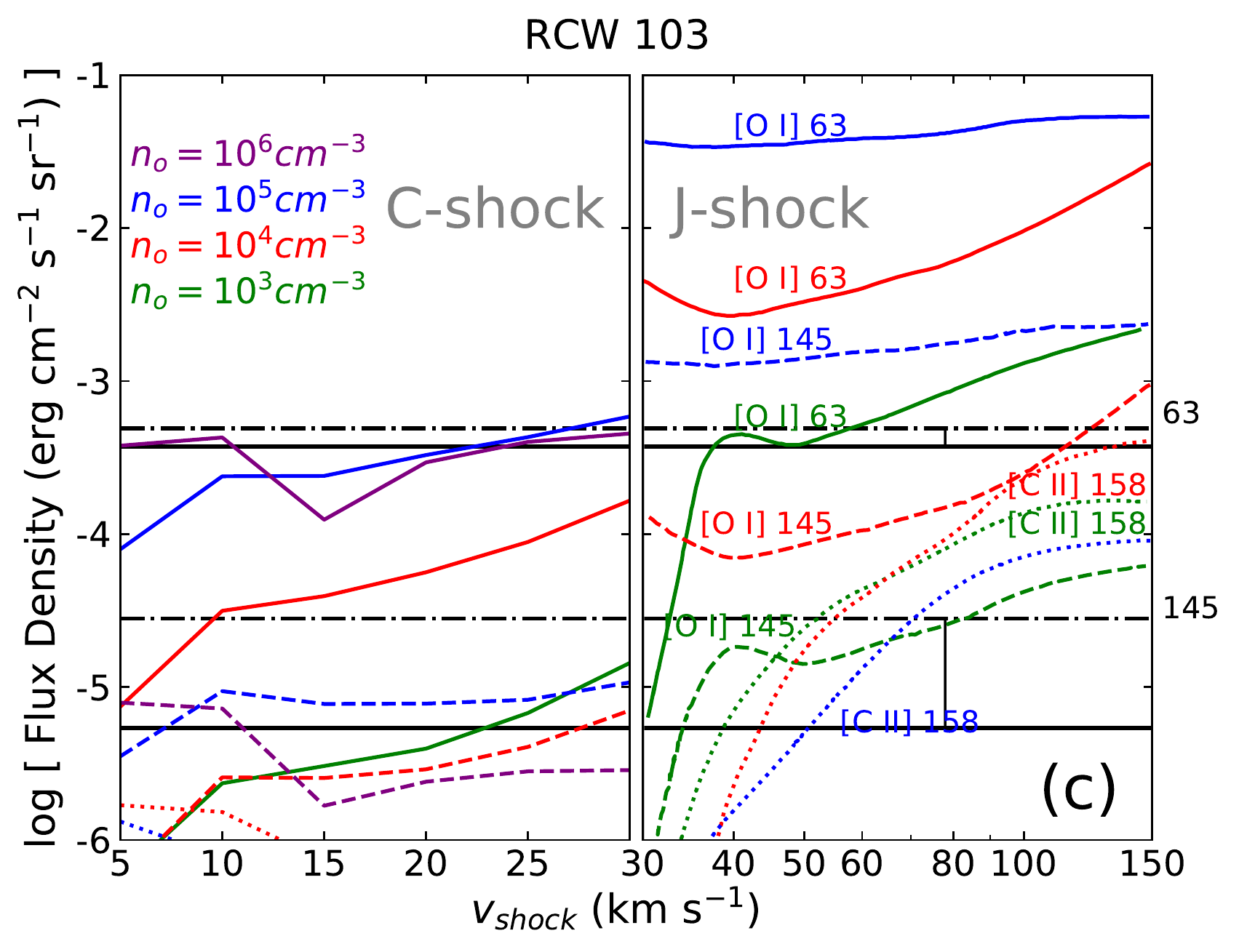} \\
\includegraphics[angle=0,height=4.5truecm]{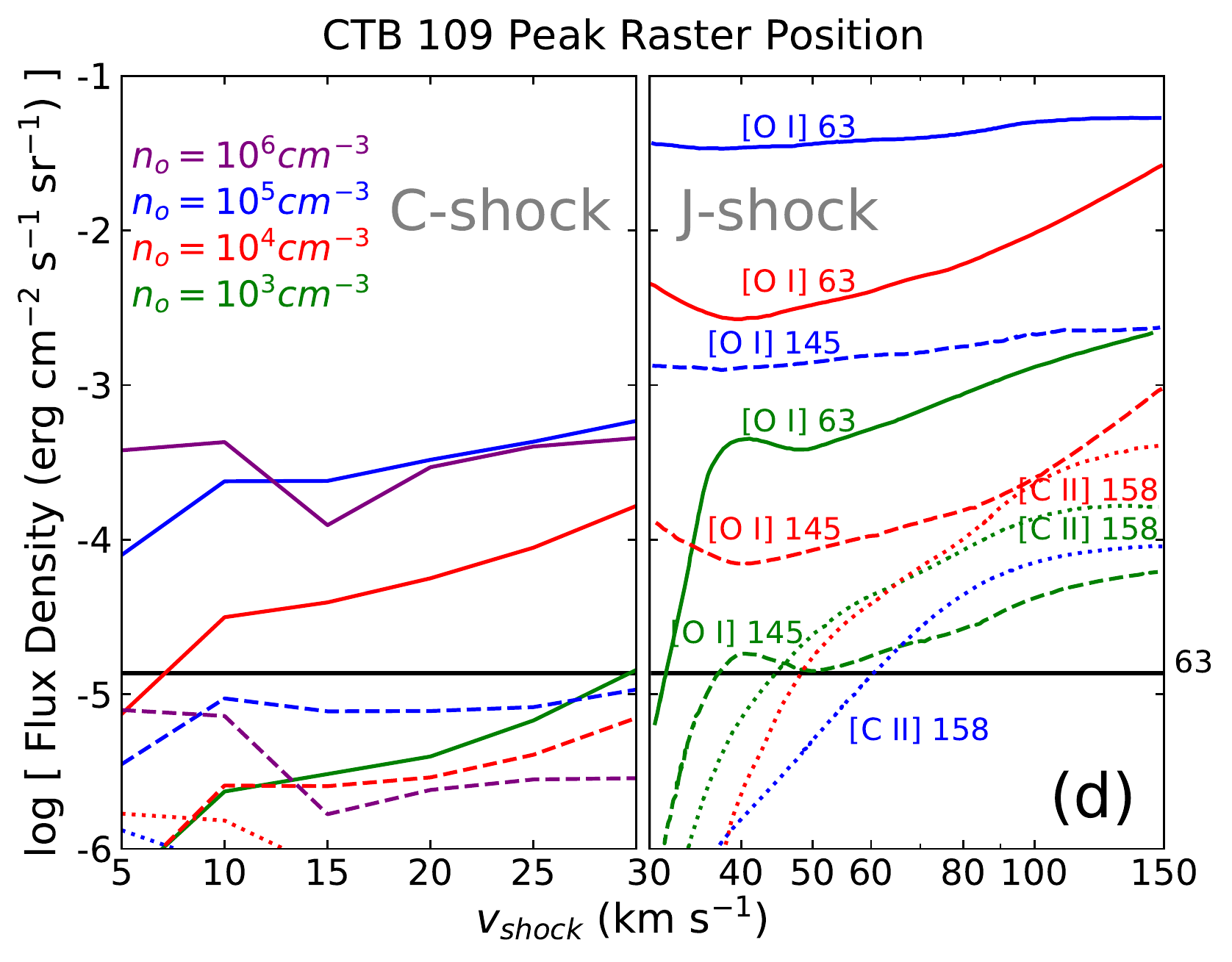} & 
\includegraphics[angle=0,height=4.5truecm]{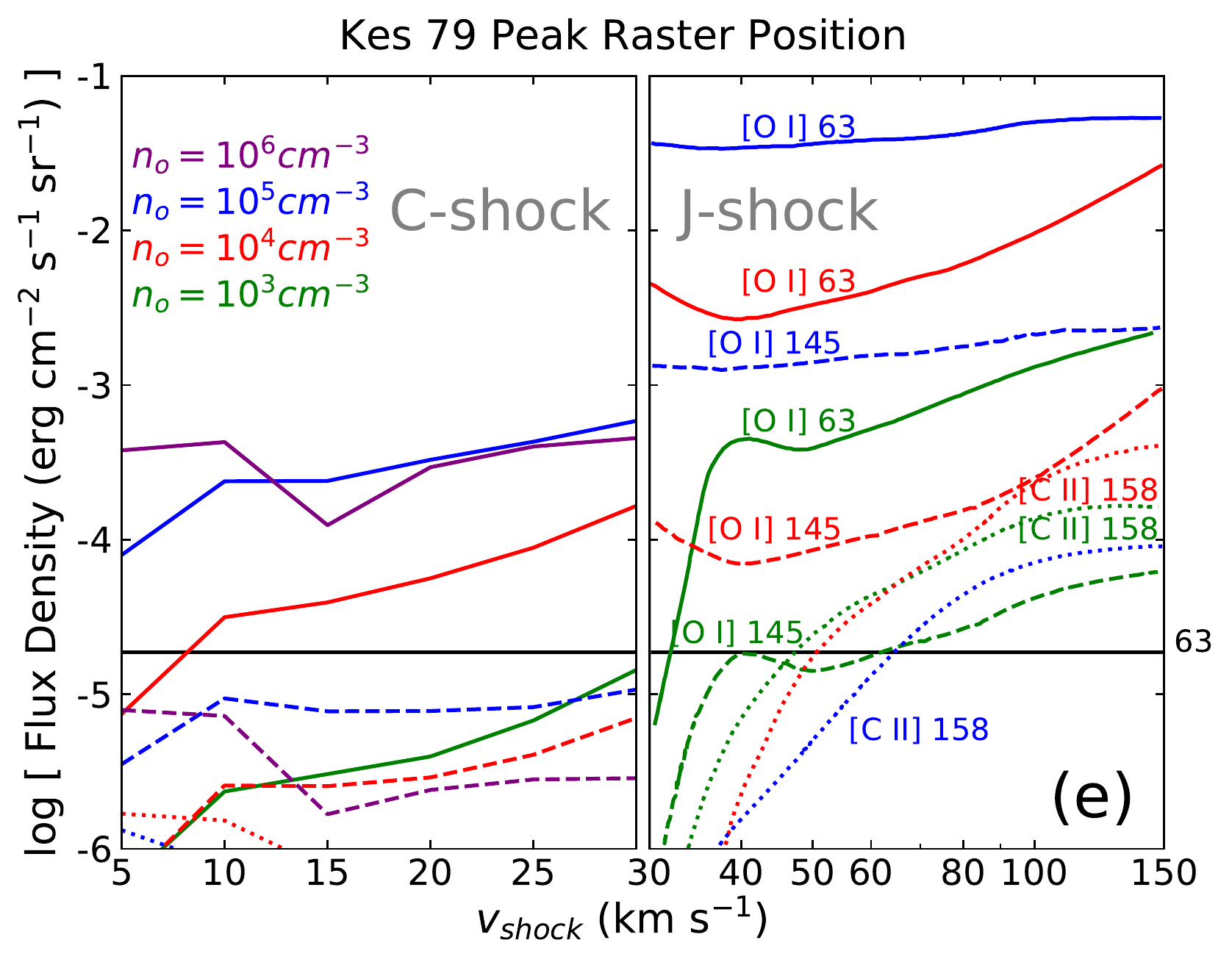}&
\includegraphics[angle=0,height=4.5truecm]{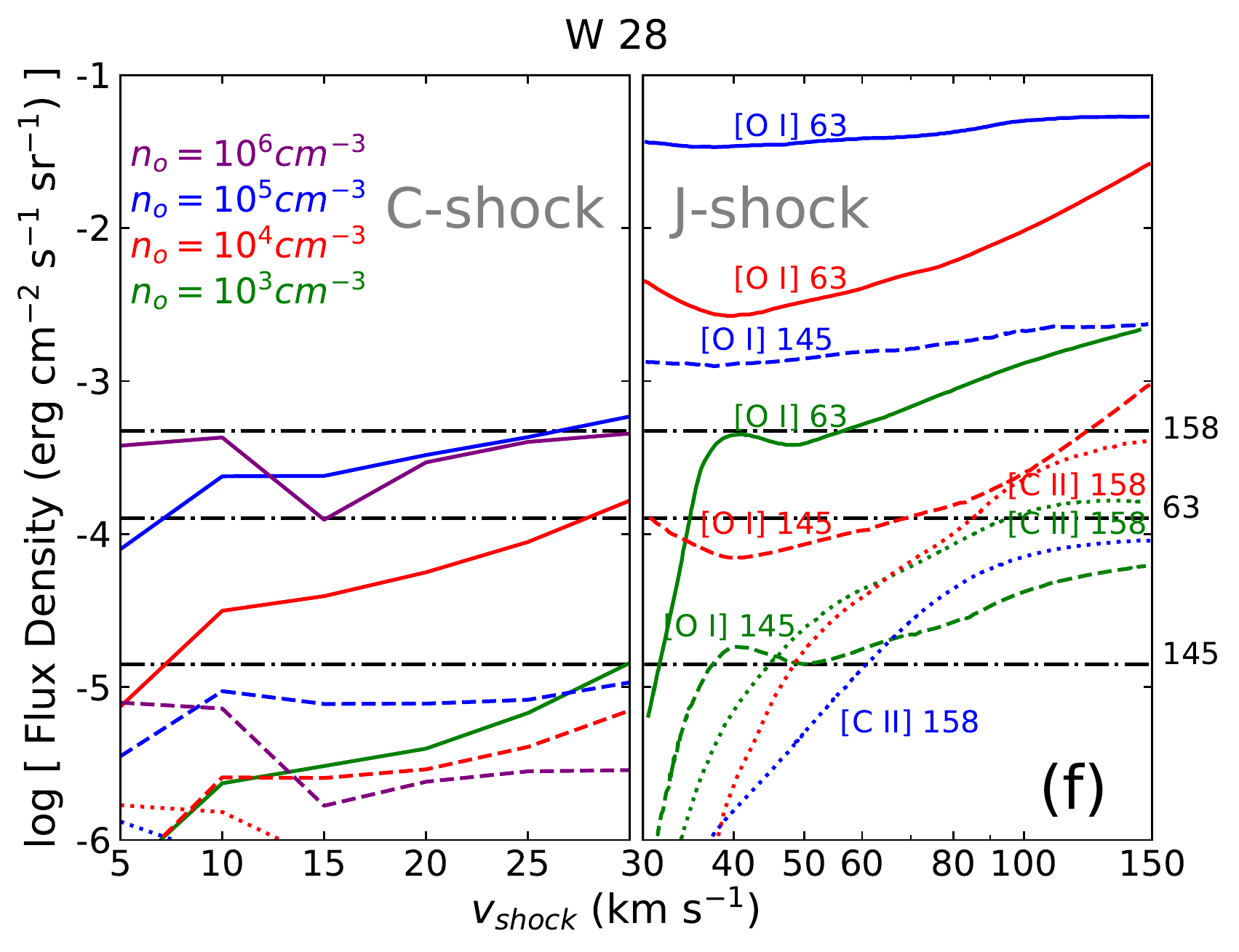}
\end{tabular}
\caption{In each panel, the predicted [O I] 63 \textmugreek{}m, [O I] 145 \textmugreek{}m, [C II] 158 \textmugreek{}m line brightness values for a slow C--shock (left side) and fast J--shock (right side) interacting with a molecular cloud. The C--shock line intensities were calculated using the Paris--Durham shock code, while the J--shock intensities are reproduced from Figure 7 of \citet{hollenbach89}. The observed background--subtracted line brightness values from $ISO$ LWS observations of W 28, IC 443, Kes 79, CTB 109, and RCW 103 are  overlaid as solid black horizontal lines. In the W 28 and RCW 103 panels, the total flux density and background--subtracted values are shown as dashed-dotted and solid lines, respectively. The W 28 observation did not include a background measurement.}  
\label{fig:hm89}
\end{figure*}

Below we briefly overview previously documented evidence of MC interactions for the relevant SNRs in our sample, and discuss our results for each remnant. Our estimated shock conditions based on the [O I] and [C II] emission for each of the MC--interacting SNRs are listed in Table \ref{tab:shocks} and the observed line fluxes are compared with shock models in Figure \ref{fig:hm89}.

{\bf IC 443 --} The MC interactions with IC 443 have been well studied. Shocked molecular gas was first detected toward the southern region of IC 443 with CO(1--0) observations that showed 20 km s\textsuperscript{--1} line widths \citep{denoyer79}. Since then, several different molecular species have been used as tracers to study the shock-cloud interactions in the SNR \citep{burton88,dickman92,vandishoeck93,snell05,reach19,kokusho20}. The importance of the FIR 63 \textmugreek{}m line of [O I] as a shock diagnostic tool was demonstrated using an observation from the Kuiper Airborne Observatory, where the  [O I] emission was found to correlate well with the near--IR H\textsubscript{2} emission \citep{burton90}. The [O I] 63 \textmugreek{}m line strength was also used to determine the nature of the shocked gas in the northeast \citep{rho01}.  \citet{rho01} observed IC 443 with $ISO$ LWS raster line spectra across the northeastern rim of the remnant and reported similar peak [O I] 63 \textmugreek{}m line brightness values ($\sim 5 \times 10^{-4}$ \ergsb) to ours. Based on strong 2MASS K\textsubscript{s}-band emission (associated with H\textsubscript{2} line emission) relative to the J and H bands, they suggested that the southern shell is dominated by a slow continuous C--shock, with a shock speed $v_s \sim$ 30 km s\textsuperscript{--1} and preshock hydrogen density, $n_o \sim$ 10\textsuperscript{4} cm\textsuperscript{-3} (as also concluded by \citet{cesarsky99}).

We compare our observed line brightness with C- and J- shock models.
We reproduce the J--shock model calculations of line fluxes based on Figure 7 of \citet{hollenbach89}\footnote{https://automeris.io/WebPlotDigitizer/}. For the C--shock models, we run the Paris--Durham code \citep{godard19}\footnote{https://ism.obspm.fr/shock.html} and produce the predicted line intensities depending on density and shock velocities. We overview the Paris--Durham code in Appendix B, and present our input parameters in Table \hyperlink{tabB1}{B.1}. Our C--shock models assume a young shock age of $10^3$ years, while the J--shock models assume the shock has reached a steady-state. Increasing the shock age of the C--shock models to $10^4$ years (nearly reaching a steady-state) enhances the predicted [O I] and [C II] intensities by a factor of a few, but does not affect our conclusions. Figure \ref{fig:hm89} shows the calculated line intensities for various preshock densities ($10^3 - 10^6$ cm\textsuperscript{-3}) and shock speeds (5 -- 150 km s\textsuperscript{--1}) for C--shock and J--shock models.

Our observed [O I] line brightness values in IC 443 are consistent with the lower density ($10^3$ cm\textsuperscript{-3}) fast (80 \kms) J--shock models from \citet{hollenbach89}. However, when we account for the large beam of the $ISO$ LWS the true surface brightness may be a factor of a few smaller than the observed brightness. In this case, we find that a C--shock solution for the higher density ($10^5$--$10^6$ cm\textsuperscript{-3}) and slow (30 or 10 \kms) shock models from the Paris--Durham code fits our measured [O I] line flux.  The observed [C II] 158 \textmugreek{}m line emission can be reproduced only in the fast J--shock models, suggesting that at least some of the observed emission is due to this type of shock. It is difficult to firmly constrain the nature of the shock-cloud interaction in IC 443, and the presence of multiple shocks in the SNR is often invoked (see \citet{snell05}  for a comprehensive review of possible shock-types). Recent IR line maps of IC 443 show that H\textsubscript{2} line emission (indicative of slower shocks) is mainly distributed along the southern ridge, while [Fe II] emission (representative of faster shocks) is present throughout the entire remnant \citep{kokusho20}. Thus, both slow shocks moving into a dense gas and fast shocks into a thinner gas may be present within the large aperture of the LWS in the southern portion of remnant.

Based on the allowed ranges of $n_o$ and $v_s$, we estimate the ram pressure to be $p_{ram} = \rho{}v_{s}^{2} \sim n_{o}v_{s}^{2} \sim 10^{-7}$ dyne cm\textsuperscript{--2} at the bright [O I] and [C II] regions. From X-ray measurements, we can estimate the thermal pressure of the remnant interior, $p_{th} = 2n_{e}kT_{e}$, where $k$ is the Boltzmann constant, and $n_e$ and $T_e$ are the postshock electron density and temperature, respectively. Using the $n_e$ and $kT_e$ values reported by \citet{troja06}, the thermal pressure is $\sim 10^{-9}$ dyne cm\textsuperscript{--2}. 
The ram pressure is more than two orders of magnitude higher than the thermal pressure, indicating that the shock is radiative. A similarly large difference in the ram and thermal pressure has been observed in other MC--interacting SNRs, W44 and 3C391 \citep{reach96}. \citet{reach96} suggested that the higher ram pressures in regions with bright [O I] and [C II] suggests the presence of dense molecular clumps, while the comparatively low thermal pressures may be from X-ray gas at the interior where the density is relatively low.

{\bf RCW 103 --} Observations of H\textsubscript{2} emission \citep{oliva89,burton93} and \textsuperscript{12}CO $J$ = 1--0  and HCO\textsuperscript{+} lines \citep{paron06} at the southern shell of RCW 103 indicate an interaction with an MC.  The [O I] 63 \textmugreek{}m line emission at the southern ridge has been observed with the IRS aboard $Spitzer$ \citep{andersen11}, and previously with the $ISO$ LWS \citep{oliva99}, with estimated intensity values of $3.2 \times 10^{-4}$ and $6.0 \times 10^{-4}$ ergs s\textsuperscript{--1} cm\textsuperscript{--2} sr\textsuperscript{--1}, respectively, roughly equivalent to our estimated value at region \# 3 (5.3 $\times 10^{-4}$ ergs s\textsuperscript{--1} cm\textsuperscript{--2} sr\textsuperscript{--1}). Based on the strong [O I] 63 \textmugreek{}m and 145 \textmugreek{}m intensities at the southern ridge, the shock conditions are similar to those in IC 443, and may be the result of either a fast shock -- low density or slow shock -- high density scenario. However, the level of [C II] 158 \textmugreek{}m emission in region \# 2 is comparable to region \#3, suggesting that most of the [C II] 158 \textmugreek{}m flux may originate from background. If the background--subtracted [C II] 158 \textmugreek{}m emission is negligible, this is more consistent with the C--shock model with an initial shock speed $v_s \sim$ 10 or 30 km s\textsuperscript{--1}, and preshock density $n_o = 10^5 - 10^6$ cm\textsuperscript{-3} (Figure \ref{fig:hm89}c). The background region around RCW 103 is complicated and the ambient [C II] 158 \textmugreek{}m emission may vary across the remnant. If the [C II] line flux is truly significantly above the background flux, then there must exist a fast J-type component, similar to the scenario discussed in the case of IC 443. The lower right panel of Figure \ref{fig:rcw103} shows the 2MASS J, H, and K\textsubscript{s} band images of RCW 103. Filamentary emission from all three bands is visible at the southern boundary, with the K\textsubscript{s} emission located to the exterior of J and H band emission. Region \# 3 more closely coincides with the J and H band emission, suggesting that the region may be dominated more by C--shock conditions -- a slower shock traversing into a dense medium.  The estimated ram and thermal pressures are roughly equal at the southern ridge ($\sim 10^7$ dyne cm\textsuperscript{--2}), possibly due to dense CSM knots.

{\bf Kes 79 --} Kes 79 is a several $10^3$ yr--old \citep{zhou16b,kuriki18} double-shelled SNR, with a central compact object \citep{giacani09}. A broad OH absorption feature \citep{green89} and bright HCO+ emission  have been observed at the east and southeast boundary of Kes 79 near the local standard of rest velocity $V_{LSR} \sim$ 105 km s\textsuperscript{--1}, consistent with the estimated distance to the SNR of 7.1 kpc based on the Galactic rotation curve \citep{green92}. The OH absorption and emission lines are detected both against Kes 79 and the associated pulsar B1849+00 \citep{stanimirovic03}. The detection of OH lines in Kes 79 suggests an interaction with dense clouds. The detection of nearby CO clouds supports such SNR-cloud interactions \citep{kilpatrick16}. \textcolor{black}{\citet{zhou16b} found a broadened \textsuperscript{12}CO $J$ = 3--2 line ($\Delta{}v = 12$ km s\textsuperscript{--1}) along the east, south, and west of the SNR, suggesting MC interactions with the SNR shock.
While the broad CO emission coincides with the southern and western radio shell (e.g., around 111 \kms\ component), the broad CO emission (e.g., at 103 \kms) in the east is a few tens of arcseconds to the west of the faint radio shell \citep[Fig. 6d of][]{zhou16b}. It indicates the interaction is at the front (around 103 \kms) of Kes 79 in the eastern direction.}
The $ISO$ pointings are slightly off the peak interacting sites (by up to a few tens of arcseconds) where the broad CO lines are detected. Thus, these $ISO$ raster observations may cover only parts of the shock--clouds interacting regions. This is probably why the \oif\ 63 \textmugreek{}m emission does not show strong enhancement at a particular position but still relatively bright (a factor of 2--3 brighter than those of background emission in IC 443 raster or those of CTB 109). We consider that the $ISO$ raster observations may not trace the strongest shock-cloud interacting regions.

{\bf CTB 109 --} CTB 109 is a $\sim 10^4$ yr-old \citep{sanchez-cruces18} SNR. It is well-known for its semi-circular shape as seen in radio and X-rays, which has been attributed to the presence of a giant MC at its western boundary \citep{heydari-malayeri81, tatematsu87, kothes02}.  However, a search for broad \textsuperscript{12}CO $J$ = 1--0 emission, which would indicate interactions between the SNR shock and the cloud, resulted in no detection \citep{tatematsu90}. Our measurements show that the [O I] 63 \textmugreek{}m emission is enhanced by a factor of $\sim$ 2 (compared with the background emission to the north) in two regions along the western boundary (3 and 4). The emission is relatively weak, and not clearly indicative of a strong shock interaction (see Figure \ref{fig:flux_v_radius}b).

{\bf W 28 --} W 28 is a few $10^4$ yr--old SNR \citep{velazquez02} which exhibits both center-filled X-ray emission and a shell-like structure in radio, and thus belongs to the mixed-morphology class of SNRs \citep{wootten81,rho02,pannuti17}. Several molecular lines from H\textsubscript{2}O, OH, and CO, and H\textsubscript{2} lines are detected in W 28, and are evidence of a shock passing through molecular gas \citep{reach98,arikawa99,reach00}.  We detect [O I] 63 and 145 \textmugreek{}m emission lines which appear which appears consistent with our C--shock model with a high preshock density ($10^4 - 10^6$ cm\textsuperscript{-3}) (see Figure \ref{fig:hm89}f). We detect a [C II] 158 \textmugreek{}m line that is exceptionally strong, which is consistent with a fast J--shock model (Table \ref{tab:shocks}). The difference in ram and thermal pressures is not as significant in W 28 (a factor of $\sim$ 2) as in IC 443 (a factor of $\sim$ 100). We note that, considering its projected position close to the Galactic plane ($b$ $\sim$ -- 0.1\degr{}), the contamination in our measured line flux from the strong Galactic background emission may be significant.

%


\begin{table*}
\caption{Shock Conditions in SNRs Interacting with Molecular Clouds}
\tabletypesize{\small}
\centering
\begin{tabular}{lllllllll}
\hline \hline
SNR & $L$\tablenotemark{*} & J--shock ($n_o$, $v_s$) & C--shock ($n_o$, $v_s$) & Pressure ($p_{ram}$, $p_{th}$) \\
  &  (L\textsubscript{\(\odot\)}) & (cm\textsuperscript{-3}, km s\textsuperscript{--1}) & (cm\textsuperscript{-3}, km s\textsuperscript{--1}) & (dyne cm\textsuperscript{--2}) &  \\
\hline
W 28 & 2  &  $10^3$, 30 -- 40 & {\bf $10^4 - 10^6$, $\sim$ 25} & $5\times{}10^{-8}$, $10^{-9}$ \\
IC 443 \#1 & 10 & {\bf $10^3$, $\sim$ 100} & $10^5 - 10^6$, $\sim$ 30 & $10^{-7}$, $10^{-9}$ \\
IC 443 \#2 &  5 & {\bf $10^3$, $\sim$ 60 } & $10^5 - 10^6$, $\sim$ 15 or $\sim$ 30 & | \\
IC 443 \#3 & 2 & {\bf $10^3$, 30 -- 40} & $10^5 - 10^6$, 10 -- 20 & | \\
IC 443 Raster (5) & 8 & {\bf  $10^3$, $\sim$ 80 } & $10^5 - 10^6$,  $\sim$10 or $\sim$30 & | \\
RCW 103 & 15 & $10^3$, 30 -- 50 & $10^5 - 10^6$, $\sim$10 or $\sim$30  &  $10^{-7}$, $10^{-7}$ \\
\hline \hline
\end{tabular}
\tablenotetext{*}{Luminosity values within the $ISO$ LWS beam. Thermal pressure calculated using values from \citet{rho02}, \citet{troja06}, and \citet{frank15} for W 28, IC 443, and RCW 103, respectively. The most favorable model is marked in bold.}  
\label{tab:shocks}
\end{table*}


%

\section{Continuum Emission in SNR G320.4--1.2 }\label{sec:continuum}


In the LWS spectra of several SNRs, we clearly detect the continuum emission in the FIR band (40 -- 200 \textmugreek{}m), originating from cold to warm dust grains. If the observed FIR continuum is associated with the SN ejecta gas, the dust emission spectrum is useful to constrain the dust production in these CC SN explosions. A $Herschel$ survey identified FIR emission from SNRs in the Galactic plane, and in a few cases, estimated the dust mass associated with SN ejecta \citep{chawner19, chawner20}. Here we focus on SNR G320.4--1.2, which was not covered by the $Herschel$ survey. The remnant shows a clearly broadened [O III] 88 \textmugreek{}m line (Section \ref{sec:ejecta}, Figure \ref{fig:velocity}) and a bright continuum (Figure \ref{fig:bbFits}). The presence of dust emission associated with high--velocity ejecta and the relatively reliable detector calibrations make this SNR the best candidate to analyze the continuum spectrum in order to constrain the dust mass produced by the SN.

To model the dust emission in this SNR, we fit the continuum spectrum (after the flux normalizations among the individual subdetectors were applied, see Section \ref{sec:obs}) with a blackbody (BB) model in the form of

\begin{equation}
   F_{\lambda} = \dfrac{\kappa_{\lambda}B_{\lambda}(T)M_{dust}}{D^{2}},
\end{equation}

where $F_{\lambda}$ is the flux at each wavelength, $\lambda$, $\kappa$ is the dust mass absorption coefficient, $B_{\lambda}$ is the Planck function at temperature $T$, and $M_{dust}$ and $D$ are the dust mass and distance to the source, respectively. We assumed the distance of $D$ =  5.2 kpc for G320.4-1.2 \citep{gaensler99}. The dust mass coefficient, $\kappa_\lambda$, is a function of wavelength:

\begin{equation}
   \kappa_{\lambda} = \kappa_{\lambda_0}\left(\dfrac{\lambda_0}{\lambda}\right)^{\beta}.
\end{equation}

We assume that the emission can be described by $\kappa_{\lambda_0}$ = 0.07 m\textsuperscript{2} kg\textsuperscript{-1} for $\lambda_0$ = 850 \textmugreek{}m \citep{james02}, and $\beta$
= 1.9, similar to bulk interstellar dust \citep{planck14}. 

Because our spectrum is not background--subtracted, the observed continuum includes dust emission components from both the SNR and ISM, and thus we apply a two--component BB model. The warm component represents emission associated with the SNR, and the cold component accounts for cold dust in the ISM background emission \citep{reach95,boulanger96,lagache98}. Before fitting, we excised the emission lines from  the spectrum, leaving only the underlying continuum.  We found that, while the observed spectrum is overall well-described by the best--fit model (see Figure \ref{fig:bbFits}:Upper), the fit is statistically poor ($\chi^{2}_{\nu}$ $\sim$ 200). 
We realized that the statistical uncertainties on a small fraction of the flux values in the spectrum ($\sim$ 5\% of the total data points) are exceptionally small (typically $\ll$ 1\% of the flux values),  which are more than an order of magnitude smaller than the average fractional uncertainties on the rest of the flux values ($\sim 15$\%). This small fraction of data points contributes the total $\chi^{2}_{\nu}$ almost entirely (by $\sim$ 98\%).  To mitigate this issue, we manually assign statistical uncertainties on these flux values, adopting the mean uncertainty values for all flux measurements in each subdetector. After this adjustment, the best--fit model significantly improves ($\chi^{2}_{\nu} \sim$ 4).

Our best--fit dust temperatures and masses are 46 -- 52 K and 0.03 -- 0.06 M\textsubscript{\(\odot\)} for the warm component, and 17 -- 20 K and 4 -- 15 M\textsubscript{\(\odot\)} for the cold component. The quoted ranges of these estimates represent the systematic uncertainties due to flux normalizations among subdetectors. The statistical uncertainties are small, on the order of $\lesssim$ 2\%. We note that when the SW1 subdetector is used as a baseline, the resulting fit is poorer, and the estimated dust mass is at least 2 times greater than those estimated with other subdetectors as the baseline. Because of this anomaly, we exclude it in our discussion hereafter. The warm component most likely represents the emission spectrum of the SN-created dust, for which we estimate the mass $\sim$ 0.03 -- 0.06 M\textsubscript{\(\odot\)}. Our best--fit temperature for the cold component is consistent with that for the background ISM dust temperature as measured in literature \citep{reach95,boulanger96,lagache98}.

\begin{figure}
\centering
\includegraphics[angle=0,width=0.49\textwidth]{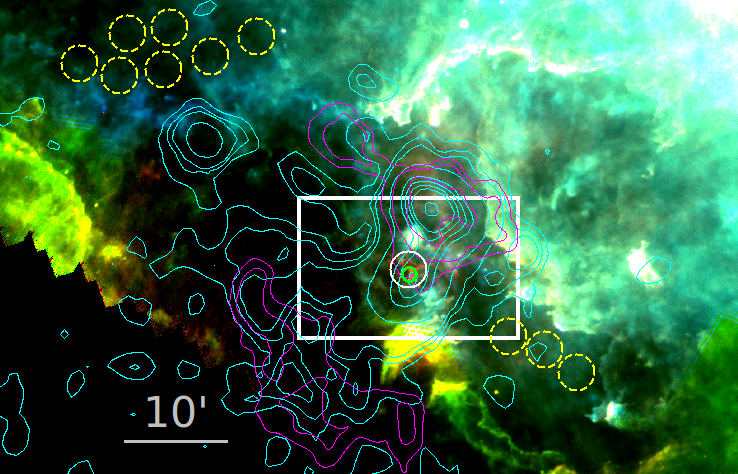} 
\includegraphics[angle=0,width=0.49\textwidth]{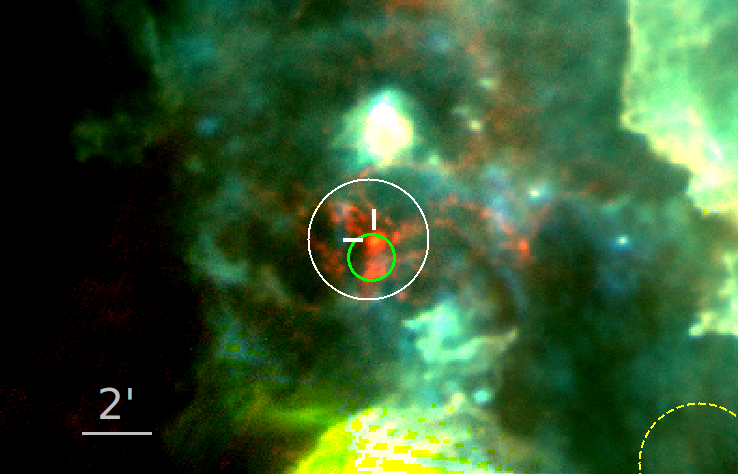} 
\caption{ $Herschel$ 3-color image (red: 70 \textmugreek{}m, green: 160 \textmugreek{}m, blue: 250 \textmugreek{}m) with the $ISO$ LWS beam shown as a green circle (0.67\arcmin{} diameter), while the solid white and dashed-yellow circles (1.7\arcmin{} diameter) shows the $Herschel$ photometric target and background regions, respectively. Contours are the same as in Figure \ref{fig:g320p4}. Upper: Zoomed-out view showing the overall extent of SNR G320.4--1.2. The white box shows the field of view of the lower panel.  Lower: Zoomed-in with contours removed to highlight the bright (in red) filamentary 70 \textmugreek{}m emission (from IRAS 15099) near the center of the SNR.  The white markers indicate the position of the compact source (IRS1).}
\label{fig:G320.4-1.2_images}
\end{figure}

\begin{figure}
\centering
\includegraphics[angle=0,width=0.49\textwidth]{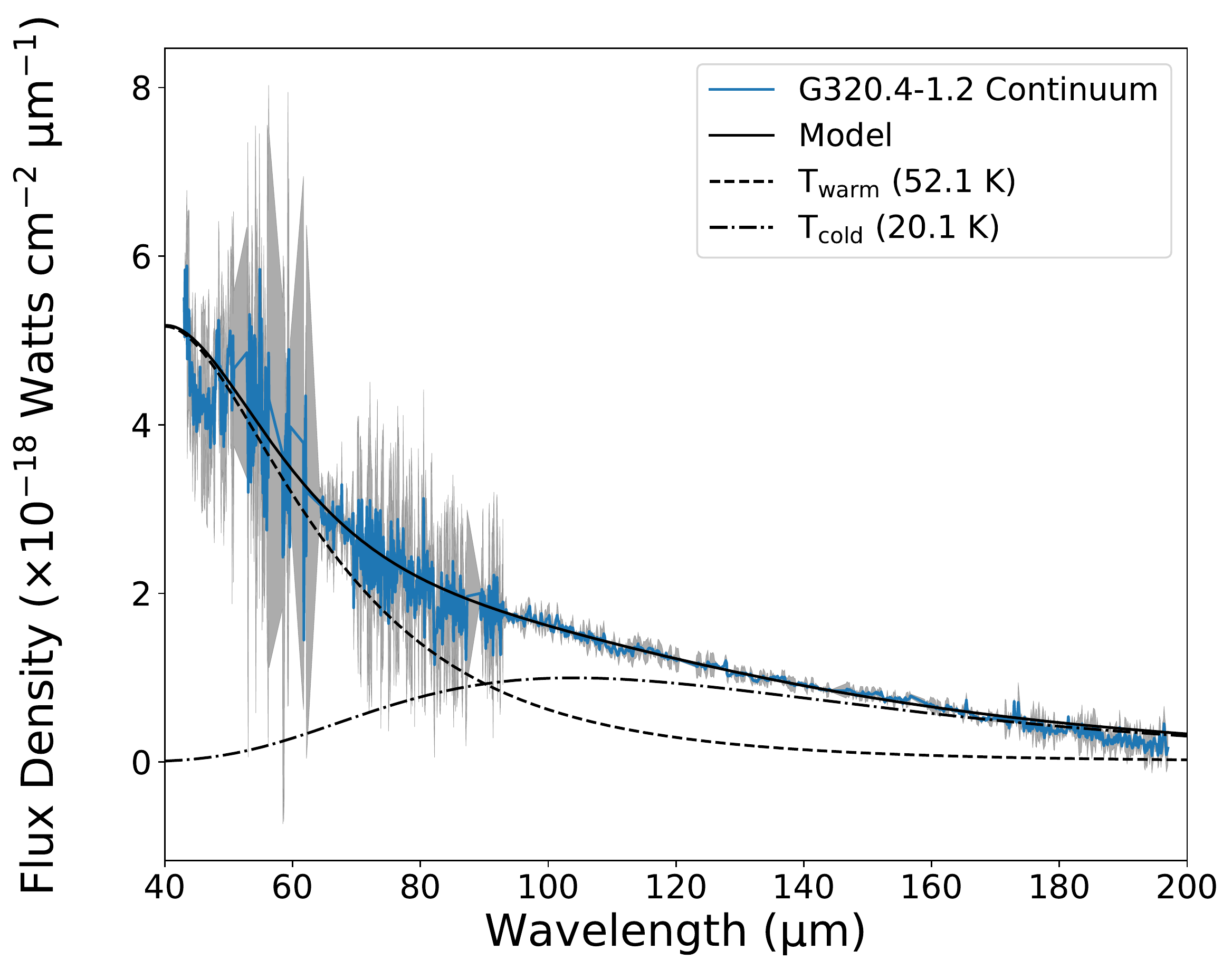} 
\includegraphics[angle=0,width=0.49\textwidth]{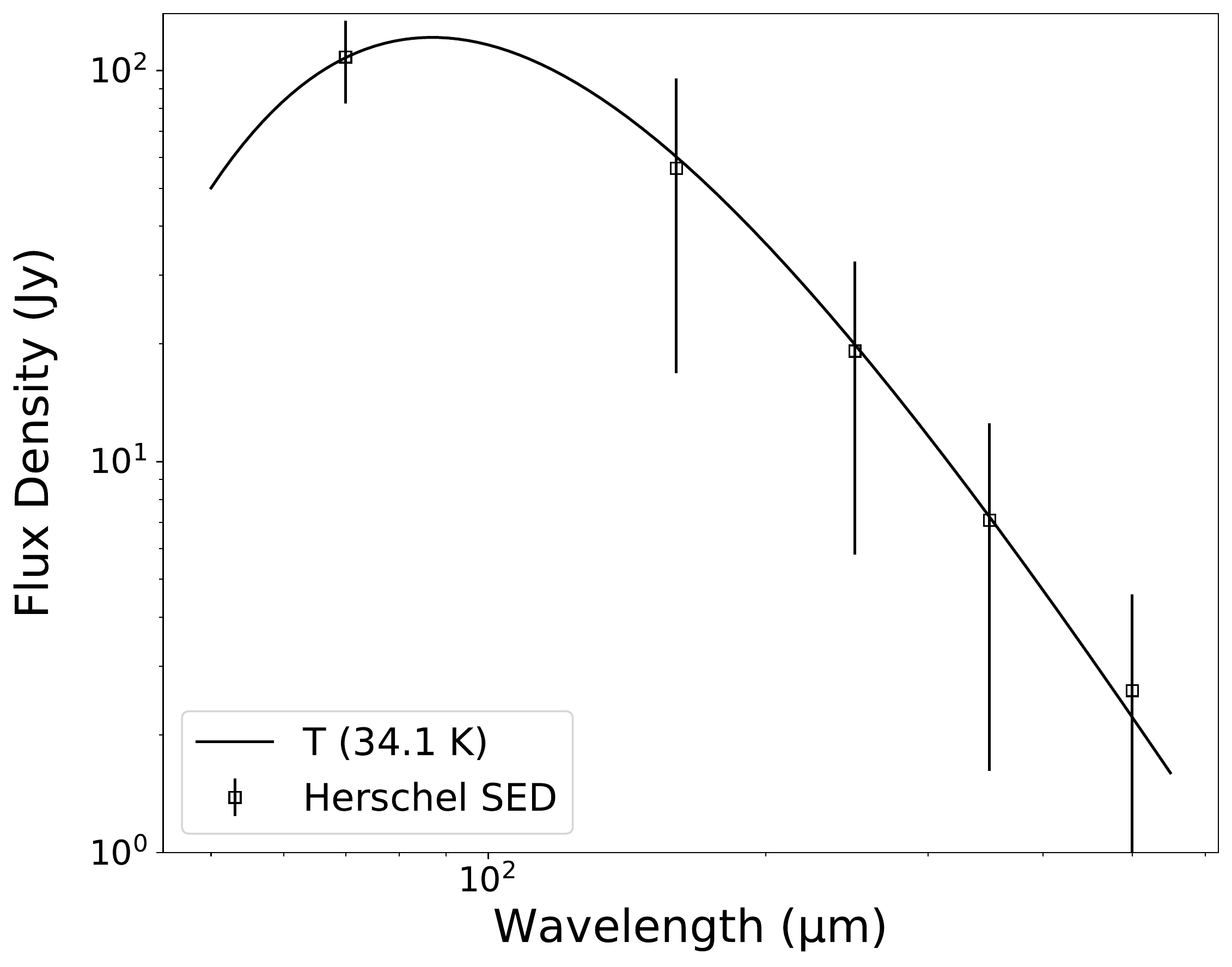}  
\caption{Upper: Continuum spectrum of G320.4--1.2 obtained with the $ISO$ LWS, with emission lines removed. Our best--fit two--component blackbody model is overlaid. A warm component ($T \sim$ 52 K) dominates the spectrum at  $\lambda \lesssim$ 90 \textmugreek{}m, while the cold component ($T \sim$ 20 K) dominates at $
\lambda$ $\gtrsim$ 90 \textmugreek{}m. Lower: $Herschel$ background--subtracted SED of the center of G320.4--1.2 fit with our blackbody model, $T \sim$ 34 K.  }
\label{fig:bbFits}
\end{figure}

Based on the $ISO$ LWS spectrum, G320.4--1.2 is a promising candidate that shows dust emission originating from grains formed in SN ejecta. We clearly detect the [O III] 88 \textmugreek{}m with a velocity dispersion of $\sim$ 2000 km s\textsuperscript{--1} indicative of high--velocity ejecta. We also clearly detect a warm (46 -- 52 K) dust continuum associated with this SN ejecta feature. The $ISO$ beam partially covers the intriguing MIR source, IRAS 15099--5856 (IRAS 15099 hereafter) located at ($\alpha{}_{2000}$, $\delta{}_{2000}$) = (15\textsuperscript{h}13\textsuperscript{m}56\fs{}32, --59\degr{}07\arcmin{}40\farcs{}9) \citep{koo11}. IRAS 15099 contains a bright central source, IRS1, surrounded by diffuse fainter emission, with spiral filaments extending out to a few arcminutes (see Figure \ref{fig:G320.4-1.2_images}. While the origin of IRAS 15099 is unclear, \citet{arendt91} suggested that the IR emission may result from dust heated by a hot plasma, or by the nearby O star, Muzzio 10.  
	
\citet{koo11} fit the $Spitzer$ IRS spectrum of IRS1 with several dust components, and found that most of the emission arises from crystalline olivine and amorphous silicate at $T$ = 54 -- 58 K. They found a total dust mass of 0.009$\pm$0.001 $d^{2}_{4kpc}$ M\textsubscript{\(\odot\)} for IRS1.  At a distance of 5.2 kpc, their estimated total dust mass becomes 0.015 M\textsubscript{\(\odot\)}. We note that, due to the large angular diameter of the beam, it is unclear if the broad [O III] line that we detect is associated with the compact central emission of IRS1, or the surrounding diffuse emission. Thus, our estimated dust mass range is in plausible agreement with that by \citep{koo11} roughly within a factor of $\sim$ 2. 

To estimate the background--subtracted dust mass of IRAS 15099, we use the archival $Herschel$ PACS \citep{poglitsch10} HPDP images at 70 \textmugreek{}m and
160 \textmugreek{}m, and SPIRE \citep{griffin10} Level 2.5 images at 250 \textmugreek{}m,
350 \textmugreek{}m, and 500 \textmugreek{}m images of SNR G320.4--1.2 (Observation IDs: 1342203291, 1342203292). Since our $Herschel$ extraction aperture diameter (206\arcsec{}) of our regions is larger than the beam size of the maps (FWHM $\leq$ 35.2\arcsec{}), we do not apply aperture corrections to the extracted spectral energy distribution (SED).

The emission associated with IRAS 15099 is most clearly visible at 70 \textmugreek{}m (see Figure \ref{fig:G320.4-1.2_images}). The brightest feature is located at the position of IRS1, with spiral filaments that reach out several arcminutes. Emission associated with IRS1 is clearly detected at 160 \textmugreek{}m. \citet{koo11} identified emission at 65 \textmugreek{}m and 90 \textmugreek{}m, but not at 140 \textmugreek{}m or 160 \textmugreek{}m with the $AKARI$ FIS (Far Infrared Surveyor). Thus, we identify clear 160 \textmugreek{}m emission from IRS1 for the first time. To measure the net flux from IRAS 15099, we chose a 206\arcsec{} diameter aperture centered at ($\alpha{}_{2000}$, $\delta{}_{2000}$) = (15\textsuperscript{h}13\textsuperscript{m}57\fs{}1980, -59\degr{}07\arcmin{}39\farcs{}78) which covers the bulk of the 70 \textmugreek{}m emission apparently associated with IRAS 15099. We chose ten background regions to the east and west of the remnant, exterior to the radio and X-ray shell. These regions are covered in the PACS and SPIRE images, and avoid the bright ISM emission to the north (see Figure \ref{fig:G320.4-1.2_images}). We estimate net flux densities of 108 $\pm$ 26.0 Jy, 56.3 $\pm$ 39.4 Jy, 19.2 $\pm$ 13.4 Jy, 7.1 $\pm$ 5.5 Jy, and 2.6 $\pm$ 2.0 Jy at 70 \textmugreek{}m, 160 \textmugreek{}m, 250 \textmugreek{}m, 350 \textmugreek{}m, and 500 \textmugreek{}m, respectively.  The flux from the 206\arcsec{} diameter aperture increases by about a factor of 3 -- 4 over our $ISO$-measured flux. Scaling the aperture sizes between $ISO$ and Herschel, we estimate a total dust mass of 0.1 -- 0.2 M\textsubscript{\(\odot\)} for our Herschel-measured fluxes. 

We fit the $Herschel$ background--subtracted SED with a one-component BB model and estimate a total dust mass of 0.66 $\pm$ 0.06 M\textsubscript{\(\odot\)} at a temperature of 34 $\pm$ 0.6 K. This dust mass is highly dependent on background subtraction, which varies significantly in the region around G320.4--1.2. The discrepancy in the dust mass estimates based on the $Herschel$ and $ISO$ data may be attributed to these uncertainties. Follow--up observations with higher spatial and spectral resolutions, (e.g., with $JWST$ and $SOFIA$), are required to verify if the dust and ejecta emission are clearly correlated. Our estimate for the dust mass associated with ejecta in G320.4--1.2 is in line with the growing number of previously published dust mass estimates in CC SNRs (see Table \ref{tab:snrinfo} for values and references).

%


\section{SNR Comparison with HII Regions}\label{sec:hii}

We compare the FIR atomic line emission of the SNRs with the HII regions in our sample, W51, NGC 6334, and G159.6--18.5 (Table \ref{tab:hiispec} and Figure \ref{fig:lineHIIFigs}). In W51 and NGC 6334, we detect a bright continuum and several emission lines, [O III] 52, 88 \textmugreek{}m, [N III] 57 \textmugreek{}m, [O I] 63, 145 \textmugreek{}m, [N II] 122 \textmugreek{}m, and [C II] 158 \textmugreek{}m. In G159.6--18.5, we detect only the 158 \textmugreek{}m [C II] line and weak continuum emission. We note that the [N III] 57 \textmugreek{}m line is detected only in the HII regions, and not in the SNRs, although \citet{reach00} found hints of a faint [N III] 57 \textmugreek{}m line flux in W 28, W44, and 3C391 ($<$ 0.25 $\times 10^{-4}$ 
ergs s\textsuperscript{--1} cm\textsuperscript{-1} sr\textsuperscript{--1}).  In an $ISO$ spectral survey of 45 compact HII regions, \citet{peeters02} reported clear [N III] 57 \textmugreek{}m detections on $\sim$ 60\% of the sample. The [N III] 57 \textmugreek{}m/[N II] 122 \textmugreek{}m ratio ranges from $\sim$ 1 -- 10, suggesting higher ionization states of nitrogen in the HII regions than in the SNRs. The lack of [N III] 57 \textmugreek{}m detections in our SNR sample compared with the HII region sample of \citet{peeters02} indicates that the presence of strong [N III] 57 \textmugreek{}m emission may be a discriminator between SNRs and HII regions.


%

\section{Summary}

We have studied the previously unpublished $ISO$ LWS spectra of 20 SNRs in the Galaxy and Magellanic Clouds. We detect a number of FIR atomic fine-structure lines, including [O III] at 52 \textmugreek{}m and 88 \textmugreek{}m, [O I] at 63 \textmugreek{}m and 145 \textmugreek{}m, [N II] at 122 \textmugreek{}m,  and [C II] at 158 \textmugreek{}m, as well as a bright continuum in several SNRs . We find that

\begin{itemize}

\item[--] In several SNRs, we find evidence for broad [O I], [O III], and [N II] lines suggesting emission from high--velocity SN ejecta. We present for the first time a firmly detected broad [O III] 88 \textmugreek{}m line in G320.4--1.2, indicative of fast-moving SN ejecta in this SNR. In G21.5--0.9, G29.7--0.3, and the Crab Nebula, we confirm previous detections of broad-line emission and present the lines that were not previously identified as broad in some cases.

\item[--] We find marginal evidence of broad lines in G54.1+0.3, RCW 103, E0102.2-7219, N132D, N49, and 0540--69.3. Follow--up high resolution imaging spectroscopy with modern instruments (e.g., $SOFIA$ or $JWST$) is required to verify the presence of high velocity ejecta in these SNRs.

\item[--] Based on the [O I] emission from IC 443 and RCW 103, we find regions along the southern ridge in both SNRs where the SNR shock is likely interacting with a dense molecular cloud. The strong [C II] 158 \textmugreek{}m emission in IC 443 indicates the presence of fast J-type shocks.

\item[--] We apply a two--component blackbody model fit to the LWS continuum of G320.4--1.2, and estimate a dust mass $\sim$ 0.1 -- 0.2 $M_\odot$ and temperature of $\sim$ 50 K associated with SN ejecta.

\item[--] We detect the [N III] 57 um line only in our HII region sample.  This distinctive spectral characteristic may serve as a discriminator in the identification of HII regions and SNRs.


\end{itemize}

\acknowledgements
The version of the \textit{ISO} data presented in this paper correspond to the Highly Processed Data Product (HPDP) set called ``Uniformly processed LWS L01 spectra'' by C. Lloyd, M. Lerate and T. Grundy, available for public use in the \textit{ISO} Data Archive \url{http://nida.esac.esa.int/nida-cl-web/}. The authors thank L.N. Tram, B. Godard, and A. Gusdorf for their helpful advice on running the Paris--Durham shock code. We also thank A. Herbst for compiling the initial dataset. We thank P. Zhou for providing the CO map of Kes 79. We thank G. Sloan, K. Kraemer, and W. Reach for discussions on $ISO$ data and IR emission from supernova remnants. This work was in part supported by NASA 2ADAP grant 80NSSC20K0449.

\newpage

\twocolumngrid

\appendix

\section{Archival LWS SNR Data Not Included in This Work}

In Table \hyperlink{tabA1}{A.1} we list the SNRs observed with the $ISO$ LWS that were not included in our sample.  In most cases, we did not choose these data because they were previously published. In other cases, the data quality was poor and suffered from inadequate sampling of emission line profiles.

\setcounter{table}{0} \renewcommand{\thetable}{A.\arabic{table}}
\startlongtable
\begin{deluxetable*}{lllllclllc}
\tablecaption{$ISO$ LWS SNR Observations Not in Our Sample} 
\tablehead{\colhead{Name} & \colhead{Target Name*} & \colhead{RA, DEC (J2000)} & \colhead{Date} & \colhead{t$_{on}$ (s)} & \colhead{Publications} }
\startdata
%
%
G6.4--0.1 & W28masA & 18\textsuperscript{h}01\textsuperscript{m}52\fs{}32 --23\degr{}19\arcmin{}25\farcs{}7 & 1997-03-17 & 1346 & 1,2,12 & \\
G10.0--00.3 & SGR1806-20 & 18\textsuperscript{h}08\textsuperscript{m}40\fs{}30 --20\degr{}24\arcmin{}40\farcs{}9 & 1997-03-23 & 730 & -- \\
G10.0--00.3 & SGR1806-20 Bkg & 18\textsuperscript{h}08\textsuperscript{m}23\fs{}72 --20\degr{}29\arcmin{}42\farcs{}1 & 1997-03-23 & 729 & -- \\
G31.9+0.0 & 3C391 & 18\textsuperscript{h}49\textsuperscript{m}20\fs{}95 --00\degr{}55\arcmin{}48\farcs{}4 & 1996-04-28 & 1414 &  1,2,10,12 \\ 
G31.9+0.0 & 3C391mol & 18\textsuperscript{h}49\textsuperscript{m}21\fs{}93 --00\degr{}57\arcmin{}22\farcs{}1 & 1997-03-10 & 1300 &  1,2,12 \\
G31.9+0.0 & 3C391mol (L04) & 18\textsuperscript{h}49\textsuperscript{m}21\fs{}93 --00\degr{}57\arcmin{}22\farcs{}1 & 1997-03-10 & 962 &  2,12 \\
G31.9+0.0 & 3C391H20 (L02) & 18\textsuperscript{h}49\textsuperscript{m}21\fs{}95 --00\degr{}57\arcmin{}22\farcs{}9 & 1997-10-26 & 1720 &  1,2,12 \\
G31.9+0.0 & 3C391cut2 (L02) & 18\textsuperscript{h}49\textsuperscript{m}26\fs{}80 --00\degr{}55\arcmin{}48\farcs{}7 & 1997-03-10 & 892 & 1,2,12 \\ 
G31.9+0.0 & 3C391cut3 (L02) & 18\textsuperscript{h}49\textsuperscript{m}31\fs{}79 --00\degr{}55\arcmin{}20\farcs{}4 & 1997-03-10 & 824 & 1,2,12 \\ 
G31.9+0.0 & 3C391cut4 (L02) & 18\textsuperscript{h}49\textsuperscript{m}17\fs{}32 --00\degr{}57\arcmin{}04\farcs{}4 & 1997-03-10 & 754 & 1,2,12 \\
G34.7--0.4 & W44 (L02) 1  & 18\textsuperscript{h}56\textsuperscript{m}32\fs{}41 +01\degr{}28\arcmin{}15\farcs{}6 & 1996-04-28 & 1680 & 1,2,10,11,12 \\ 
G34.7--0.4 & W44 (L02) 2  & 18\textsuperscript{h}56\textsuperscript{m}28\fs{}37 +01\degr{}29\arcmin{}59\farcs{}3 & 1996-04-28 & 1462 & 1,2,10,11,12 \\ 
G34.7--0.4 & W44cut3 (L02)  & 18\textsuperscript{h}56\textsuperscript{m}00\fs{}36 +01\degr{}12\arcmin{}49\farcs{}00 & 1997-03-10 & 1520 & 2,12 \\ 
G34.7--0.4 & W44masA  & 18\textsuperscript{h}56\textsuperscript{m}28\fs{}33s +01\degr{}29\arcmin{}58\farcs{}90 & 1997-03-10 & 962 & 2,12 \\ 
G34.7--0.4 & W44masA (L04)  & 18\textsuperscript{h}56\textsuperscript{m}28\fs{}33 +01\degr{}29\arcmin{}59\farcs{}00 & 1997-03-10 & 1300 & 2,12 \\ 
G35.2--1.7 & W48 (L04) & 19\textsuperscript{h}01\textsuperscript{m}47\fs{}60 +01\degr{}13\arcmin{}6\farcs{}10 & 1996-10-24 & 2312 & -- \\
G042.8+00.6 & SGR1900+14 & 19\textsuperscript{h}07\textsuperscript{m}15\fs{}22 +09\degr{}19\arcmin{}21\farcs{}6 & 1996-09-27 & 978 & -- \\
G042.8+00.6 & SGR1900+14 (Bkg) & 19\textsuperscript{h}07\textsuperscript{m}09\fs{}18 +09\degr{}15\arcmin{}21\farcs{}3 & 1997-03-29 & 1029 & -- \\
G111.7--2.1 & Cas A \#1 & 23\textsuperscript{h}23\textsuperscript{m}27\fs{}75 +58\degr{}48\arcmin{}47\farcs{}0 & 1996-06-24 & 1054 & 13 \\ 
G111.7--2.1 & Cas A \#2 & 23\textsuperscript{h}23\textsuperscript{m}26\fs{}74 +58\degr{}50\arcmin{}09\farcs{}9 & 1996-06-24 & 1052 & 13 \\ 
G111.7--2.1 & Cas A \#3 & 23\textsuperscript{h}23\textsuperscript{m}18\fs{}70 +58\degr{}47\arcmin{}30\farcs{}8 & 1996-06-24 & 1054 & 13 \\ 
G111.7--2.1 & Cas A \#4 & 23\textsuperscript{h}23\textsuperscript{m}53\fs{}94 +58\degr{}50\arcmin{}24\farcs{}3\farcs{} & 1996-06-24 & 1054 & 13 \\ 
G111.7--2.1 & Cas A \#5 & 23\textsuperscript{h}23\textsuperscript{m}38\fs{}33 +58\degr{}49\arcmin{}22\farcs{}1 & 1997-06-09 & 1612 & 13 \\ 
G111.7--2.1 & Cas A \#6 & 23\textsuperscript{h}23\textsuperscript{m}35\fs{}12 +58\degr{}47\arcmin{}38\farcs{}0 & 1997-06-09 & 1612 & 13 \\ 
G111.7--2.1 & Cas A \#7 & 23\textsuperscript{h}23\textsuperscript{m}16\fs{}47 +58\degr{}49\arcmin{}07\farcs{}8 & 1997-06-02 & 1614 & 13 \\ 
%
%
G184.6--05.8 & Crab Nebula \#1 & 05\textsuperscript{h}34\textsuperscript{m}34\fs{}27 +22\degr{}01\arcmin{}02\farcs{}4 & 1997-10-11 & 1124 & 14,17 \\
G184.6--05.8 & Crab Nebula \#2 & 05\textsuperscript{h}34\textsuperscript{m}31\fs{}94 +22\degr{}02\arcmin{}04\farcs{}8 & 1997-10-09 & 1126 & 14,17 \\
G184.6--05.8 & Crab Nebula \#3 & 05\textsuperscript{h}34\textsuperscript{m}29\fs{}31 +22\degr{}00\arcmin{}37\farcs{}0 & 1997-10-09 & 1124 & 14,17 \\
G184.6--05.8 & Crab Nebula \#4 & 05\textsuperscript{h}34\textsuperscript{m}34\fs{}19 +21\degr{}59\arcmin{}54\farcs{}7 & 1997-10-09 & 1630 & 14,17 \\
G184.6--05.8 & Crab Nebula (L02 Narrow) & 05\textsuperscript{h}34\textsuperscript{m}32\fs{}02 +22\degr{}02\arcmin{}04\farcs{}5 & 1998-02-19 & 1184 & -- \\
G184.6--05.8 & Crab Nebula (L02 Narrow) & 05\textsuperscript{h}34\textsuperscript{m}32\fs{}02 +22\degr{}02\arcmin{}04\farcs{}6 & 1998-04-04 & 1184 & -- \\
%
%
%
G189.1+3.0 & IC443cut & 06\textsuperscript{h}17\textsuperscript{m}49\fs{}69 +22\degr{}45\arcmin{}28\farcs{}4 & 1998-03-13 & 1218 & 3, 12 \\
G189.1+3.0 & IC443 -- C FP Scan (L04) 1. & 06\textsuperscript{h}17\textsuperscript{m}42\fs{}80 +22\degr{}21\arcmin{}37\farcs{}7 & 1998-03-13 & 1574 & -- \\
G189.1+3.0 & IC443 -- C FP Scan (L04) 2. & 06\textsuperscript{h}17\textsuperscript{m}07\fs{}64 +22\degr{}25\arcmin{}34\farcs{}7 & 1998-04-04 & 2230 & -- \\
%
%
G315.4--2.3 & RCW 86 & 14\textsuperscript{h}43\textsuperscript{m}04\fs{}20 --62\degr{}27\arcmin{}44\farcs{}4 & 1996-02-14 & 812 & -- \\ 
G332.4--0.4 & RCW103 H2+ion & 16\textsuperscript{h}17\textsuperscript{m}35\fs{}78 --51\degr{}06\arcmin{}18\farcs{}6 & 1996-02-23 & 4160 & 4,5 \\ 
J0047.2--7308 & SMC0045 (L04) & 00\textsuperscript{h}47\textsuperscript{m}17\fs{}06 --73\degr{}08\arcmin{}29\farcs{}3 & 1997-05-11 & 1694 & -- \\
J0048.5--7319 & SMC0046 0 (L04) & 00\textsuperscript{h}48\textsuperscript{m}25\fs{}43 --73\degr{}19\arcmin{}16\farcs{}3 & 1997-05-11 & 1020 & -- \\
J0059.4--7210 & SMC N66 & 00\textsuperscript{h}59\textsuperscript{m}8\fs{}02	--72\degr{}10\arcmin{}25\farcs{}97 & 1996-04-01 & 815 & 6,8,9 \\
J0535--6916 & SN 1987A  & 05\textsuperscript{h}35\textsuperscript{m}28\fs{}04 --69\degr{}16\arcmin{}11\farcs{}7 & 1998-02-04 & 3428 & 15,16 \\
J0537.8--6910 & LMC-N157B  & 05\textsuperscript{h}37\textsuperscript{m}51\fs{}76 --69\degr{}10\arcmin{}22\farcs{}0 & 1996-04-29 & 1054 & 8,9 \\ 
\enddata
\tablecomments{{\scriptsize{1. \citep{reach98} 2. \citep{reach00} 3. \citep{rho01} 4. \citep{oliva98} 5. \citep{oliva99} 6. \citep{liseau06} 7. \citep{he05} 8. \citep{vermeij02b} 9. \citep{vermeij02} 10. \citep{reach96} 11. \citep{cox99} 12. \citep{yuan11} 13. \citep{docenko10} 14. \citep{green04} 15. \citep{lundqvist99} 16. \citep{sollerman02} 17. \citep{gomez12} }}}
\tablenotetext{*}{\hypertarget{tabA1}{The} observing mode is in parentheses. L01--L02: Grating, L03--L04: Fabry-P\'erot.}
\end{deluxetable*}

\section{Paris--Durham Models}

The line intensities of the C--shock models presented in Figure \ref{fig:hm89} are computed using the Paris--Durham shock code, which simulates the progression of a shock wave through a gaseous medium \citep{godard19}. By incorporating relevant heating, cooling, and grain processes, as well as a network of chemical reactions, the code finds the kinematic, thermodynamic, and chemical properties of the shocked layer.  These models assume that the shock is irradiated by an external radiation field.  Thus, the model contains a pre-shock radiative buffer created by the coupling of the radiation field with the inferred molecular cloud.

To account for an irradiated shock, we run the code in three stages.  In the first stage, the code calculates the chemical and thermal conditions at the border of the radiative buffer and the diffuse interstellar medium. In the second stage, the conditions inside the buffer are evolved until the desired extinction ($A_v$) value is reached. Here, we choose the standard $A_v$ value of $10^{-1}$ \citep{godard19}. In the final stage, the propagation of the shock is computed. In this stage, we performed multiple runs to calculate the model line fluxes for a range of shock velocities ($v_s$ = 5 -- 30 km s\textsuperscript{--1}) and pre-shock densities ($n_o =$ $10^3$ -- $10^6$ cm\textsuperscript{-3}).  We compared the resulting [O I] 63 and 145 \textmugreek{}m and [C II] 158 \textmugreek{}m line intensities for each model with our measured values (see Figure \ref{fig:hm89}). The relevant input parameters are listed in Table \hyperlink{tabB1}{B.1}.  



%
\setcounter{table}{0} \renewcommand{\thetable}{B.\arabic{table}}

\begin{table}[!h]

\caption{Paris--Durham Input Parameters }
\label{tab:pdinputs}
\tabletypesize{\small}
\centering
\begin{tabular}{lll}
\hline \hline
Parameter & Value \\
\hline
$n_H$, pre-shock proton density & 10\textsuperscript{3} -- 10\textsuperscript{6}  cm\textsuperscript{-3} \\
$G_o$, radiation field scaling factor & 1  \\
$A^o_V$, pre-shock visual extinction & 0.1  \\
$V_s$, shock speed & 5 -- 30 km s\textsuperscript{--1}   \\
$b$, magnetic field parameter & 1\textsuperscript{a}\\
$\zeta_{H_2}$, H\textsubscript{2} cosmic ray ionization rate & $3 \times 10^{-17}$ s\textsuperscript{--1}\\
$u_{turb}$, turbulent velocity & 1 km s\textsuperscript{--1}\\
$timeJ$, shock age & $10^3 - 10^4$ yr \\
\hline \hline
\end{tabular}
\tablenotetext{a}{\hypertarget{tabB1}{Dimensionless} value of the initial magnetic field transverse to the plane of the shock, $b = B_o[\mu{}G]/(n_H)^{1/2}$. }
\end{table}

\clearpage

\bibliography{msrefsall}{}

\end{document}